\RequirePackage{tikz}
\documentclass[pdflatex,sn-basic]{sn-jnl}

\usepackage[utf8]{inputenc}
\usepackage{caption}
\usepackage{subcaption}
\usepackage{graphicx}
\usepackage{tikz}
\usepackage{amsmath}
\usepackage{amsfonts}
\usetikzlibrary{quantikz}
\usepackage{adjustbox}

\makeatletter 
\def\@acknow{}%
\long\def\EarlyAcknow#1 \par{%
\def\@acknow{\abstractfont\abstracthead*{Acknowledgments}
#1\par}}%

\def\printabstract{\ifx\@acknow\empty\else\@acknow\fi\par%
    \ifx\@abstract\empty\else\@abstract\fi\par}

\makeatother

\jyear{2023}%

\theoremstyle{thmstyleone}%
\theoremstyle{thmstyletwo}%

\theoremstyle{thmstylethree}%

\begin{document}

\title[Hybrid Genetic Optimisation for Quantum Feature Map Design]{Hybrid Genetic Optimisation for Quantum Feature Map Design}


\author*[1,2]{\fnm{Rowan} \sur{Pellow-Jarman}}\email{rowanpellow@gmail.com}

\author[1,3]{\fnm{Anban} \sur{Pillay}}\email{pillayw4@ukzn.ac.za}

\author[1,2]{\fnm{Ilya} \sur{Sinayskiy}}\email{sinayskiy@ukzn.ac.za}

\author[4,2]{\fnm{Francesco} \sur{Petruccione}}\email{petruccione@sun.ac.za}

\affil*[1]{\orgname{University of KwaZulu-Natal}}

\affil[2]{\orgname{National Institute for Theoretical and Computational Science (NITheCS)}}

\affil[3]{\orgname{Centre for Artificial Intelligence Research (CAIR)}}

\affil[4]{\orgname{Stellenbosch University}}


\EarlyAcknow{We would like to thank the National Institute of Theoretical and Computational Sciences (NITheCS) and the Center for Artificial Intelligence Research (CAIR) for funding this research.}

\abstract{Kernel methods are an import class of techniques in machine learning.  To be
  effective, good feature maps are crucial for  mapping non-linearly separable
  input data into a higher dimensional (feature) space, thus allowing the data to be linearly separable in feature space.
Previous work has shown that quantum feature map design can be automated for a
given dataset using NSGA-II, a genetic algorithm, while both
minimizing circuit size and maximizing classification accuracy. However, the
evaluation of the accuracy achieved by a candidate feature map is costly. 
In this work, we demonstrate the suitability of kernel-target alignment as a
substitute for accuracy in genetic algorithm based quantum feature map design.
Kernel-target alignment is faster to evaluate than accuracy and doesn't require
some data points to be reserved for its evaluation.  To further accelerate
the evaluation of  genetic fitness, we provide a method to approximate
kernel-target alignment.  To improve kernel-target alignment and root mean
squared error,  the final trainable parameters of the generated circuits are further trained using COBYLA 
 to determine whether a hybrid approach applying conventional circuit parameter
 training can easily complement the genetic structure optimization approach. A
 total of  eight new approaches are compared to the original across nine varied binary
classification problems from the UCI machine learning repository, showing that 
kernel-target alignment and its approximation produce feature map circuits
enabling comparable accuracy to the previous work but with larger margins on 
training data (in excess of 20\% larger)
that improve further with circuit parameter training.}




\maketitle

\section{Introduction}\label{sec1}

Quantum computers leverage quantum properties such as entanglement, and
promise the potential of a speed advantage
over classical algorithms when applied to specialized problems. Some algorithms
such as Shor’s algorithm for factorization \citep{shor} and Grover’s search
algorithm \citep{grover} have been shown theoretically to outperform all known
classical algorithms applied to the same tasks. Quantum algorithm design is made
difficult by the unintuitive nature of quantum entanglement which must be used
effectively to achieve an advantage over classical algorithms.
Quantum machine learning seeks to apply quantum computation to machine learning
tasks to achieve a quantum advantage over classical machine learning.

Quantum machine learning and classical machine learning show promise for automating many practical
tasks that would otherwise require human intelligence, including disease diagnosis
\citep{disease_diagnosis_article}, natural language processing
\citep{nlp_article}, and image classification \citep{image_recognition_overview}.
Machine learning algorithms are designed to learn functions from data
\citep{goodfellow2016deep}. These algorithms can be separated into three
categories: supervised learning, unsupervised learning, and reinforcement
learning. Supervised learning algorithms make use of prelabeled training data
while unsupervised learning algorithms do not
\citep{supervised_vs_unsupervised_distinction}.  In reinforcement learning,
agents learn behaviour by interacting with an environment that provides rewards
and punishments to guide them.

Kernel methods are an important class of techniques in both classical and
quantum machine learning.  The Support Vector Machine (SVM) \citep{initial_svm} is an important
classical supervised kernel method, due to its theoretical relations to other learning models
\citep{neural_tangent_kernel, SQMLMAKM} and results regarding its
generalisation ability \citep{Vapnik1998Book, VCBound}. It is built around an
optimization algorithm for finding an optimal linear hyperplane that separates
data points into two classes. The hyperplane is selected to maximize the minimum
margin of any point in the training dataset, where the margin of a point is
defined as the distance of the point from the separating hyperplane. Larger
margin sizes have been theoretically linked to improved generalisation
performance \citep{Vapnik1998Book, VCBound}. Non-linear decision boundaries can
be achieved by mapping non-linear data to a higher dimensional feature
space. The mapping function used is called a feature map, and the range of a
feature map is called a feature space. By use of a technique known as 
the \textit{``kernel trick”}, the decision boundary optimization
problem can be reformulated in terms of a kernel function that computes the
similarity of a pair of data points in the feature space \citep{initial_svm}.
This obviates the need to explicitly compute feature map outputs for data
points, so long as the corresponding kernel function can be computed.

The feature map must be carefully selected for effective separation of the data.
For any kernel function and labelled training set combination, a quantity known
as the kernel-target alignment of the kernel can be calculated. This indicates the
degree of agreement between the kernel function and a hypothetical oracle kernel induced
by the training labels that is well suited to the training data
\citep{on_kernel_target_alignment}. A high kernel-target alignment has been shown
in other works to correlate with improved classification performance
\citep{on_kernel_target_alignment} and it has been proposed for use as a metric
for selecting suitable kernels for a dataset in classification problems
\citep{on_kernel_target_alignment, training_kernel_target_alignment}.

The QSVM algorithm enhances the SVM algorithm by implementing the feature map function as a quantum circuit (see section \ref{sec:QSVM_explanation}).
While quantum feature map circuits are parameterized in the feature values of a
single data point, they can also contain additional trainable parameters which
can be optimized to improve the suitability of the kernel circuit for a specific
dataset \citep{training_kernel_target_alignment}. A quantum circuit containing
trainable parameters is called an \textit{ansatz}. Prior work has used classical
optimizers on trainable parameterized quantum kernels to maximize their
kernel-target alignment, which resulted in positive effects on the
classification accuracy of the resulting SVM models
\citep{training_kernel_target_alignment}. 

 Quantum feature map circuits of fixed structure that make use of trainable
 parameter values are reported in \cite{training_kernel_target_alignment} .  The trainable
parameter values are optimized using Stochastic Gradient Ascent to maximize the
kernel-target alignment of the corresponding kernel functions. This was
performed to test whether kernel-target alignment optimization could improve the
performance of a fixed-structure quantum feature map on a given dataset.
Increased kernel-target alignment had previously been shown in
\cite{on_kernel_target_alignment} to correlate with improved classification
ability. The technique described can be applied either to tailor an existing feature map
to a dataset or to fully generate a feature map for a dataset from a feature map
ansatz that has a predetermined circuit structure and parameter placement. The
work made use of both classical noise-free simulations of quantum computers and
real NISQ computers to run quantum circuits, reporting improvements in
classification accuracy after kernel-target alignment maximization
\citep{training_kernel_target_alignment}. 

A second model training metric applicable to quantum kernel classifiers is a
classifier's root mean squared error (RMSE). This is often optimized to train
parameterized models for classification tasks. It can be better suited to
training than direct evaluation of accuracy since it accounts for the magnitudes
of misclassification errors rather than simply the number of errors that occur.
Accuracy can be too insensitive to circuit parameter changes to show when a
circuit has slightly improved if a data set is not sufficiently large.


Another approach to optimizing the choice of kernel circuit for a dataset is to
optimize the selection of circuit gates used in the feature map in addition to
the values of trainable circuit parameters. This approach has been applied to
optimizing circuits applied to other problems
\citep{rotosolve_rotoselect_circuit_structure_optimization}. Genetically
inspired algorithms have also been applied to circuit structure optimization
since they are capable of combinatorial optimization
\citep{evolving_quantum_circuits_using_genetic_algorithm,
  quantum_circuit_design_genetic_programming,
  genetic_algorithm_quantum_circuit_compilation}. 

\cite{Altares_L_pez_2021} detailed the implementation of a genetic algorithm for
automated feature map circuit design for use with QSVM classifiers that both
maximizes classification accuracy and minimizes circuit size. The optimization
was performed using a variation of genetic algorithm named NSGA-II \citep{nsga2}
which customizes the usual genetic selection and fitness evaluation operations
of the genetic algorithm (as outlined in section \ref{sec:GA_explanation}) to
evaluate multiple fitness functions and preferentially select non-dominated
solutions for crossover. 

In a minimization problem with two fitness functions, a solution $s$ with
fitness values $(a, b)$ is considered non-dominated with respect to another set
of $n$ solutions with fitness
values $\{(f_i, g_i) \vert  i \in \{1, 2, ..., n\}\}$ if and only
if $\forall i \in \{1, 2, ..., n\},  (f_i < a)\implies(g_i > b)$, i.e. if and
only if there are no solutions in the set with all fitness values superior to
the corresponding fitness values of $s$. NSGA-II also makes use of elitism to
guarantee the preservation of the solutions that best optimize at least 1 of the
individual fitness functions. 

In order to apply a genetic algorithm to quantum feature map design, the work
also defined a binary string representation for encoding feature map circuits.
The encoding strategy can be found in \cite{Altares_L_pez_2021} and is
summarised here. 

\begin{table}
\centering
\begin{tabular}{|c|c|}
\hline
Bits & Gate \\
\hline
000 & H \\
001 & CNOT \\
010 & I \\
011 & Rx \\
100 & Rz \\
101 & I \\
110 & I \\
111 & Ry \\
\hline
\end{tabular}
\begin{tabular}{|c|c|}
\hline
Bits & Parameter\\
\hline
00 & $\pi$ \\
01 & $\pi/2$ \\
10 & $\pi/4$ \\
11 & $\pi/8$ \\
& \\
& \\
& \\
& \\
\hline
\end{tabular}
\caption{Mapping used on each consecutive 5-bit sequence of bits encoding a feature map gate to determine the gate type and a proportionality parameter value used in the case of parameterised gates. The available gates for the encoding to select from are Hadamard (H), CNOT, identity (I), and parameterised rotations around the X, Y, and Z axis of the Bloch sphere which are used to encode data point feature values into the circuit. We define the Rx($\theta$) gate as cos$(\theta / 2)$I$ - i$sin$(\theta/2)$X, the Ry($\theta$) gate as cos$(\theta / 2)$I$ - i$sin$(\theta/2)Y$, and the Rz($\theta$) gate as cos$(\theta / 2)$I$ - i$sin$(\theta/2)Z$. If a parameterised rotation gate $R_a$ around axis $a$ is selected, the parameter selected by the last two of the five gate bits will be used to selected a proportionality parameter $p$ . When the gate $R_a$ is applied to encode a feature value $x_i$, the gate applied will be $R_a(px_i)$. In the case of a CNOT gate being selected and this gate being applied to qubit $i$, qubit $i$ will be used as the control qubit and qubit $(i+1)\mod M$ will be used as the target qubit.}
\label{table:gate_bits_mapping}
\end{table}

Each circuit gate is encoded in a sequence of 5 bits, the first three of which encode the type of gate applied and the last two of which encode a proportionality parameter for use in the case that a parameterised rotation gate was selected by the first three bits. The mapping of bits to gates and proportionality parameters is shown in Table \ref{table:gate_bits_mapping}.

Hyperparameters $M$ and $N$ which designate the maximum number of qubits and maximum number of gate layers respectively must be chosen before applying the genetic algorithm and are fixed for the duration of the genetic optimization. A single solution is represented as a bit string of length $5MN$, which holds the concatenation of the encodings of the individual gates in the feature map. 

The gates in the encoding are applied successively with the target qubit and feature value to potentially encode being selected in a round-robin fashion. Stated explicitly, for each consecutive group of 5 gate bits, the $i$th gate description will be applied to qubit $i\mod M$ and will encode feature $i\mod N$ if it performs a parameterised rotation. This encoding strategy was selected for simplicity, although other strategies could also feasibly be attempted.

Two fitness functions were optimized in the work: accuracy on a test set was maximized and a weighted size metric was simultaneously minimized. The unweighted size metric \textit{SM} was calculated in terms of the number of qubits $M$, the number of single qubit gates $N_{\text{local}}$, and the number of entangling gates $N_{\text{CNOT}}$, by the expression
\[
\text{SM} = \frac{N_{\text{local}} + 2N_{\text{CNOT}}}{M}.
\]
The weighted size metric \textit{WS} supplied to the genetic algorithm was given by the expression
\[
\text{WS} = \text{SM} + \text{SM} \cdot \text{accuracy}^2.
\]

The work was able to demonstrate the effectiveness of using NSGA-II with the devised feature map binary string encoding strategy, accuracy maximization, and weighted size minimization to automatically produce quantum feature map circuits for QSVM classification using only a dataset and a few hyperparameters as input. The generated circuits were also experimentally shown to generalise to unseen data. In addition to using few qubits and quantum gates, the circuits produced by the approach were observed to make little to no use of entanglement, meaning they could be efficiently simulated classically and the approach could constitute a quantum-inspired classical machine learning algorithm.

Some attempts have been made to enhance the genetic optimization and compare the approach to others.
The work done in \cite{single_objective_genetic_extension_work} is based on the algorithm put forward in \cite{Altares_L_pez_2021}. They used a modified encoding scheme which encoded the proportionality parameter values using three bits instead of only two, doubling the number of encodable parameter values. A restricted choice of parameter values was one of the potential limitations of the algorithm designed in \cite{Altares_L_pez_2021}. The algorithm was also further modified to optimize gate cost and classification accuracy in a single objective expression, using a single objective genetic algorithm instead of a multi-objective genetic algorithm such as NSGA-II. Notably, this introduced a new hyperparameter used to weight the focus of the optimization between circuit size and accuracy, which is not done with NSGA-II.

The feature map generated by the genetic algorithm was compared with two other choices of ansatz: a hardware efficient ansatz proposed in \cite{he_ansatz}, and a unitary decomposition ansatz proposed in \cite{unitary_decomposition_ansatz}. Each ansatz was trained with COBYLA \citep{cobyla_book}, directly evaluating classification accuracy as a cost function. It was found that the feature map circuits generated by the genetic algorithm variation used in the work performed similarly to the hardware efficient ansatz, depending on the circuit depth hyperparameter selected when generating the hardware efficient ansatz. However, both were beaten by the unitary decomposition ansatz in accuracy, which achieved the highest accuracy at the cost of having a large, fixed size.

Our work investigates kernel-target alignment as a metric for automating
quantum feature map design for the Quantum-Enhanced Support Vector Machine
(QSVM) algorithm \citep{QMLFHS, QSVMBDC}. 
Our work has two goals:  investigate the suitability of using alternative cost
functions to accuracy in the genetic optimization process described in
\cite{Altares_L_pez_2021} and secondly,  investigate whether the problem of limited circuit parameter choices in the genetic algorithm can be addressed by a hybrid process of genetic and circuit parameter training.

We address the first goal by evaluating two alternative  cost functions to
the accuracy metric in \cite{Altares_L_pez_2021} : firstly,
kernel-target alignment for the genetic optimization step   and secondly, a
heuristic estimation of the kernel-target alignment performing a fraction of the
kernel evaluations. For the second goal, a hybrid method involving further
optimising the final choice of trainable circuit parameter values after the
genetic algorithm terminates for each of the above  approaches is evaluated.
This final optimization uses COBYLA \citep{cobyla_book} to maximize either
kernel-target alignment or RMSE. The new approaches are compared to the original
across several binary classification problems of varying difficulties. We show
that even though the kernel-target alignment metric is less computationally
expensive to compute in terms of quantum kernel evaluations and avoids the
training of an SVM classifier, the performance of the constructed classifiers
is comparable to the original approach and often achieves a better margin
distribution on training data.  It has been demonstrated Theoretically that   increased margin
sizes indicates better generalisation ability \citep{Vapnik1998Book, VCBound}.
The kernel-target alignment approximation heuristic is shown to perform
marginally worse than exact kernel-target alignment optimization but at a
fraction of the computational cost. The hybrid approaches are shown to improve
margin sizes over the original. The original approach is also shown to sometimes
overfit to the test data used to evaluate its accuracy metric, particularly on
difficult problems.

In the following section, we give a more detailed explanation of the background topics involved in understanding this work and related works, and explain our experimental setup. This is followed by a section covering our findings and interpretations. In the final section we give an overview of the contributions made and suggest ideas for further research.

\section{Methods}\label{sec2}
\subsection{Binary classifiers using quantum kernels}
\subsubsection{Support Vector Machine (SVM)}\label{sec:SVM_explanation}
The SVM algorithm is a classical supervised machine learning algorithm for binary classification problems that works by finding an optimal separating hyperplane between two classes of data points. The SVM algorithm is applicable when the data points can be represented by real-valued feature vectors. For simplifying definitions, the class labels are usually replaced with positive and negative one.

The \textit{margin} of a single data point is defined as the distance from the data point to the SVM's chosen hyperplane. The margin of an SVM classifier refers to the minimum of the margins of the data points. The data points with minimum margin are known as the \textit{support vectors}. The hyperplane chosen by the SVM is optimal in the respect that it maximises the minimum of the margins of the training set data points by solving a quadratic programming optimization problem.

\begin{figure}
    \centering
    \includegraphics{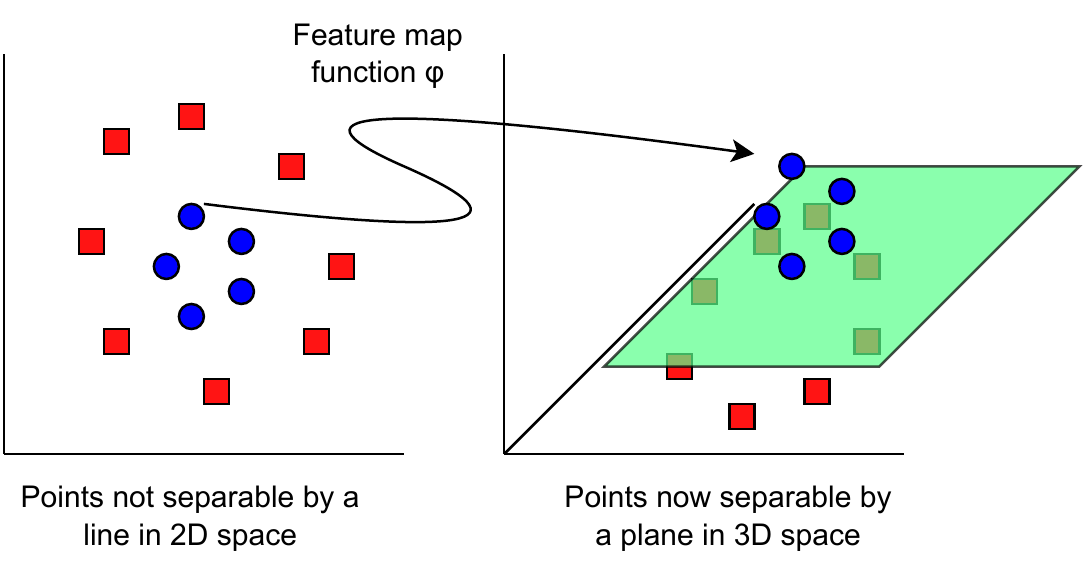}
    \caption{An example illustrating how a feature map function could be used to make non-linearly separable points linearly separable in a higher dimensional space. In this case, the feature map could be implemented as a function that adds a third dimension to the points with decreasing value as distance from the central region of the points increases.}
    \label{fig:feature_map_diagram}
\end{figure}

The SVM algorithm is also capable of classifying datasets with classes that are not linearly separable. This can be achieved by first mapping the data points to a higher dimensional space in such a way that they become linearly separable in the higher dimensional space (see Figure \ref{fig:feature_map_diagram} for an illustration). A function used to perform this mapping is called a \textit{feature map} and the range of the function is called the \textit{feature space}. The choice of feature map must be suited to the dataset in order to classify it well, since it determines whether the data will become linearly separable after transformation.

A feature map \(\phi(x)\) that maps a point into feature space has a corresponding \textit{kernel function} \(\kappa(x_i,x_j) = \langle \phi(x_i) , \phi(x_j) \rangle\) which computes the inner product of a pair of data points in the feature space. The margin optimization problem can be equivalently reformulated as a dual problem in terms of the kernel function \citep{initial_svm}, which can sometimes avoid the explicit computation of the feature map. This advantage is often referred to as the \textit{``kernel trick"}. Seeing that in many cases where a feature map can not be efficiently computed but the corresponding kernel function can be, the dual formulation of the problem increases the number of potential feature maps that can be applied to a dataset. 

In the dual form of the SVM, the classifier output for a given class is determined using a coefficient sequence \(\alpha = \{\alpha_1, \alpha_2, ..., \alpha_n\}\) and offset sequence \(b = \{b_1, b_2, ..., b_n\}\) chosen by the SVM algorithm during the hyperplane optimization. The decision function \textit{df} outputs an indication of the distance of its input point from the hyperplane after mapping into feature space. It is defined in terms of the kernel function $\kappa$, the training samples \(\{x_1, x_2, ..., x_n\}\), the $\alpha$ coefficients, and the $b$ offsets as follows
\[
	\text{df}(x) = \sum_{i=1}^n (\alpha_i \kappa(x, x_i) + b_i).
\]
The sign of $\text{df}(x)$ is used to determine the predicted class of the argument point $x$:
\[
\text{Class}(x) = \text{sgn}(\text{df}(x)).
\]
\subsubsection{Quantum-enhanced Support Vector Machine}\label{sec:QSVM_explanation}
The Quantum-enhanced Support Vector Machine (QSVM) algorithm extends the SVM algorithm by performing the kernel computation on a quantum computer \citep{QMLFHS, QSVMBDC}. A quantum circuit parameterised in the values of a single data point is used as a feature map to map the data points to a high dimensional quantum state in a quantum Hilbert space.

For a quantum feature map encoding data points into \(q\) qubits, the dimensionality of the feature Hilbert space is \(2^q\). Although a quantum computer can efficiently compute the quantum state feature space representation of data point, in general the quantum state cannot be efficiently represented classically due to the exponentially increasing dimensionality of the feature space. To work around this classical limitation, the kernel function is computed directly on the quantum computer and the kernel-based formulation of the SVM is used. The kernel computation for a pair of data points can be efficiently performed by measuring the overlap of their corresponding states in the quantum feature space.

To train a QSVM model, the \textit{Gram matrix} $K_{n \times n}$ of the training points must be computed. For $n$ training points $\{x_1, x_2, ..., x_n\}$ and a quantum kernel function $\kappa$, the Gram matrix is defined by $K_{ij} = \kappa(x_i, x_j)$ \text{where} $i,j \in \{1, 2, ..., n\}$. In the case of a noise free quantum computer or simulator being used to execute $\kappa$, the symmetric property $\kappa(x_i, x_j) = \kappa(x_j, x_i)$ and the property that $\kappa(x_i, x_i) = 1$ can be used to reduce the number of required evaluations.

Assuming the stated properties, the $n$ main diagonal entries of $K$ ($K_{11}$, $K_{22}$, ..., $K_{nn}$) do not require kernel evaluations, since $K_{ii} = \kappa(x_i, x_i) = 1$. For the remaining $n^2-n$ entries $K_{ij}$ which are not on the main diagonal, there is a symmetric entry $K_{ji}$ with the same value, since $K_{ij} = \kappa(x_i, x_j) = \kappa(x_j, x_i) = K_{ji}$. This means that only half of the entries need to be explicitly computed by kernel evaluations. In effect, only $\frac{n^2-n}{2}$ kernel evaluations must be performed to construct $K$. In the case of a NISQ computer, this technique could potentially be applied if measures were taken to mitigate noise and correct the kernel matrix such as in \cite{training_kernel_target_alignment}.

Since the only difference between the SVM and QSVM algorithms is how the kernel computation is performed, the potential advantage of the QSVM algorithm lies in enabling the computation of kernel functions that are hard to estimate classically \citep{liu2021rigorous}. While examples of such kernels have been discovered \citep{liu2021rigorous} for artificial datasets, it is an open question how best to design quantum feature maps to achieve a useful kernel with a quantum speed advantage.
\subsection{Kernel quality metrics}
\subsubsection{Kernel-target alignment}
Kernel-target alignment is a heuristic for kernel quality that measures the degree of similarity between two kernels, or the degree of agreement between a kernel and a dataset \citep{on_kernel_target_alignment}. It is calculated using a matrix inner product between the Gram matrix constructed from the training samples and an oracle matrix constructed from the training labels, where the oracle matrix acts as a stand-in for the Gram matrix of a hypothetical kernel which is very well-suited to the data. 

For a set of training points $\{x_1, x_2, ..., x_n\}$ with corresponding labels $\{y_1, y_2,$ $..., y_n\}$ with $\forall i, y_i  \in \{-1, 1\}$, a kernel function $\kappa(x_i, x_j)$, and with the \textit{Frobenius} inner product for matrices defined as $\langle A, B \rangle_F = \sum_{i,j} A_{ij} B_{ij}$, the kernel-target alignment can be computed as follows \citep{on_kernel_target_alignment}
\begin{enumerate}
\item Compute the Gram matrix $K_{n \times n}$ using the kernel function and training points by the rule
\[
K_{ij} = \kappa(x_i, x_j).
\]
\item Compute the oracle matrix $O_{n \times n}$ using the training labels by the rule
\[
O_{ij} = y_i y_j.
\]
\item Compute the kernel-target alignment \textit{KTA} using the Frobenius inner product as
\[
\text{KTA} = \frac{\langle K, O \rangle_F}{\sqrt{\langle K, K \rangle_F \langle O, O \rangle_F}}.
\]
\end{enumerate}
A high kernel-target alignment has been shown in other works to correlate with improved classification performance \citep{on_kernel_target_alignment} and it has been proposed for use as a metric for selecting applicable kernels for a dataset in classification problems \citep{on_kernel_target_alignment, training_kernel_target_alignment}.
\subsubsection{Root Mean Squared Error}
Root Mean Squared Error (RMSE) is a common metric for measuring the error of a model. It is calculated as the square root of the mean of the squared errors of a classifier's predictions on each training set data point. In this work, the RMSE of a classifier is calculated using the errors of the decision function on training data with an adjustment to the error calculation. The adjustment is to account for there not being a definitively correct output of the SVM decision function for a given sample and label pair. The error is measured relative to a positive target decision function output $m$, which we set to one in this work.

We calculate the error for a decision function output $a$ and training label $b$ using the following rule:
\[
\text{error}(a, b) = \begin{cases}
			(m - a) & \text{if } b = 1 \text{ and } a < m \\
			(a - m) & \text{if } b = -1 \text{ and } a > -m \\
			0 & \text{otherwise}
		\end{cases}
\]
This choice of error function means that only points not classified to the desired degree of confidence $m$ contribute to the error calculation, and the errors of the considered points increase with distance from the target output.

For a set of training points $\{x_1, x_2, ..., x_n\}$ with corresponding labels $\{y_1, y_2,$ $..., y_n\}$ with $\forall i, y_i  \in \{-1, 1\}$, 
the RMSE is calculated in terms of this adjusted error function and the decision function \textit{df} by the following rule
\[
\text{RMSE} = \sqrt{\frac{\sum_{i=1}^n \text{error}(\text{df}(x_i), y_i)^2}{n}}
\]
\subsection{Overview of genetic algorithms}\label{sec:GA_explanation}
Genetic algorithms are flexible metaheuristic algorithms inspired by the real-world evolutionary principles of natural selection, genetic inheritance, and random mutation. They are a popular choice of algorithm for optimizing complex objective functions in cases where algorithms known to produce a global optimum are unknown or infeasible.

Implementing a genetic algorithm first requires designing a solution representation on which genetic operations can be performed. The solution representation often (but not always) takes the form of a binary string, which is mapped by a problem-specific decoding function to a usable solution. A genetic algorithm manages a set of these solution representations, which is called a population. The initial population can be a set of randomly generated solutions or chosen according to a problem-specific heuristic.

The optimization process is iterative and typically repeats until a suitable solution is found, a desired number of iterations has passed, or the rate of improvement of the solutions becomes low. Each iteration, genetically inspired operations are applied to the population to create a new replacement population. The population creation process involves fitness evaluation, selection, crossover, and mutation operations.

Fitness evaluation is performed by decoding a solution representation into a solution, then evaluating a numeric score of its suitability for the problem. This is performed for the entire population, after which point a selection operation is applied to select some solutions for crossover and mutation. It is important that better solutions are more likely to be selected for crossover, since this is the main mechanism driving improvement between generations. The solutions selected for crossover are referred to as \textit{``parents"}.

The crossover operation is performed between two parent solutions to produce one or more new solutions, called child solutions. This is usually performed by taking a simple combination of the solution representations of the parents. In the case of a binary string solution representation, a simple crossover can be performed by combining two non-overlapping subsequences of the parents, taking a random number of bits from the first and the remainder from the corresponding positions in the second. A mutation operation can be applied to a child solution by randomly editing its representation by a small amount. This simulates the random mutation which occurs in real life and affects the diversity of available genetic material in the population.

A strategy often employed when determining which individuals will make up the next generation is to preserve the best performing of the solutions among the current generation and the newly created children. This is known as \textit{elitism} and ensures that solutions can survive through multiple generations and potentially indefinitely, so long as they continue to outperform newer ones. This helps prevent regression of the achieved fitness due to chance as generations pass.

The general idea for a genetic algorithm is flexible enough that many variations and extensions of the discussed components have also been studied \citep{genetic_algorithm_review}.

\subsection{Experiments}
The algorithm for automated feature map design described in \cite{Altares_L_pez_2021} was reimplemented using the Julia programming language \citep{Julia-2017}, the Yao quantum simulator framework \citep{yao}, and the pymoo \citep{pymoo} implementation of NSGA-II. All experiments were run with the maximum qubit count and feature map depth hyperparameters set to 6. The genetic algorithm population size was set to 100, with 15 new individuals being produced every generation. 30\% of the new individuals were produced by crossover; the rest were chosen randomly from the parents. In each generation, 70\% of the population underwent mutation. When mutation occured, 20\% of the bits in the mutated solution were flipped. All experiments were run using a noise free quantum simulator provided by Yao.

\begin{table}
\centering
\begin{adjustbox}{center}
\centering
\begin{tabular}{|c|c|c|c|c|c|c|c|}
	\hline
	Dataset & Class -1 & Class 1 & \begin{tabular}{@{}c@{}}Features \\ (PCA)\end{tabular} & Train & Test & Validation \\
	\hline
	Moons & Top left & Bottom right& 2 (N/A) & 210 & 90 & 500 \\
	Cancer & Benign & Malignant & 30 (10) & 210 & 90 & 124 \\
	Iris & Versicolor & Virginica & 4 (N/A) & 42 & 18 & 40 \\
	Digits & Eight & Nine & 64 (10) & 140 & 60 & 148 \\
	Circles & Outer & Inner & 2 (N/A) & 210 & 90 & 500 \\
	Random & Red & Blue & 2 (N/A) & 210 & 90 & N/A \\
	Voice & Acceptable & Unacceptable & 309 (10) & 28 & 12 & 44 \\
	SUSY & Background & Signal & 18 (10) & 210 & 90 & 500 \\
	SUSY reduced & Background & Signal & 8 (N/A) & 210 & 90 & 500 \\
	\hline
\end{tabular}
\end{adjustbox}
\caption{Table showing the characteristics of the datasets and sample splits used. Not all points in the base datasets were used to ensure the sample split remained balanced in each of the sample sets. Other considerations in determining the data splits were experiment runtime while maintaining a sufficiently large ratio of test points to train points and a sufficiently large number of validation points. All datasets with more than 10 feature values were reduced to 10 features using Principle Component Analysis (PCA). The moons, circles, and random datasets are artificial, with the moons and circles datasets being generated with Scikit-learn \citep{sklearn}. The rest of the datasets are sourced from the UCI Machine Learning Repository \citep{UCIDATA} either directly or indirectly through Scikit-learn \citep{sklearn}.}
\label{table:datasets}
\end{table}

Three configurations of the original algorithm were run on nine different datasets of varying difficulty (see Table \ref{table:datasets}) to compare their effectiveness. The first configuration maximized accuracy on a test set and minimized weighted size, as in the original work \citep{Altares_L_pez_2021}. The second configuration maximized kernel-target alignment on the training data, ignoring the test data, and minimized the unweighted size metric also defined in \cite{Altares_L_pez_2021}. The third configuration maximized an approximation of kernel-target alignment on the training data, and minimized the same unweighted size metric.

Before performing the genetic optimization the datasets are split into three disjoint subsets, namely training data, testing data, and validation data. The training data is used to evaluate kernel-target alignment and its approximation, as well as train the QSVM model for a given feature map circuit. The testing data is only used to evaluate the accuracy metric in the first approach. The validation data is used to determine the generalisation ability of the generated models, and must be separate from the testing data since the first approach can indirectly access the testing data through the accuracy metric and potentially overfit to it.

In order to calculate the kernel-target alignment approximation for $n$ training points, the $n$ points are divided into $a$ disjoint complementary subsets of size roughly $n/a$. The number of subsets $a$ can be adjusted based on the number of training points to balance speed and precision. The kernel target alignment is calculated on each of the subsets in turn, then averaged.

Assuming the properties of the kernel function are not used to accelerate the Gram matrix computation, $n^2$ kernel evaluations are required to compute the exact kernel-target alignment, and only $a (n/a)^2 = (n^2)/a$ evaluations are required to compute the approximation, giving a factor $a$ speedup. If the kernel properties are used, then $(n^2-n)/2 = n^2/2 - n/2$ kernel evaluations are required to compute the exact kernel-target alignment. Therefore, the number of kernel evaluations required when evaluating the kernel-target alignment approximation can be derived as

\[
a ( \frac{(\frac{n}{a})^2-\frac{n}{a}}{2} )
= \frac{n^2}{2a} - \frac{n}{2},
\]
meaning the factor speedup by evaluating the approximation is larger than $a$, but should approach $a$ as $n$ increases. In this work, we use a value of $a=5$ in all experiments involving the kernel-target alignment approximation.

\begin{figure}
    \centering
    \includegraphics[width=\columnwidth]{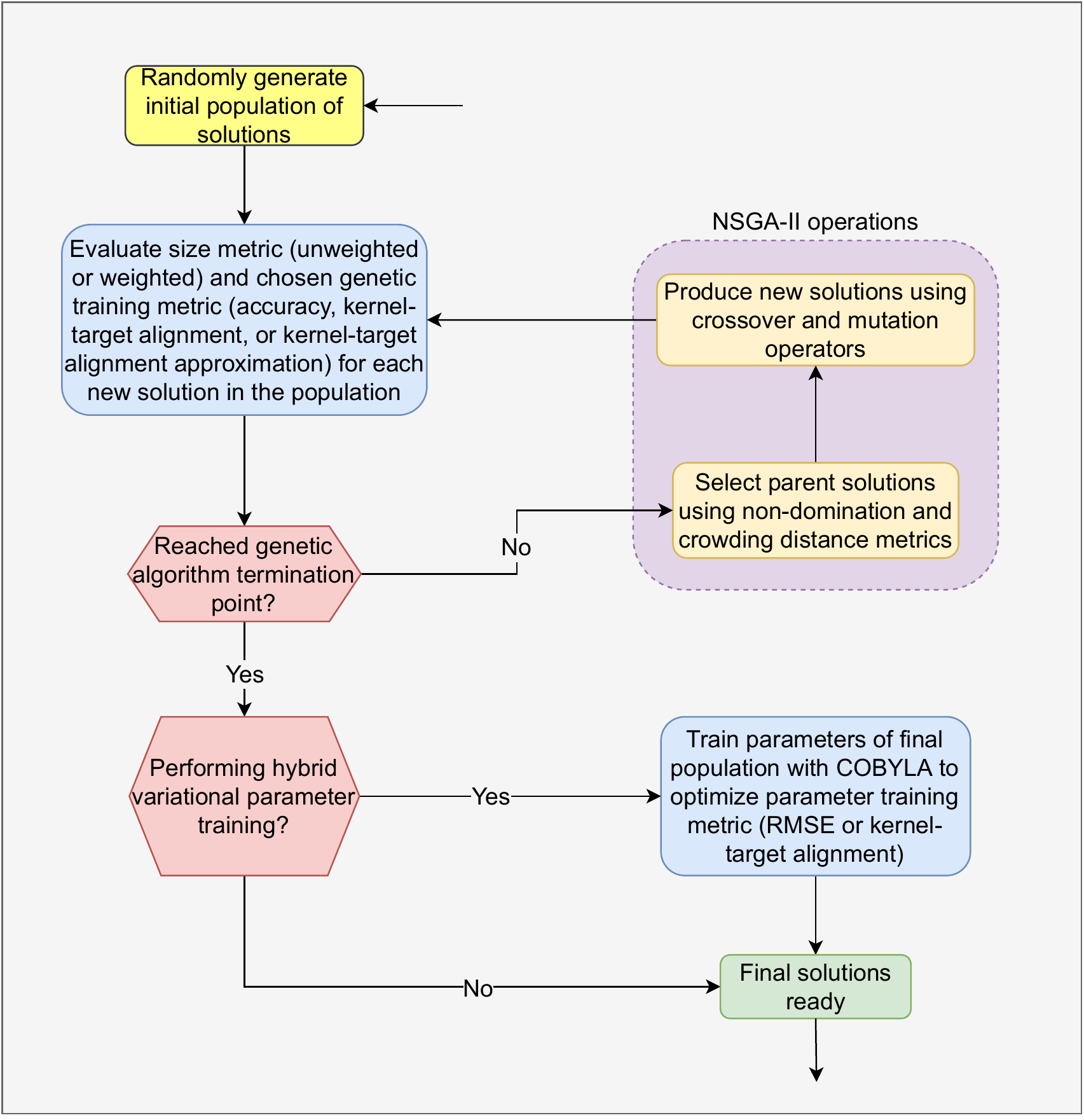}
    \caption{A flow diagram outlining the algorithm followed to genetically train quantum feature map circuits. The diagram also shows how a hybrid method involving circuit parameter training can be performed after genetic optimization.}
    \label{fig:algorithm_flow_diagram}
\end{figure}

After the genetic optimization in each configuration completes, we attempt further improvement by further optimizing just the proportionality parameters encoded in the last two bits of the gate representation using an implementation of COBYLA \citep{cobyla_book} provided by the NLopt optimization library \citep{NLopt}. This allows the parameter values to not be restricted to one of only four possibilities. This optimization aims to either minimize RMSE or maximize kernel-target alignment using the training set to evaluate the metrics. The COBYLA optimizer is allowed one hundred evaluations of the cost function to perform the optimization. A flow diagram outlining the algorithmic process can be seen in Figure \ref{fig:algorithm_flow_diagram}.

The COBYLA cost functions for RMSE and kernel-target alignment both require computing a Gram matrix each evaluation, meaning the number of kernel evaluations performed is $\frac{100(n^2-n)}{2}$. This can be contrasted with the genetic optimization of accuracy or kernel-target alignment, where at least as many kernel evaluations are performed to evaluate the fitness of just the first generation of 100 solutions in the genetic algorithm. In the subsequent 1199 generations $15 \times 1199=17 985$ more Gram matrix evaluations are performed for a total of 18085, meaning the final parameter training for the entire output population requires roughly 55\% percent of the number of kernel evaluations performed in the genetic optimization in the cases of genetically optimizing accuracy or the exact kernel-target alignment.

We name the three base approaches 1, 2, and 3, respectively. Each approach has two additional sub-approaches defined for further training of RMSE or kernel-target alignment, for a total of nine approaches. The RMSE and kernel-target alignment variations are named with a .1 and .2 suffix respectively. We graph the classification accuracies, average margins, ROC curves, feature map circuits, and confusion matrices of the best models produced by each approach, where the best model of a population is taken to be the one achieving the highest validation set accuracy. For two dimensional datasets, decision boundaries are also graphed.

The code implementing the experiments and result graphing can be found on GitHub \citep{RPellowJarman2022Software}.

\section{Results}\label{sec3}

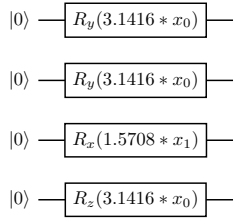
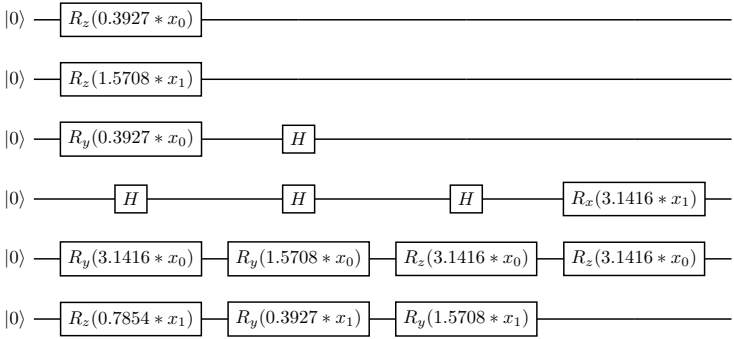
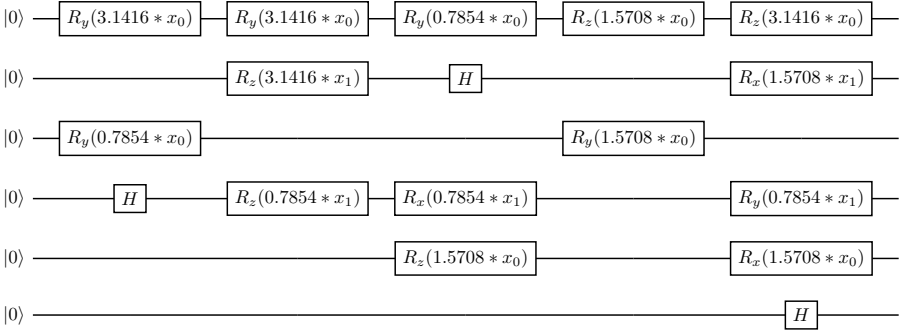
\begin{figure}
\centering
\begin{subfigure}{\textwidth}
\centering
\begin{tikzpicture}
\node[scale=0.7] {
\begin{quantikz}
    \lstick{$\ket{0}$} & \gate{R_y(3.1416 * x_{0})} & \qw \\
    \lstick{$\ket{0}$} & \gate{R_y(3.1416 * x_{0})} & \qw \\
    \lstick{$\ket{0}$} & \gate{R_x(1.5708 * x_{1})} & \qw \\
    \lstick{$\ket{0}$} & \gate{R_z(3.1416 * x_{0})} & \qw \\
\end{quantikz}
};
\end{tikzpicture}
\caption{Approach 1 - Accuracy, Weighted size}
\end{subfigure}

\begin{subfigure}{\textwidth}
\centering
\begin{tikzpicture}
\node[scale=0.7] {
\begin{quantikz}
    \lstick{$\ket{0}$} & \gate{R_z(0.3927 * x_{0})} & \qw & \qw & \qw & \qw \\
    \lstick{$\ket{0}$} & \gate{R_z(1.5708 * x_{1})} & \qw & \qw & \qw & \qw \\
    \lstick{$\ket{0}$} & \gate{R_y(0.3927 * x_{0})} & \gate{H} & \qw & \qw & \qw \\
    \lstick{$\ket{0}$} & \gate{H} & \gate{H} & \gate{H} & \gate{R_x(3.1416 * x_{1})} & \qw \\
    \lstick{$\ket{0}$} & \gate{R_y(3.1416 * x_{0})} & \gate{R_y(1.5708 * x_{0})} & \gate{R_z(3.1416 * x_{0})} & \gate{R_z(3.1416 * x_{0})} & \qw \\
    \lstick{$\ket{0}$} & \gate{R_z(0.7854 * x_{1})} & \gate{R_y(0.3927 * x_{1})} & \gate{R_y(1.5708 * x_{1})} & \qw & \qw \\
\end{quantikz}
};
\end{tikzpicture}
\caption{Approach 2 - Kernel-target alignment, Unweighted size}
\end{subfigure}

\begin{subfigure}{\textwidth}
\centering
\begin{tikzpicture}
\node[scale=0.7] {
\begin{quantikz}
    \lstick{$\ket{0}$} & \gate{R_y(3.1416 * x_{0})} & \gate{R_y(3.1416 * x_{0})} & \gate{R_y(0.7854 * x_{0})} & \gate{R_z(1.5708 * x_{0})} & \gate{R_z(3.1416 * x_{0})} & \qw \\
    \lstick{$\ket{0}$} & \qw & \gate{R_z(3.1416 * x_{1})} & \gate{H} & \qw & \gate{R_x(1.5708 * x_{1})} & \qw \\
    \lstick{$\ket{0}$} & \gate{R_y(0.7854 * x_{0})} & \qw & \qw & \gate{R_y(1.5708 * x_{0})} & \qw & \qw \\
    \lstick{$\ket{0}$} & \gate{H} & \gate{R_z(0.7854 * x_{1})} & \gate{R_x(0.7854 * x_{1})} & \qw & \gate{R_y(0.7854 * x_{1})} & \qw \\
    \lstick{$\ket{0}$} & \qw & \qw & \gate{R_z(1.5708 * x_{0})} & \qw & \gate{R_x(1.5708 * x_{0})} & \qw \\
    \lstick{$\ket{0}$} & \qw & \qw & \qw & \qw & \gate{H} & \qw \\
\end{quantikz}
};
\end{tikzpicture}
\caption{Approach 3 - Kernel-target alignment approximation, Unweighted size}
\end{subfigure}
\caption{The circuits with highest validation set accuracy produced by the three base genetic approaches when creating quantum feature maps for the Moons dataset. (a) shows the best produced circuit when training to maximize accuracy and minimize weighted size as in the original work, (b) shows the best circuit when training to maximize the exact kernel-target alignment and minimize unweighted size, and (c) shows the best circuit when training to maximize the approximation of the kernel-target alignment and minimize unweighted size. Circuits (b) and (c) are significantly larger. Unused gate layers and qubits are omitted from the diagrams.}
\label{figure:moons_circuits}
\end{figure}

As in the original work by \cite{Altares_L_pez_2021}, the feature map circuits produced by each of the approaches tend to make little to no use of entangling gates (see Figure \ref{figure:moons_circuits}). However, the circuits produced by optimizing the kernel-target alignment based metrics tend to produce significantly larger circuits overall (see Figure \ref{figure:moons_circuits}). This could be explained by the fact that the weighted size metric in the genetic optimization was replaced with an unweighted size metric in those approaches due to the weighted size metric depending on the test set accuracy, which was not evaluated. The optimization of the circuit size did not converge in the allocated 1200 generations in approaches 2 and 3, which can be inferred from the presence of redundant gates. This could possibly be addressed by allowing more generations to pass or using a size metric weighted in kernel-target alignment instead of accuracy, similarly to the original approach. Another possible explanation for the larger circuit size is that a circuit achieving perfect accuracy may still be able to improve its kernel-target alignment; in the accuracy maximization case, the genetic algorithm is able to shift focus to minimizing circuit size after achieving 100\% accuracy, but the same cannot be done as easily when maximizing kernel-target alignment since its limiting value of one is more difficult to achieve. Additionally, the high mutation rate of 70\% could be reduced to attempt to reach convergence in the allocated 1200 generations.

\begin{figure}
\centering
\includegraphics[width=0.9\columnwidth]{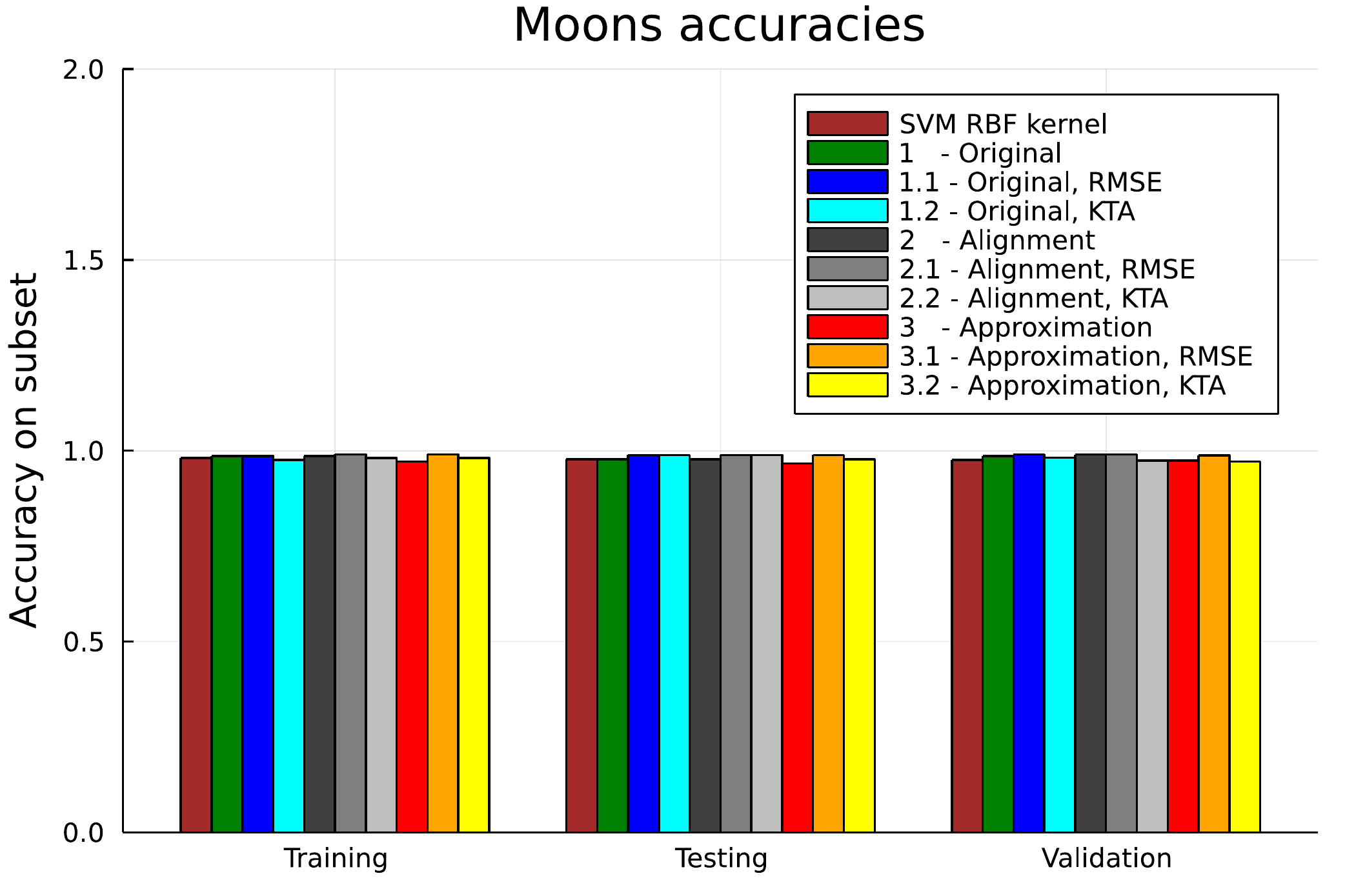}
\caption{A graph showing the classification accuracies of the best models produced by various approaches of quantum feature map design on the Moons dataset, compared with a classical RBF kernel for reference. All approaches can be seen to achieve comparable accuracy across the different subsets.}
\label{figure:moons_accuracies}
\end{figure}

\begin{figure}
\centering
\includegraphics[width=0.9\columnwidth]{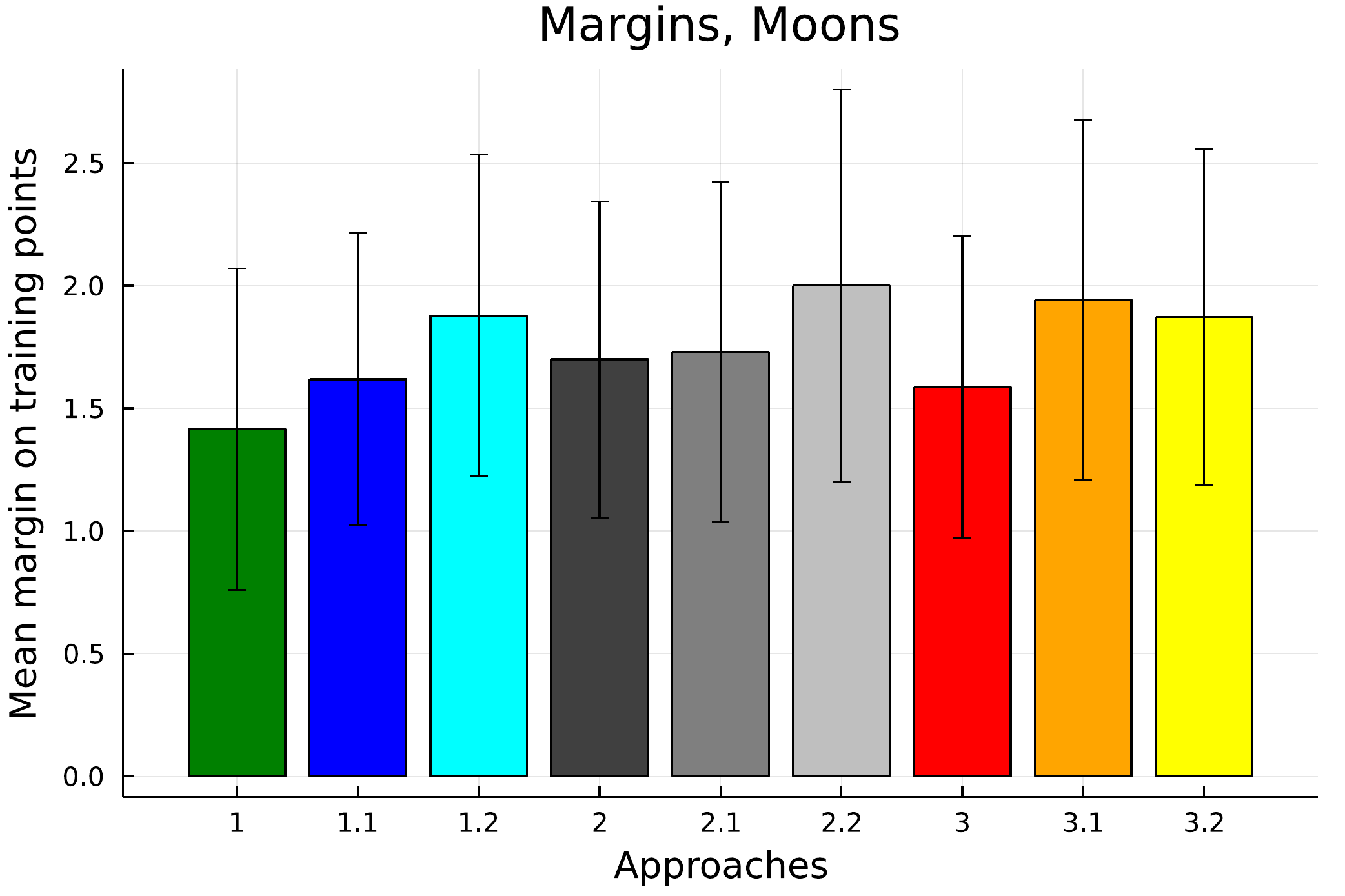}
\caption{A graph showing the mean margin of the Moons training set points for the best classifiers produced by each approach, with errors bars showing standard deviation. Circuit parameter training and genetic training of kernel-target alignment are both shown to increase the mean margin size. The approach numbering corresponds to the numbering used in Figure \ref{figure:moons_accuracies}.}.
\label{figure:moons_margins}
\end{figure}

\begin{figure}
\centering
\includegraphics[width=0.9\columnwidth]{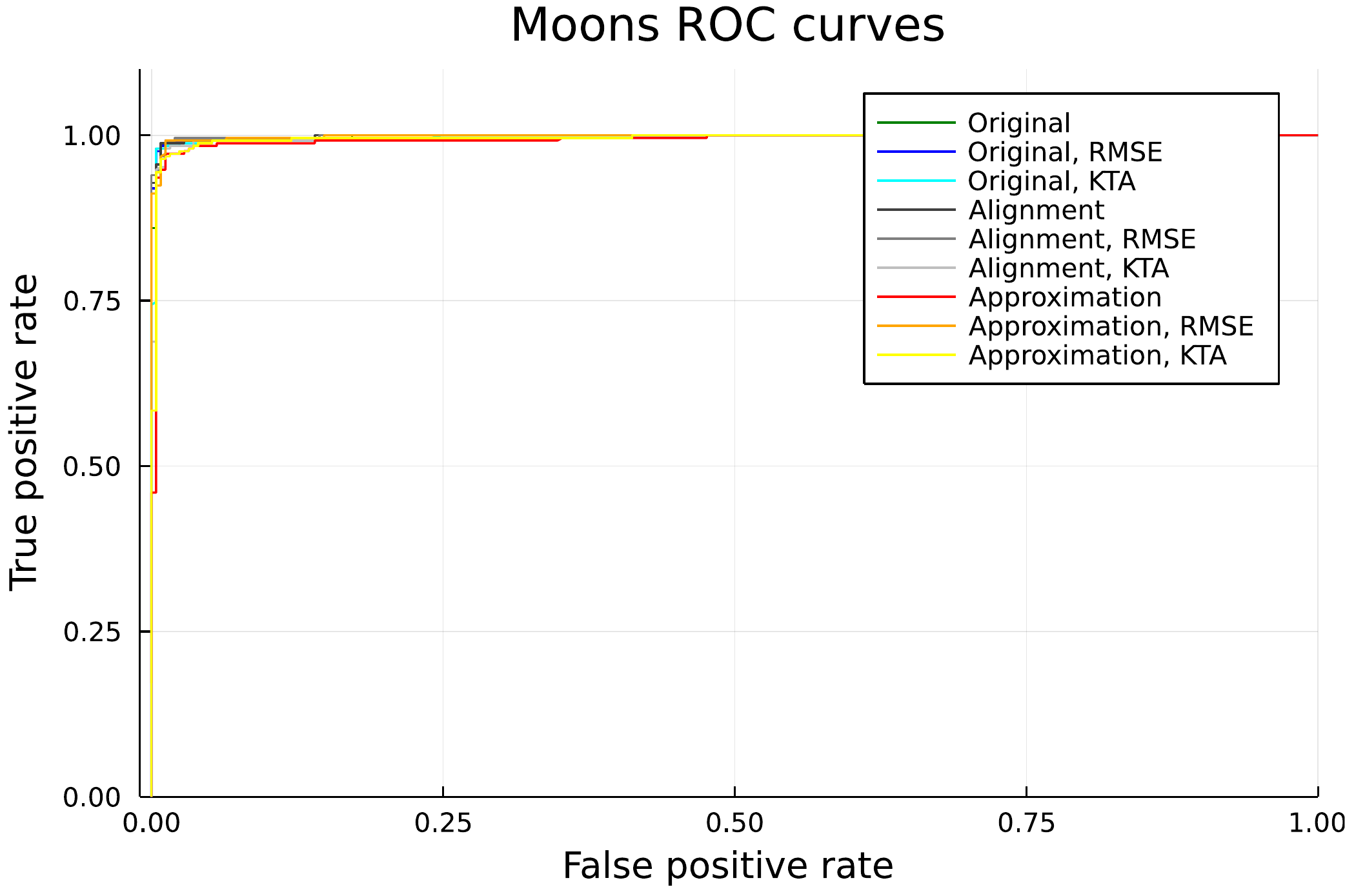}
\caption{A graph showing the ROC curves of the best models produced by various approaches of quantum feature map design on the Moons dataset. All of the produced models are shown to perform similarly on the dataset.}
\label{figure:moons_roc_curves}
\end{figure}

\begin{figure}
\centering
\begin{adjustbox}{center}
\centering
\includegraphics[width=0.7\columnwidth]{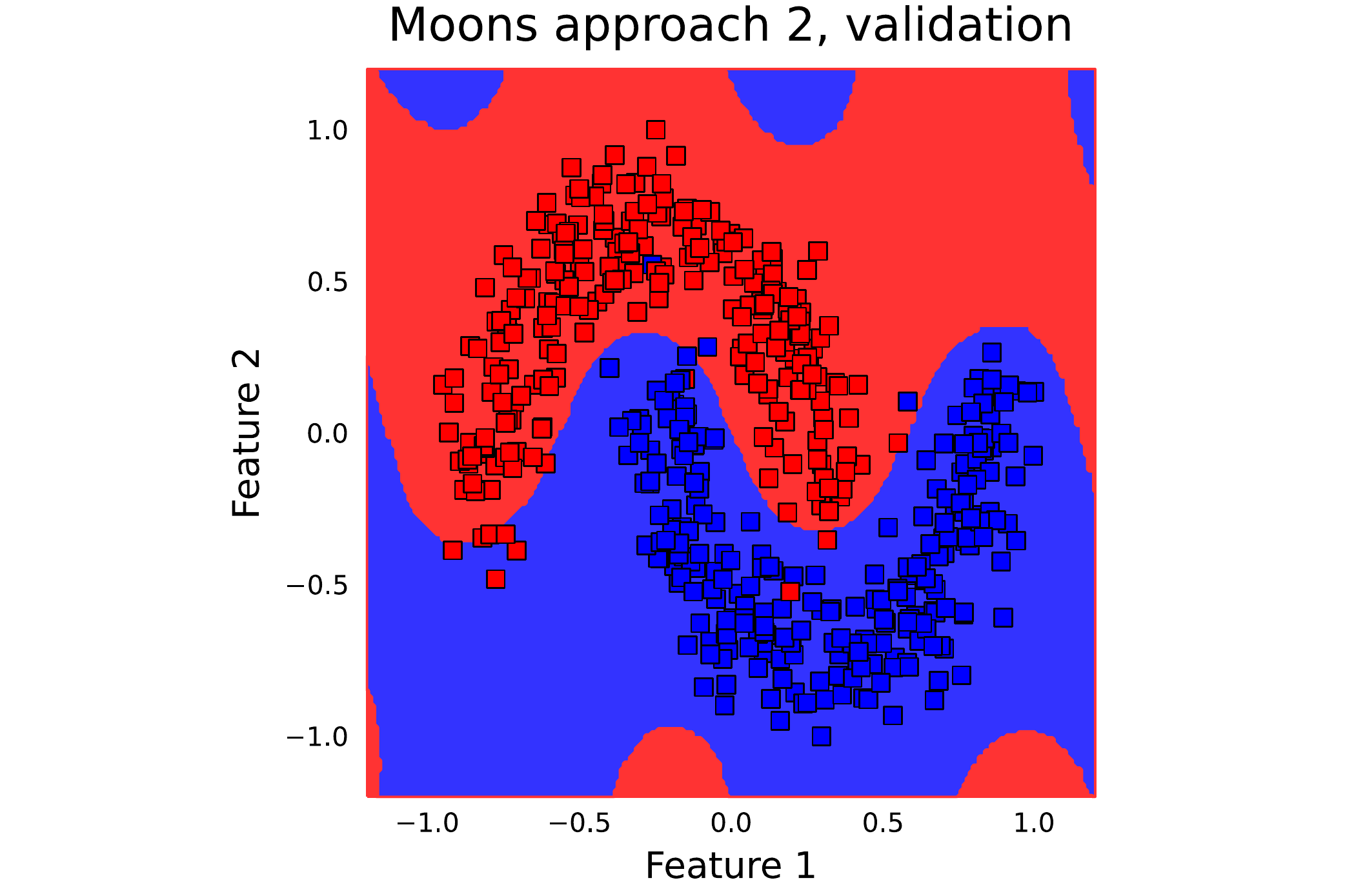}
\includegraphics[width=0.7\columnwidth]{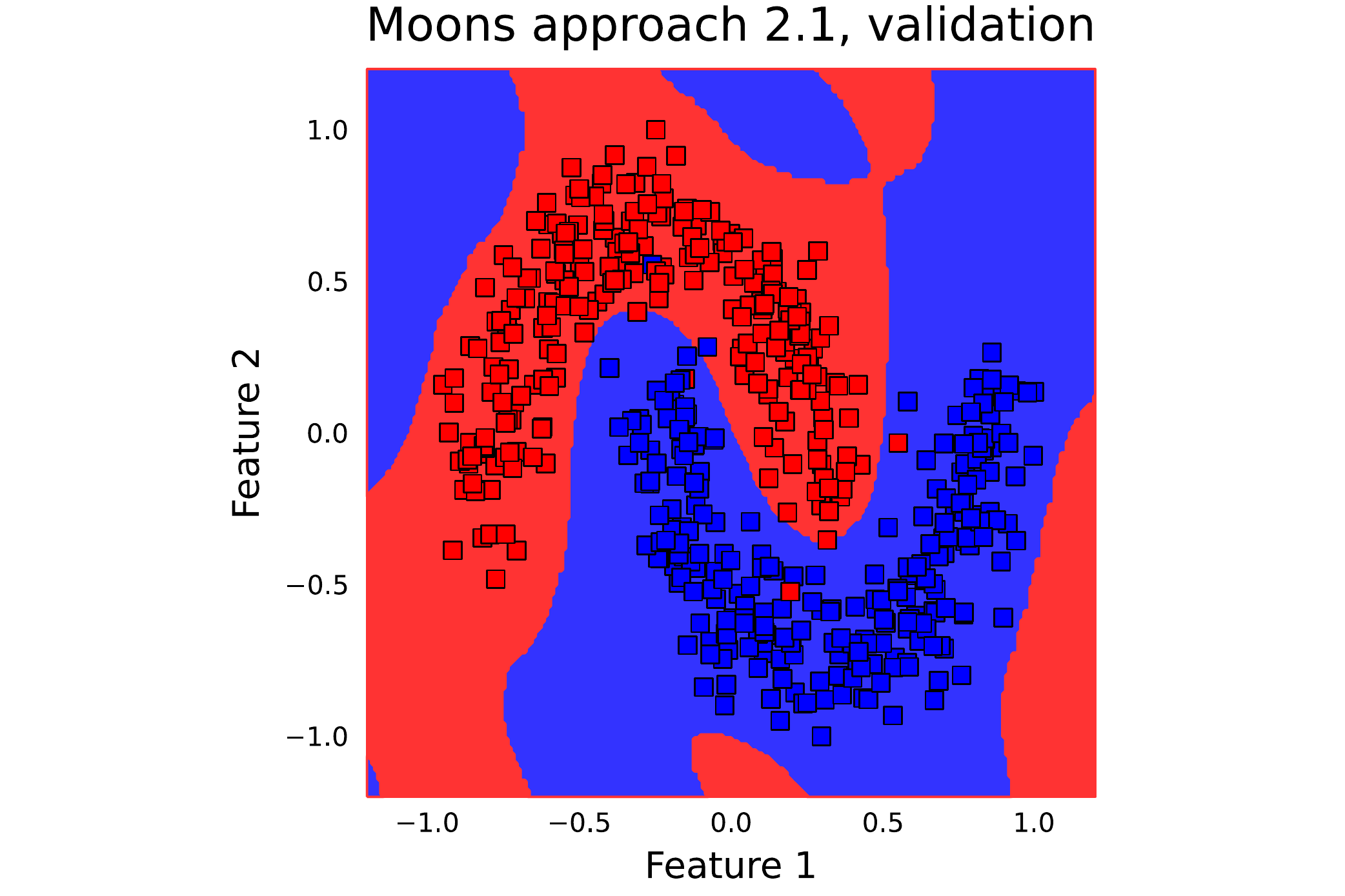}
\end{adjustbox}
\caption{Decision boundaries of the best classifiers for approaches 2 and 2.1 on the Moons validation data. Further parameter training on training data to minimize RMSE after genetically designing the feature maps to maximize kernel-target alignment is shown to improve classification ability on validation data.}
\label{figure:moons_decision_boundary_improvement}
\end{figure}

Our experiments show that substituting kernel-target alignment or approximated kernel-target alignment for accuracy in the genetic optimization process produces feature map circuits with accuracy comparable to the original approach across all datasets (see Figures \ref{figure:moons_accuracies} and \ref{figure:moons_roc_curves} for example). Further optimizing the final population's trainable parameters using COBYLA was often able to improve the average margin sizes of the classifiers (see Figure \ref{figure:moons_margins}) on training data and sometimes able to improve validation set classification accuracy (see Figures \ref{figure:moons_accuracies} and \ref{figure:moons_decision_boundary_improvement}), showing that a hybrid approach performing further optimization of the final populations parameter values is worth attempting despite the computational cost if improving accuracy is important. Training the parameters of a single solution for 100 cost evaluations requires only half a percent of the kernel evaluations as the genetic optimization process, so a smaller subset of the final solutions could be trained at a much lower cost. Additionally, the untrained parameters encoded in the solution binary strings are not lost if further training is performed and can still be used if they happen to perform better than the trained ones.

\begin{figure}
\centering
\includegraphics[width=0.85\columnwidth]{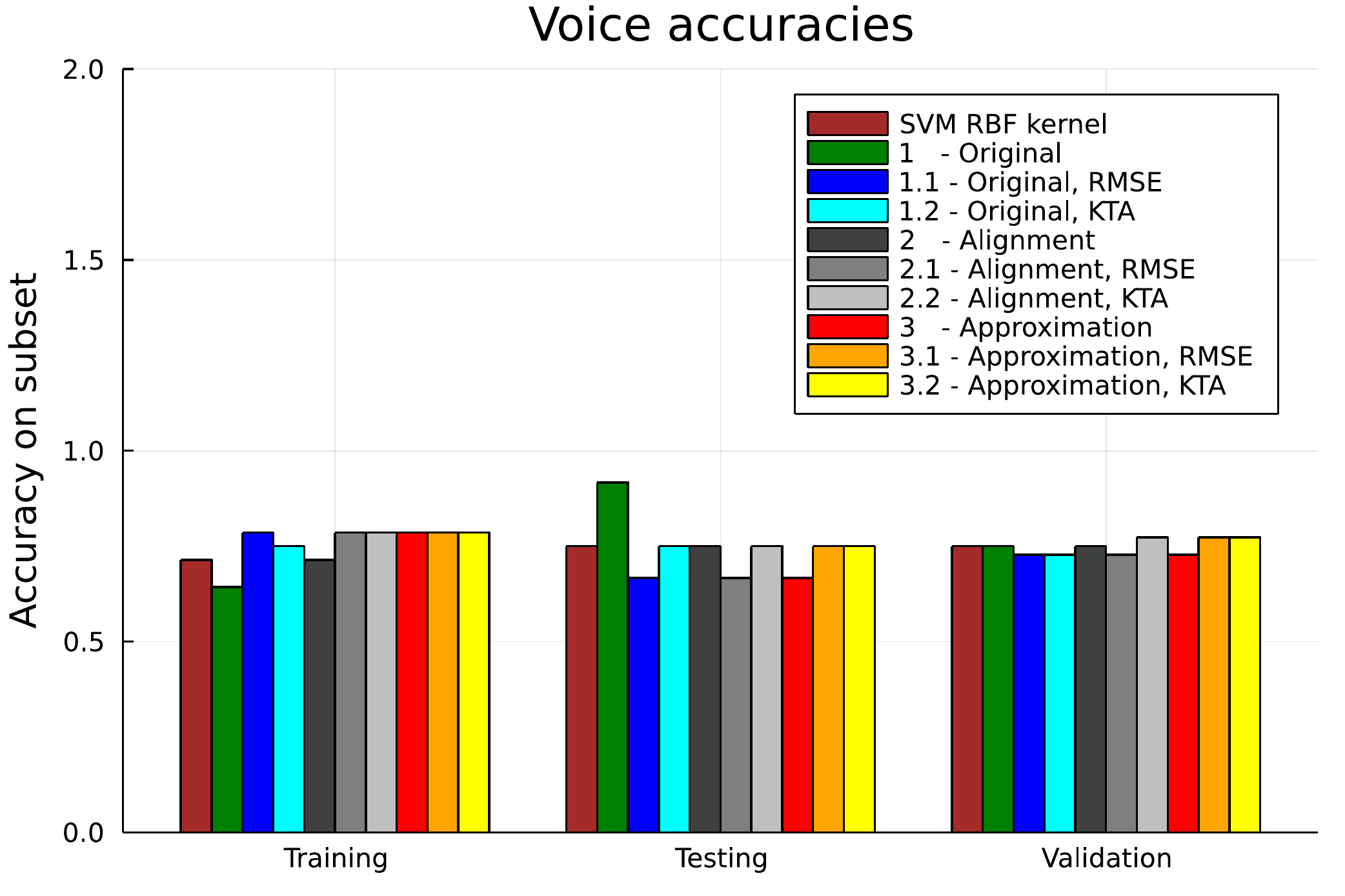}
\caption{A graph showing the classification accuracies of the best models produced by various approaches of quantum feature map design on the Voice dataset, compared with a classical RBF kernel for reference. Genetic accuracy maximization is shown to overfit to the testing data used to evaluate the accuracy metric, justifying the necessity of a separate validation set.}
\label{figure:voice_accuracies}
\end{figure}

The results demonstrate that on difficult datasets such as the SUSY, SUSY reduced features, Voice, and Random datasets, the original approach's models can overfit to the testing data used to evaluate the accuracy metric (see Figure \ref{figure:voice_accuracies}). This is likely due to the fact that test set accuracy is directly optimized in the genetic algorithm without regard to training set accuracy. Since the kernel-target alignment approaches make use of only the training data during the genetic optimization they do not suffer from the same drawback, although they do not show improvement on validation data for the difficult problems. This problem could possibly be avoided by shuffling the training and testing data each generation, although this would make the accuracy metric depend on the generation at which the accuracy was evaluated and could prevent caching of solution fitnesses in the genetic algorithm. A second possible solution is to average the accuracy over subsets of the data. Given a dataset of $n$ points, this can be performed while requiring at most $\frac{n^2-n}{2}$ kernel evaluations in the worst case, since the Gram matrix for the entire dataset can be computed once and used as a cache to look up the kernel output for any pair of points when creating models with arbitrary choices of training and testing subsets.

\begin{figure}
\centering
\includegraphics[width=0.9\columnwidth]{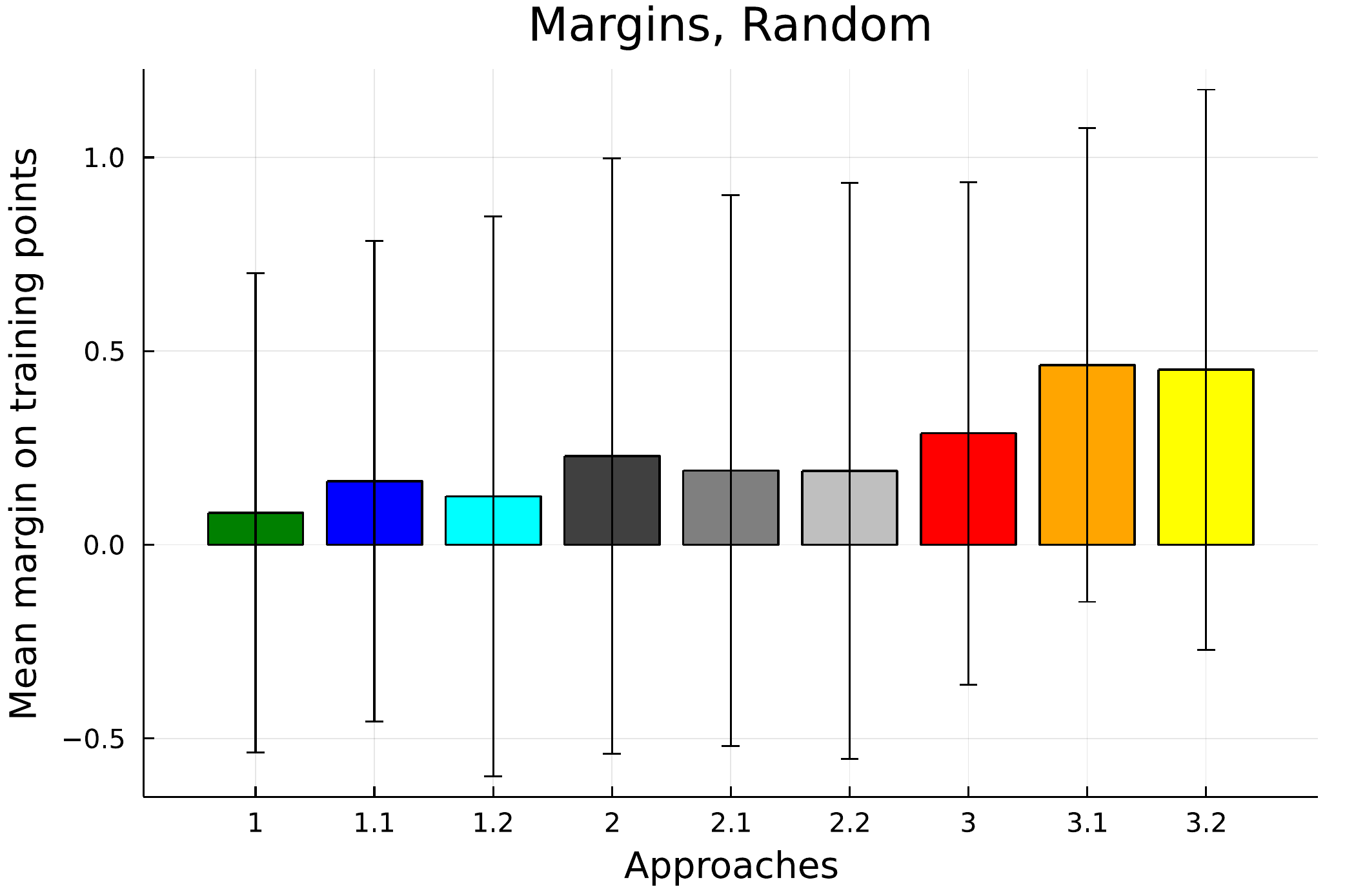}
\caption{A graph showing the mean margin of the Random training set points for the best classifiers produced by each approach, with errors bars showing standard deviation. The approach numbering corresponds to that in Figure \ref{figure:moons_accuracies}.}
\label{figure:adhoc_margins}
\end{figure}

\begin{table}
\centering
\begin{adjustbox}{center}
\centering
\begin{tabular}{|c|c|c|c|}
	\hline
	Approach & Average margin & Absolute Change & Percentage Change \\
	\hline
	1-Accuracy (Original work) & 0.838 & N/A & N/A \\
	Accuracy, RMSE training & 0.993 & +0.154 & +18.42\verb|%| \\
	Accuracy, KTA training & 0.971 & +0.133 & +15.87\verb|%| \\
    \hline
	2-Alignment & 1.043 & +0.205 & +24.46\verb|%| \\
	Alignment, RMSE training & 1.016 & -0.028 & -2.66\verb|%|\\
	Alignment, KTA training & 1.090 & +0.047 & +4.49\verb|%| \\
    \hline
	3-Approximation & 1.065 & 0.226 & +26.99\verb|%| \\
	Approximation, RMSE training & 1.145 & +0.080 & +7.54\verb|%| \\
	Approximation, KTA training & 1.124 & +0.059 & +5.59\verb|%| \\
	\hline
\end{tabular}
\end{adjustbox}
\caption{Table showing the average margin size of the best classifier produced by each approach, averaged across the nine datasets with equal weighting given to each dataset. For this purpose, the best classifier is defined as the classifier achieving the highest validation set accuracy for the target dataset. The improvement columns for the base approaches (2 and 3) show how genetic optimization of kernel-target alignment and its approximation improve on the margins achieved in the original work. For the hybrid approaches (with additional RMSE and KTA parameter training), the columns show change relative to the base approaches.}
\label{table:margins}
\end{table}

The margins of classifiers trained with the second and third approach tended to be larger than those trained with the first (see Figures \ref{figure:moons_margins} and \ref{figure:adhoc_margins}, as well as Table \ref{table:margins}). This could be due to the fact that the kernel-target alignment metric and its approximation are evaluated on the training subset as opposed to accuracy which is evaluated on the testing subset, leading to the former two approaches having higher confidence on the training subset. In either case, increased margin size is an indicator of improved generalisation ability according to theoretical works \citep{Vapnik1998Book, VCBound} showing that margin size bounds VC dimension, and VC dimension bounds expected generalisation error. Further parameter training on the final generation also tended to show some improvements in margin sizes, even on easier datasets such as the Moons and Circles datasets where there was not much effect on overall classification accuracy. This improvement in margin can be seen visually in the decision boundary graphs of the classifiers (see Figure \ref{figure:moons_decision_boundary_improvement}).

\section{Conclusion}\label{sec13}

In this paper, we compared our implementation of the approach defined in \cite{Altares_L_pez_2021} with adjustments to the genetic algorithm cost functions. These adjustments were aimed at investigating the suitability of kernel-target alignment as an alternative metric to test set accuracy and at reducing the number of kernel evaluations required by the approach. The new approaches were shown to still be effective at designing accurate classifiers with fewer kernel evaluations, although at the cost of increased circuit size. They were also shown to often produce classifiers with better margins on training data. We also put forward a hybrid approach extending the original work by applying COBYLA \citep{cobyla_book} to further optimize the trainable parameters of the produced quantum feature map circuits after the termination of the genetic algorithm to attempt further improvement, at a lower additional computational cost than the genetic algorithm's base cost. This parameter training was also shown to be capable of improving margin sizes and sometimes accuracy without increasing the circuit gate cost.

There is still more work to be done in accelerating the genetic algorithm while keeping gate costs low. A potential avenue to achieving this goal is the use of a multi-phase genetic algorithm in which the cost function is initially easy to evaluate but increases in precision after a set number of generations passes. For example, the $a$ parameter of the kernel-target alignment approximation could be made to decrease as generations pass for the approximation to become more accurate at the cost of more kernel evaluations, or the cost function could be switched from kernel-target alignment to classification accuracy to reduce gate cost once a predetermined kernel-target alignment has been achieved.

The original approach could also further be extended to have the gate encoding for parameterised gates select a classical data encoding function such as those used in \cite{ASFMKBQC} to introduce classical nonlinearity to the encoding, potentially allowing for even lower gate cost or higher accuracy circuits to be produced.

\backmatter

\bibliography{sn-bibliography}


\begin{appendices}

\section{Figures}\label{secA1}

\begin{figure}[h]
\includegraphics[width=0.9\columnwidth]{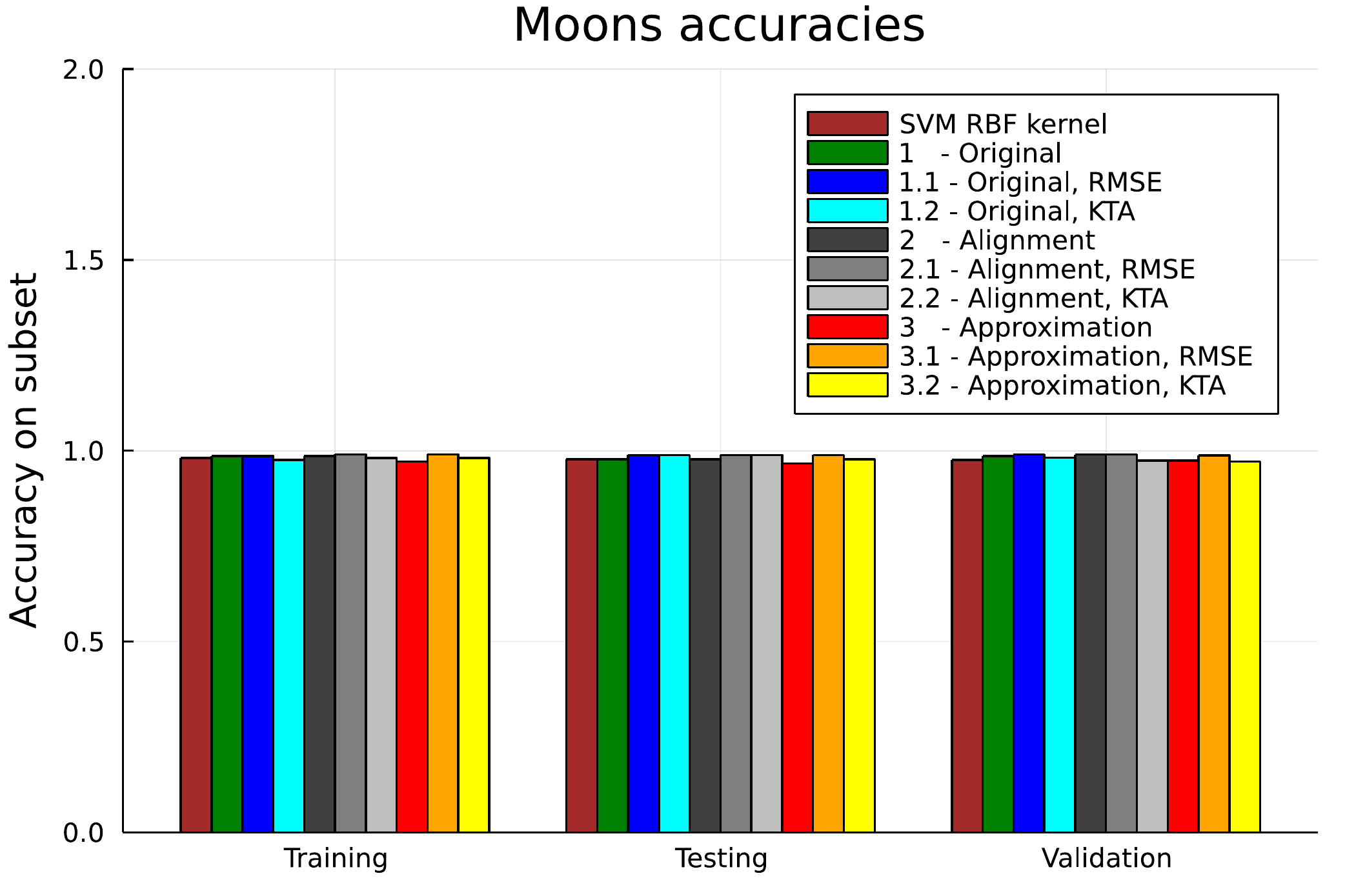}
\caption{A graph showing the classification accuracies of the best models produced by various approaches of quantum feature map design on the Moons dataset, compared with a classical RBF kernel for reference.}
\label{figure:moons_accuracies_appendix}
\end{figure}

\begin{figure}[h]
\includegraphics[width=0.9\columnwidth]{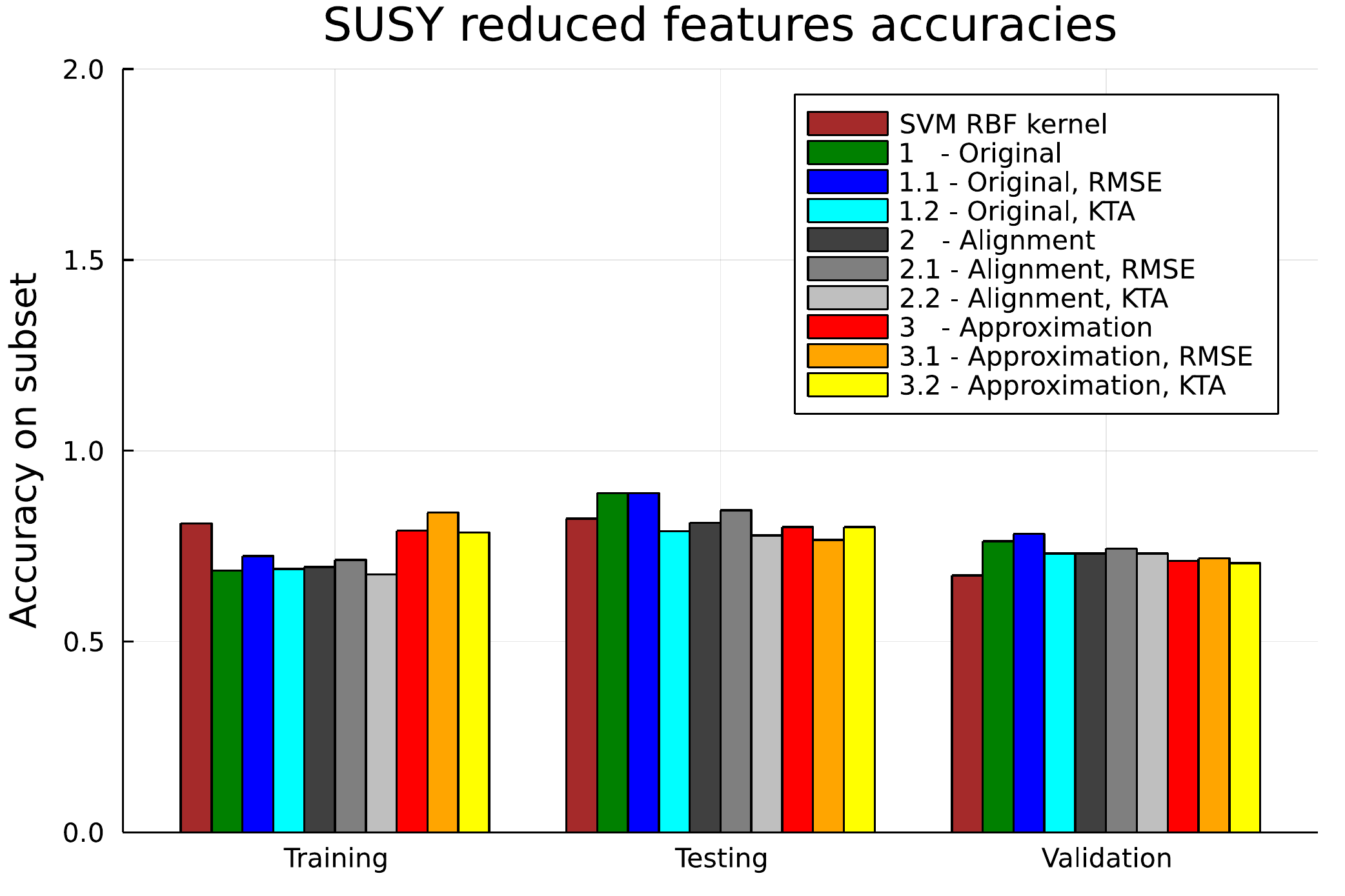}
\caption{A graph showing the classification accuracies of the best models produced by various approaches of quantum feature map design on the SUSY reduced features dataset, compared with a classical RBF kernel for reference.}
\label{figure:susy_hard_accuracies_appendix}
\end{figure}

For Figures \ref{figure:moons_accuracies_appendix} - \ref{figure:susy_hard_accuracies_appendix}, we show the accuracy of the best produced kernels on each dataset, with the classical RBF kernel also shown for comparison. These figures show that the newly proposed approaches achieve comparable classification accuracy to the original approach, despite not directly evaluating classification accuracy during training. The graphs for the difficult problems (Random, Voice, SUSY, SUSY reduced features) show that the original approach has a tendency to overfit to the testing data used to evaluate the accuracy metric, and that a separate validation set must be reserved to estimate the generalisation of its models. The approaches based on maximizing the approximation of kernel-target alignment are also seen to overfit to training data in these cases more than the other approaches do.

\begin{figure}[h]
\includegraphics[width=0.9\columnwidth]{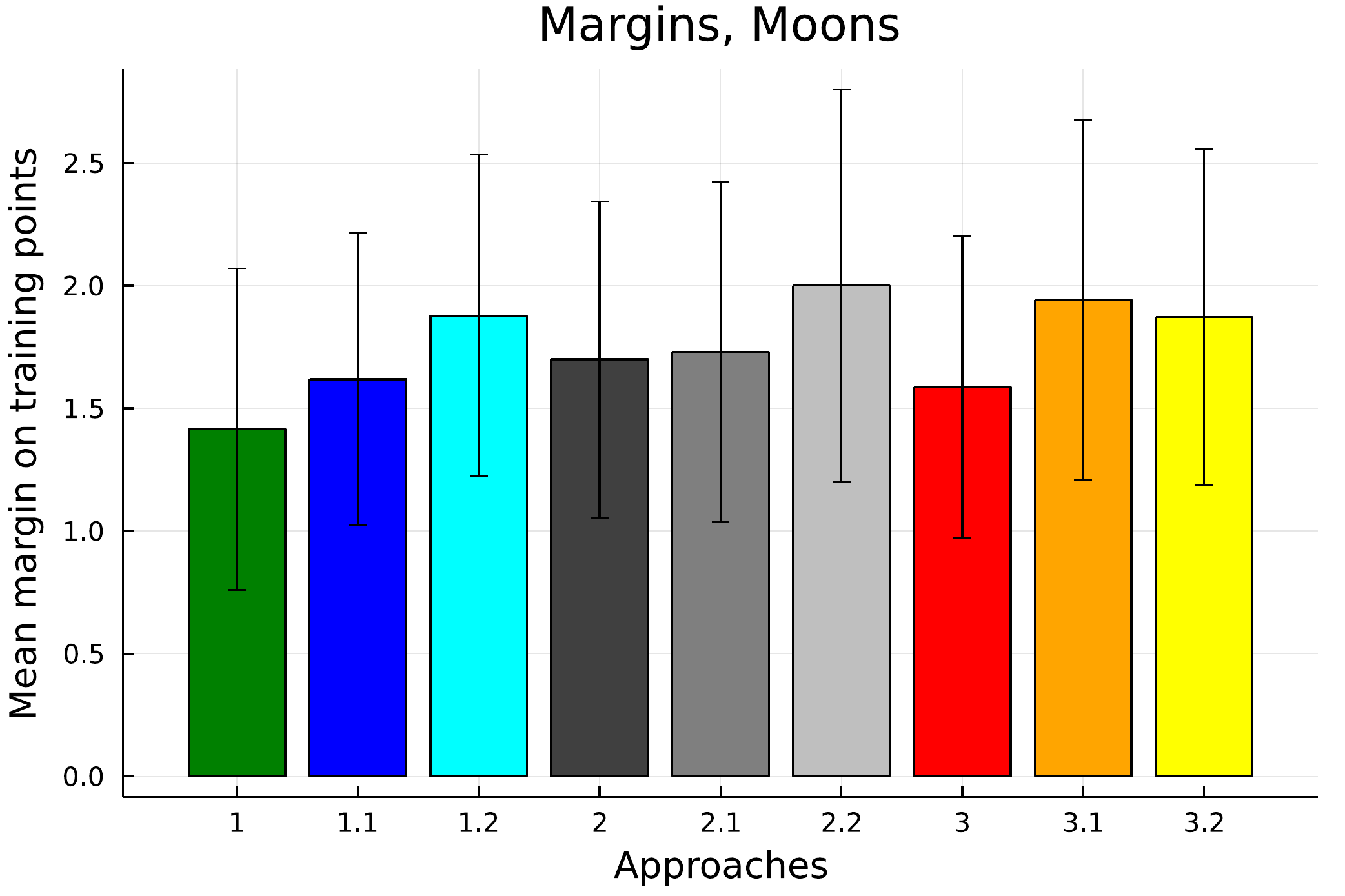}
\caption{A graph showing the mean margin of the Moons training set points for the best classifiers produced by each approach, with errors bars showing standard deviation.}
\label{figure:moons_margins_appendix}
\end{figure}

\begin{figure}[h]
\includegraphics[width=0.9\columnwidth]{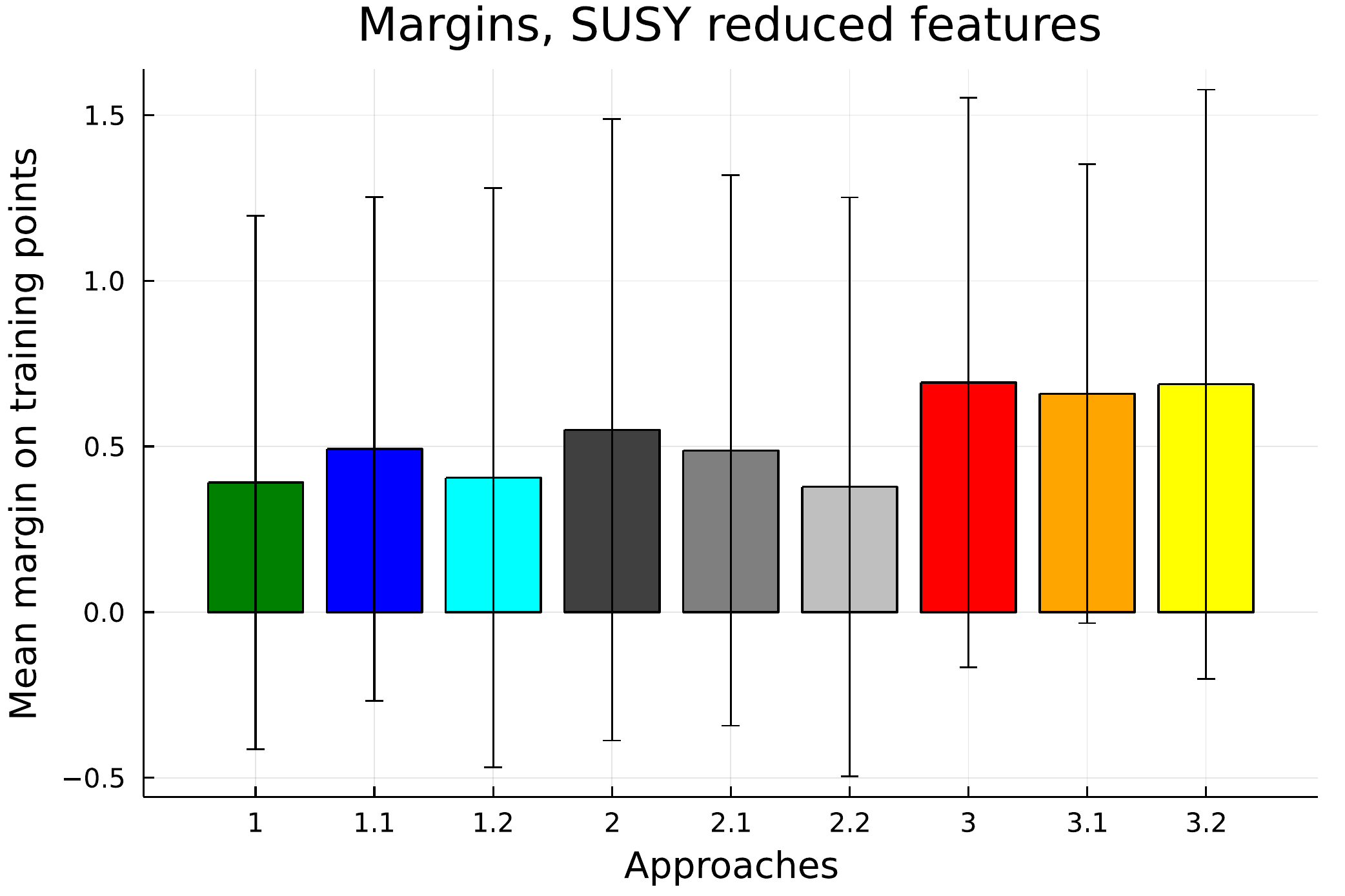}
\caption{A graph showing the mean margin of the SUSY reduced features training set points for the best classifiers produced by each approach, with errors bars showing standard deviation.}
\label{figure:susy_hard_margins_appendix}
\end{figure}

Figures \ref{figure:moons_margins_appendix} - \ref{figure:susy_hard_margins_appendix} show the margins of the same best produced kernels on the training data from each dataset. The approaches based on maximizing kernel-target alignment and its approximation tend to have larger average margin sizes, and final parameter training for RMSE and kernel-target alignment also tends to increase the margin sizes. The exception to the trend was the Iris dataset, which was also the second smallest dataset with only 42 training set points.

\begin{figure}
\includegraphics[width=0.9\columnwidth]{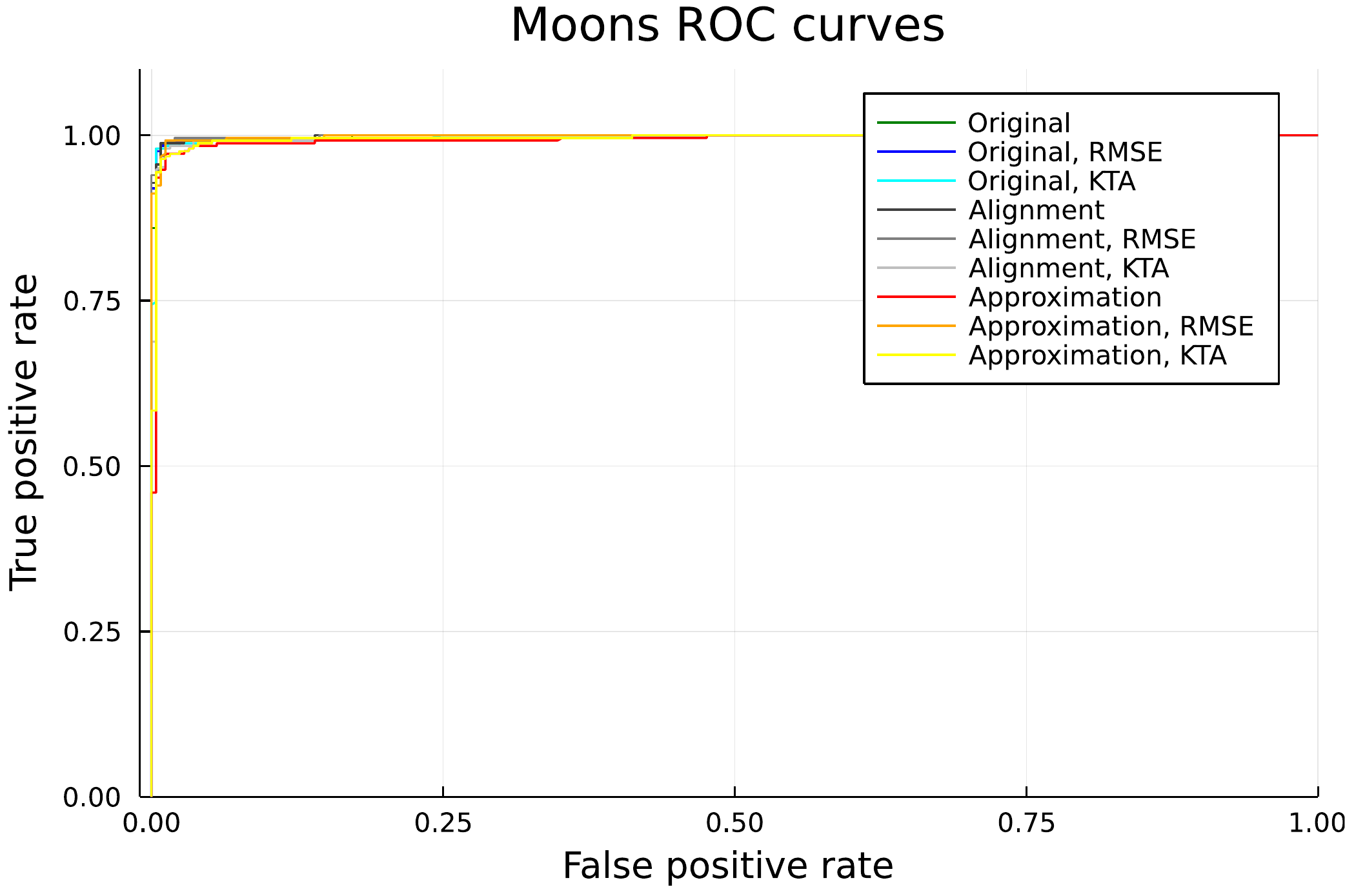}
\caption{A graph showing the ROC curves of the best models produced by various approaches of quantum feature map design on the Moons dataset.}
\label{figure:moons_roc_curves_appendix}
\end{figure}

\begin{figure}
\includegraphics[width=0.9\columnwidth]{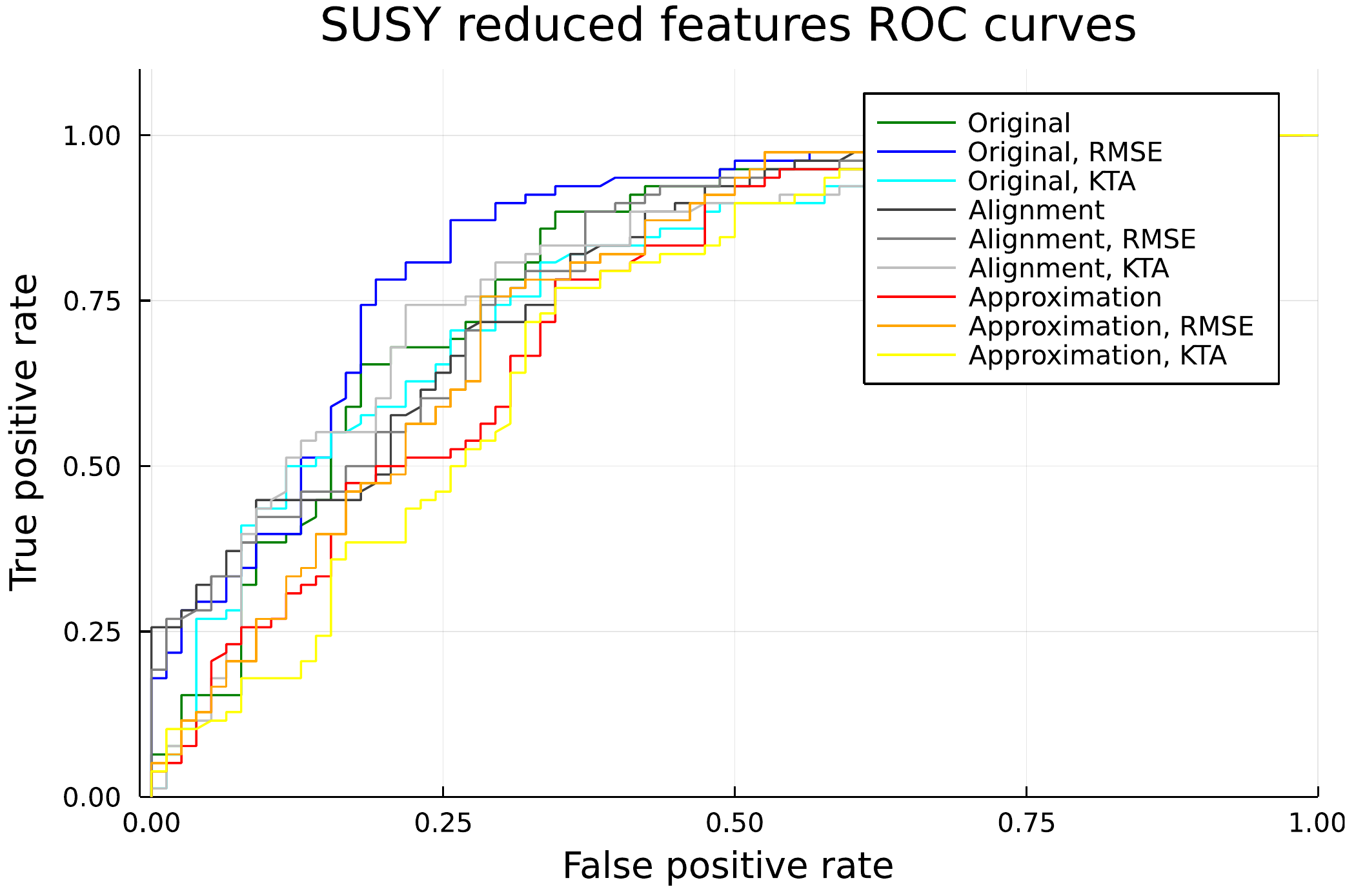}
\caption{A graph showing the ROC curves of the best models produced by various approaches of quantum feature map design on the SUSY reduced features dataset.}
\label{figure:susy_hard_roc_curves_appendix}
\end{figure}

Figures \ref{figure:moons_roc_curves_appendix} - \ref{figure:susy_hard_roc_curves_appendix} show the ROC curves of the best produced models for each dataset. Across the datasets, the curves are mostly similar when comparing approaches.

\begin{figure}
\begin{adjustbox}{center}
\begin{subfigure}{0.33\paperwidth}
	\includegraphics[width=\columnwidth]{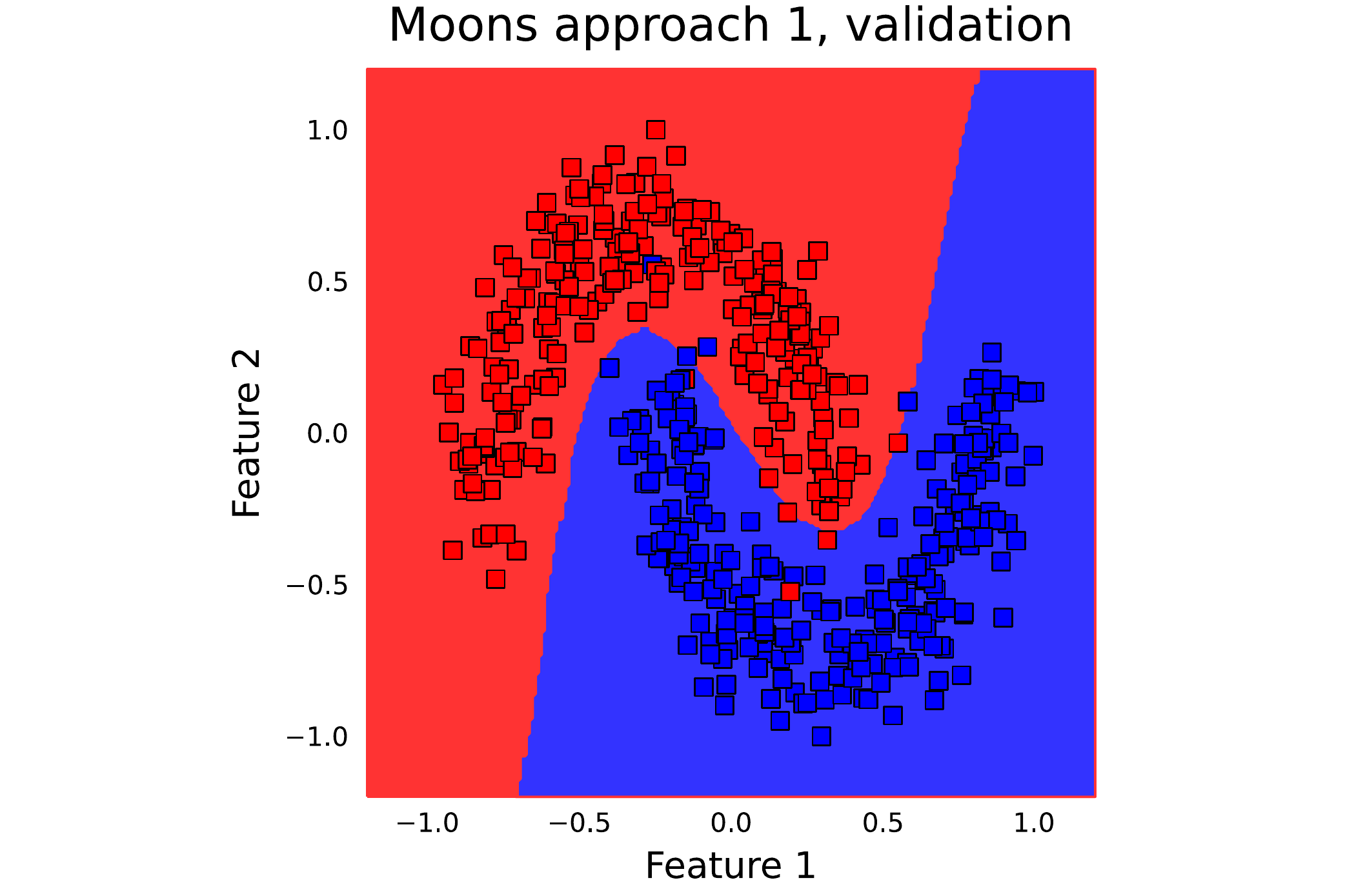}
\end{subfigure}%
\begin{subfigure}{0.33\paperwidth}
	\includegraphics[width=\columnwidth]{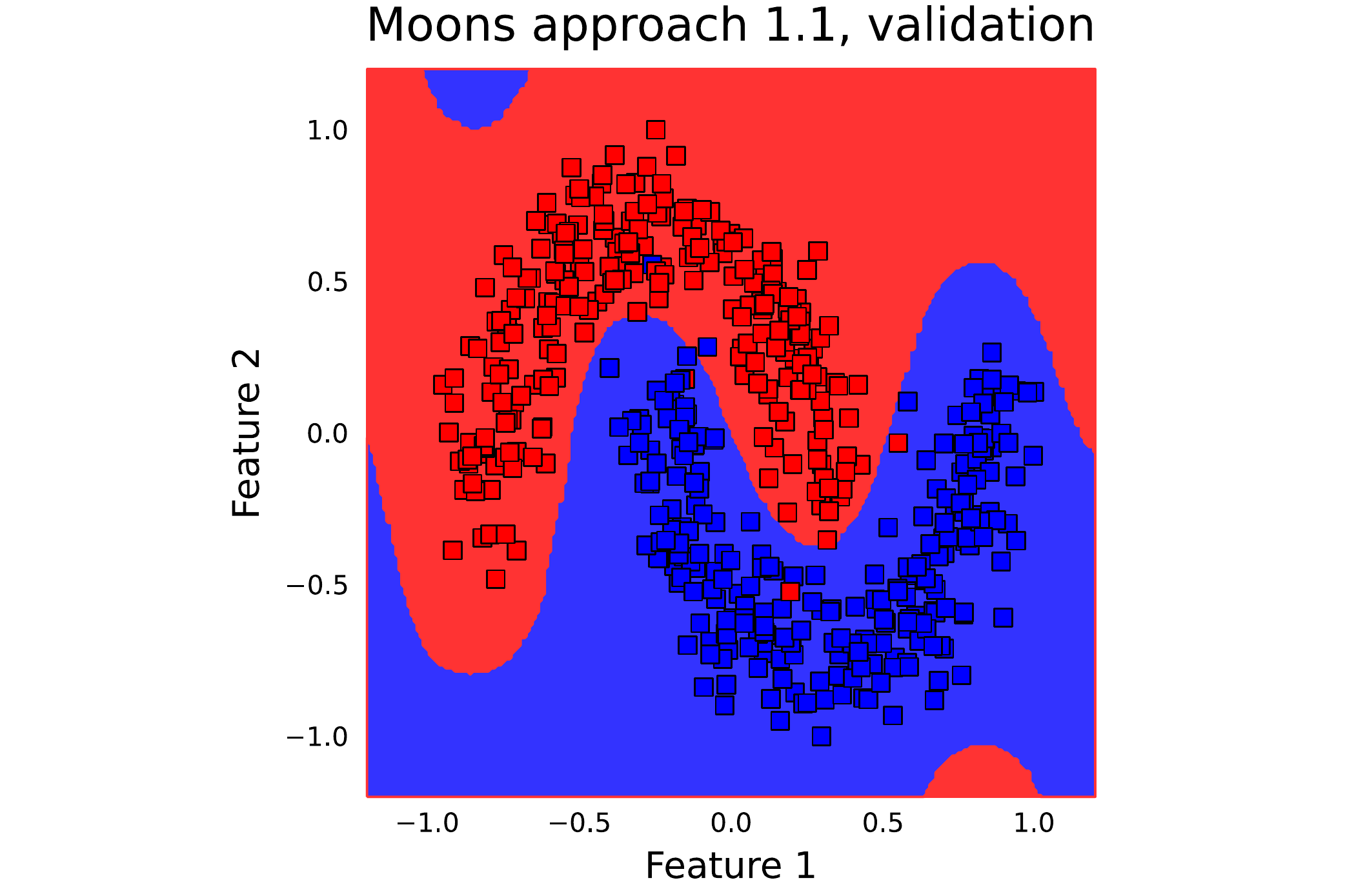}
\end{subfigure}%
\begin{subfigure}{0.33\paperwidth}
	\includegraphics[width=\columnwidth]{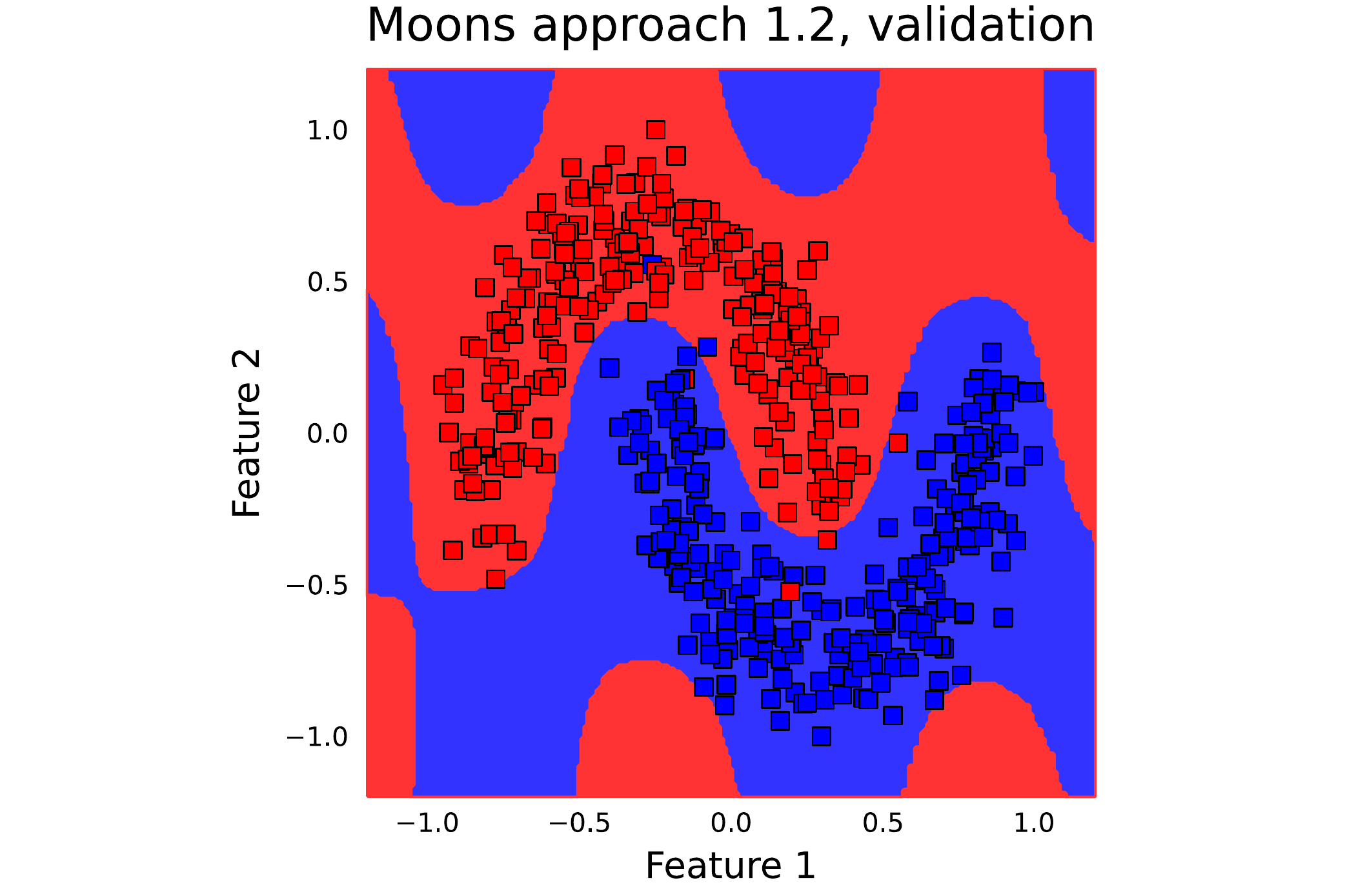}
\end{subfigure}
\end{adjustbox}

\begin{adjustbox}{center}
\begin{subfigure}{0.33\paperwidth}
	\includegraphics[width=\columnwidth]{Fig9_1.pdf}
\end{subfigure}%
\begin{subfigure}{0.33\paperwidth}
	\includegraphics[width=\columnwidth]{Fig9_2.pdf}
\end{subfigure}%
\begin{subfigure}{0.33\paperwidth}
	\includegraphics[width=\columnwidth]{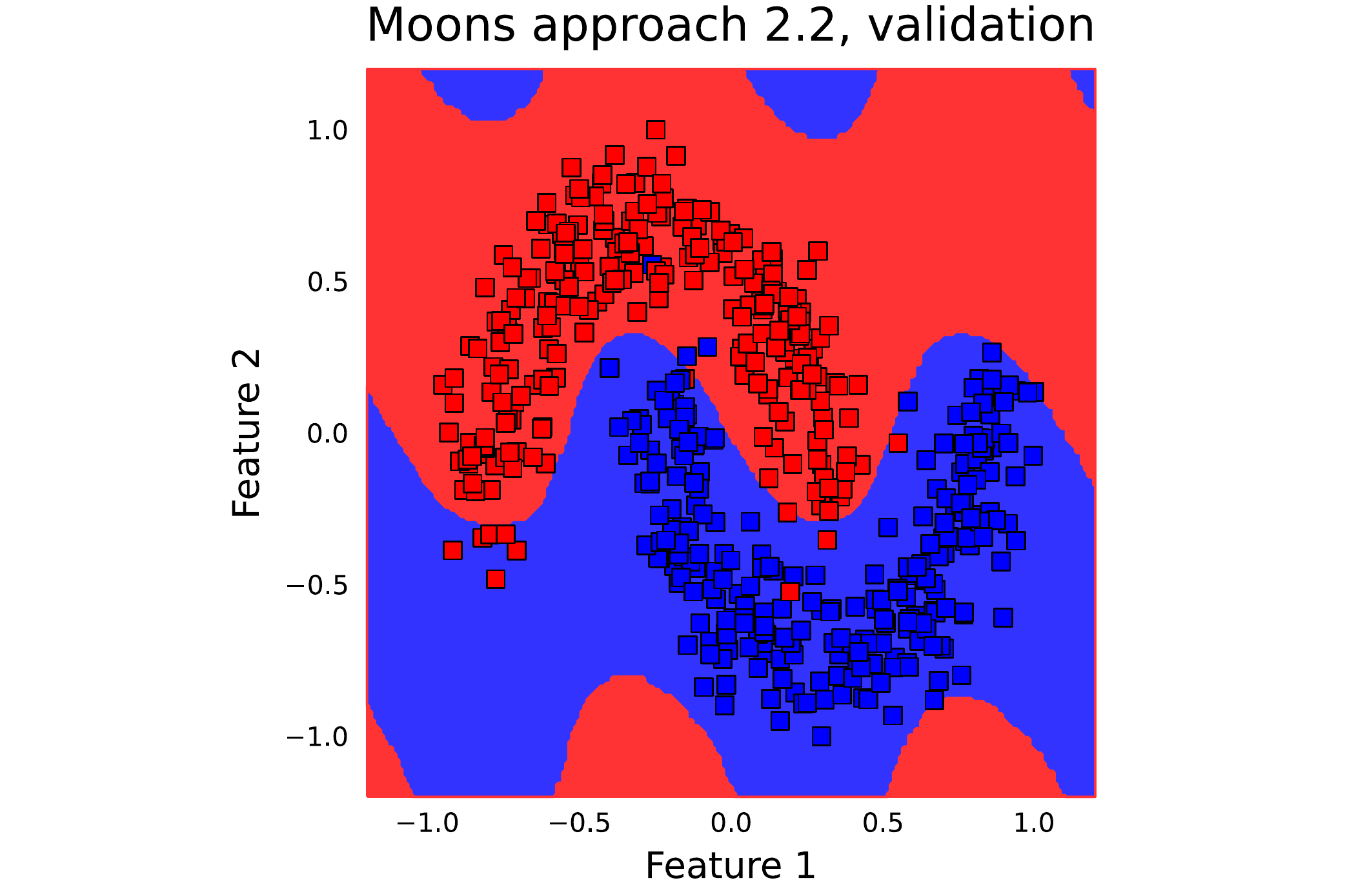}
\end{subfigure}
\end{adjustbox}

\begin{adjustbox}{center}
\begin{subfigure}{0.33\paperwidth}
	\includegraphics[width=\columnwidth]{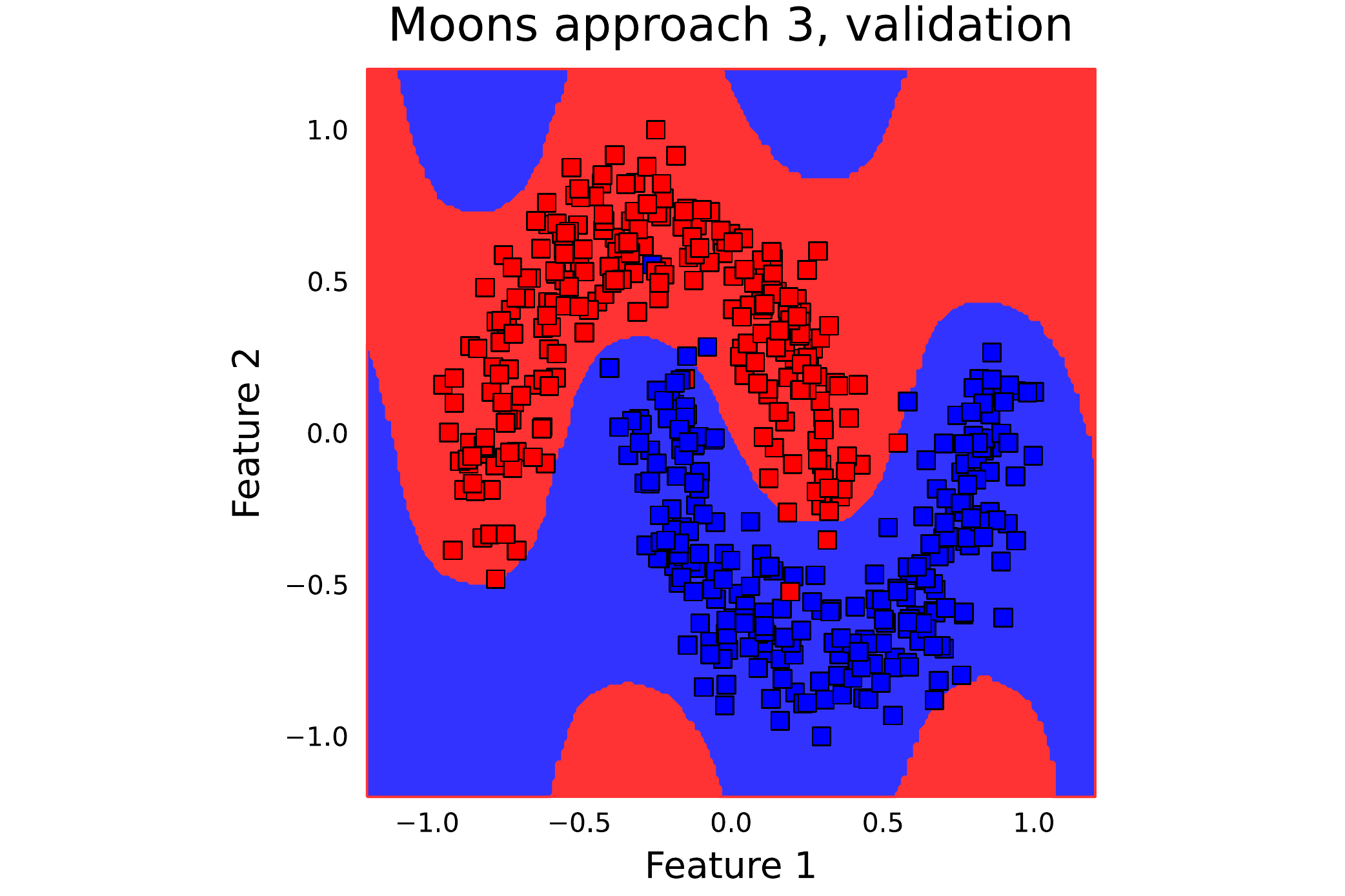}
\end{subfigure}%
\begin{subfigure}{0.33\paperwidth}
	\includegraphics[width=\columnwidth]{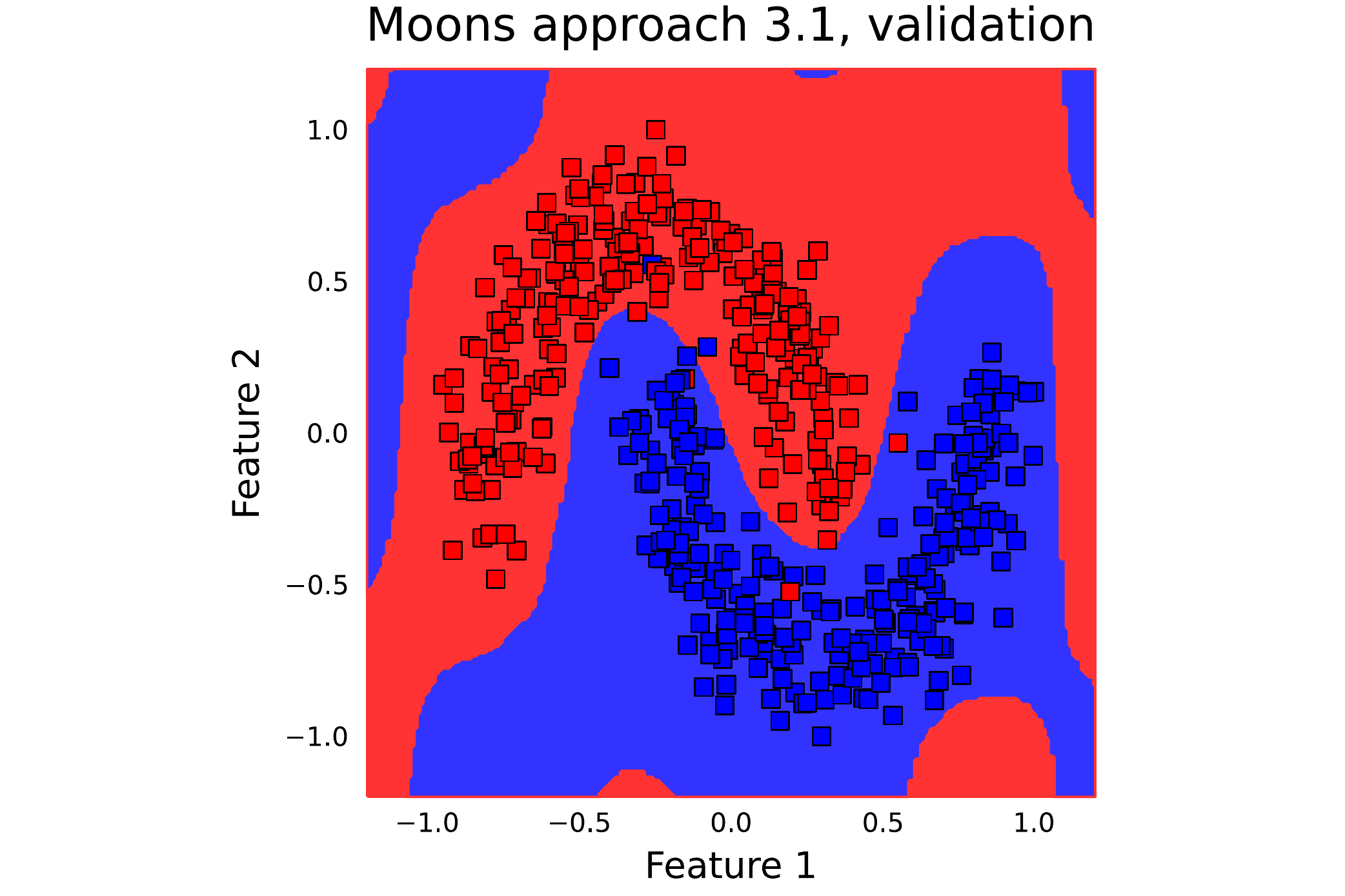}
\end{subfigure}%
\begin{subfigure}{0.33\paperwidth}
	\includegraphics[width=\columnwidth]{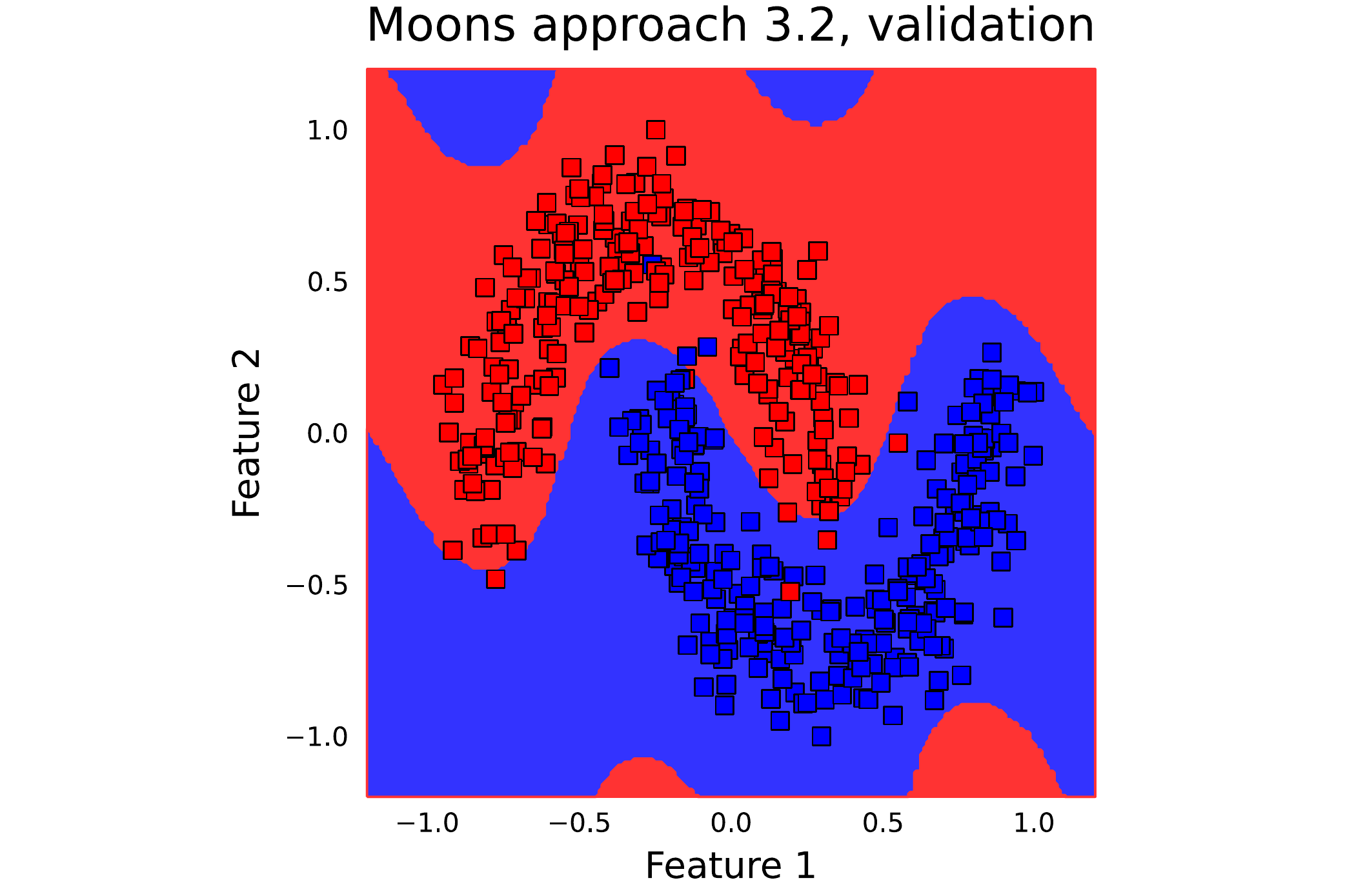}
\end{subfigure}
\end{adjustbox}
\caption{Decision boundaries for each approach on the Moons validation data.}
\label{figure:decision_boundaries_moons_appendix}
\end{figure}

\begin{figure}
\begin{adjustbox}{center}
\begin{subfigure}{0.33\paperwidth}
	\includegraphics[width=\columnwidth]{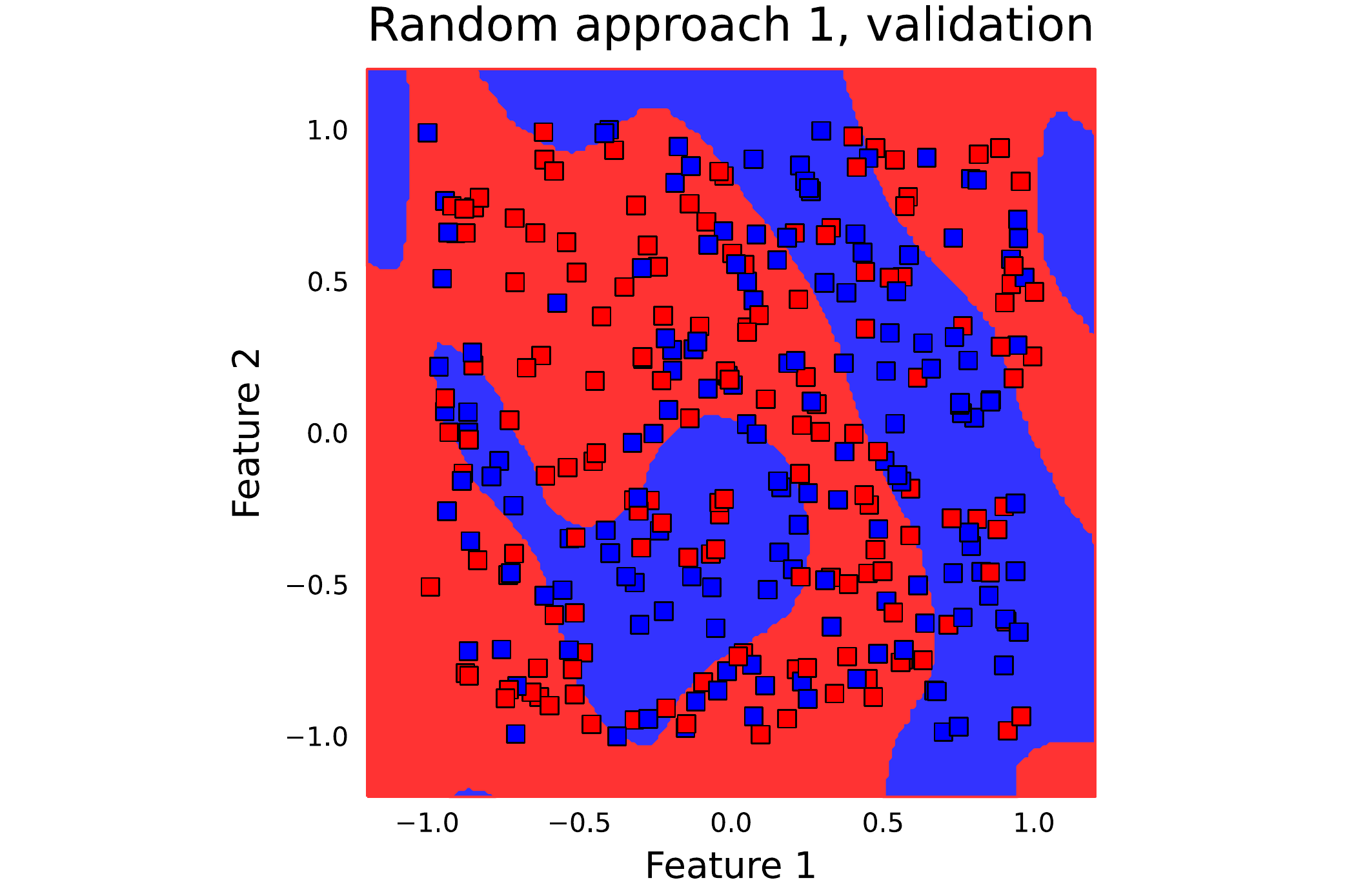}
\end{subfigure}%
\begin{subfigure}{0.33\paperwidth}
	\includegraphics[width=\columnwidth]{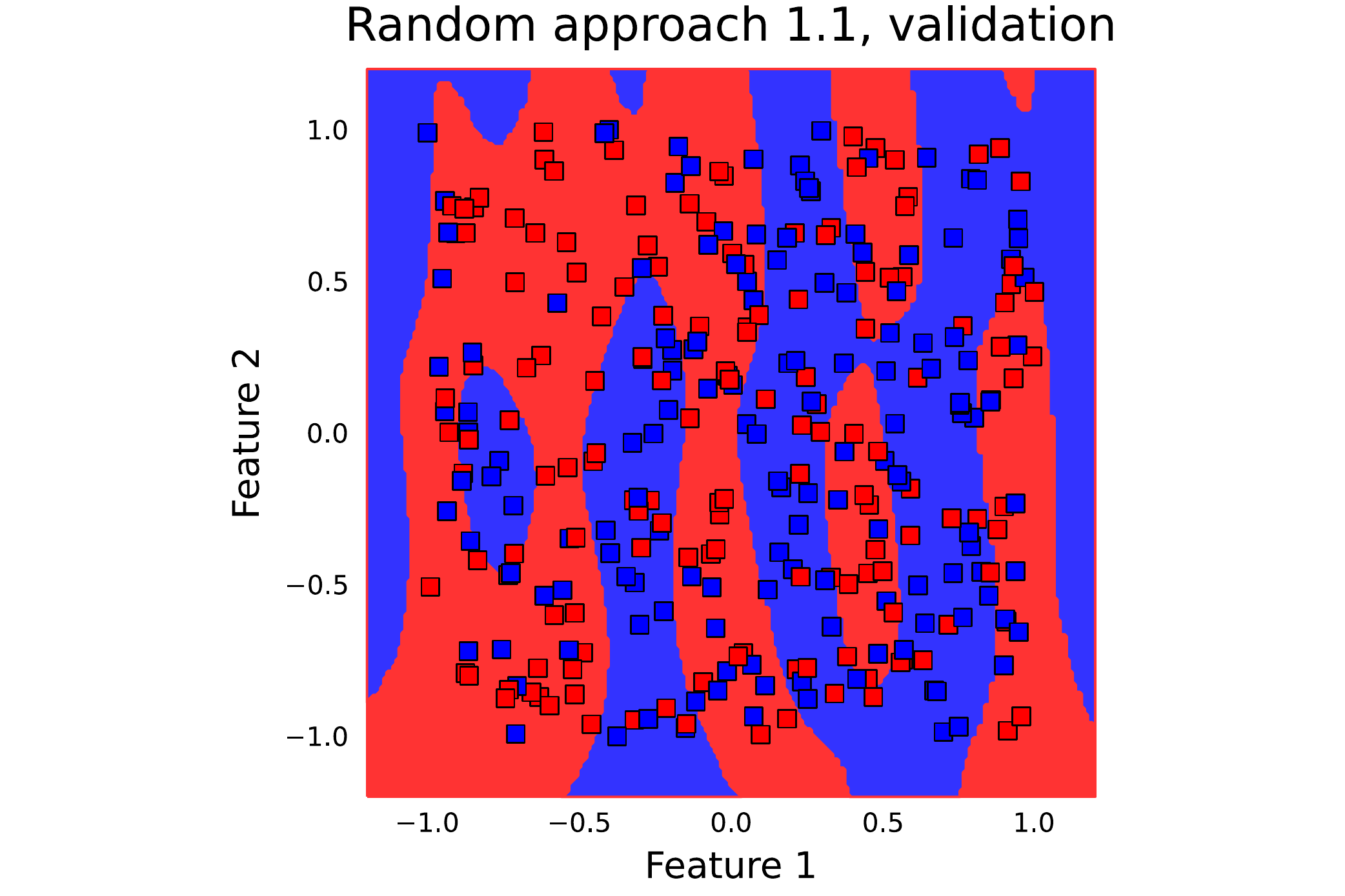}
\end{subfigure}%
\begin{subfigure}{0.33\paperwidth}
	\includegraphics[width=\columnwidth]{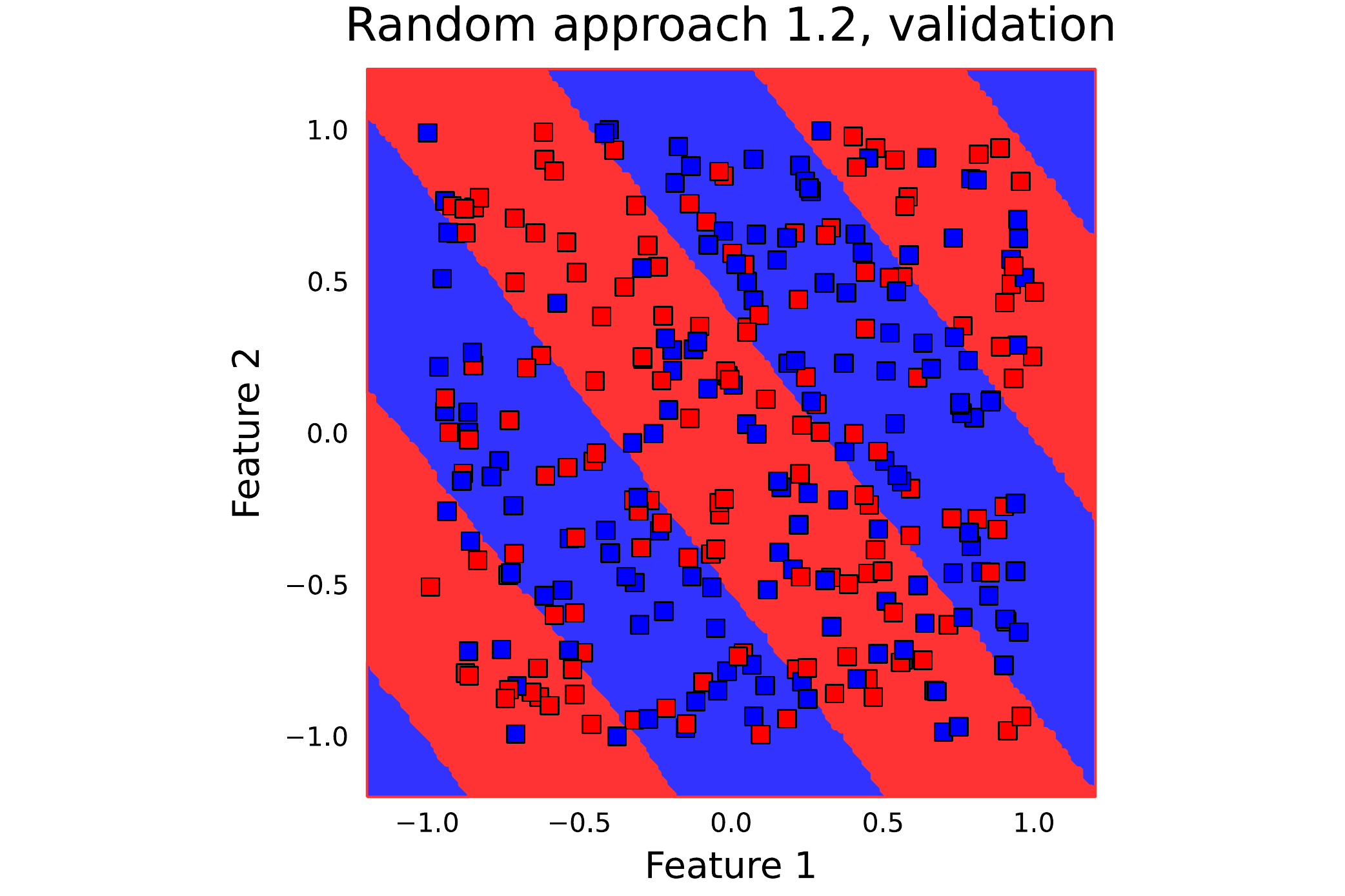}
\end{subfigure}
\end{adjustbox}

\begin{adjustbox}{center}
\begin{subfigure}{0.33\paperwidth}
	\includegraphics[width=\columnwidth]{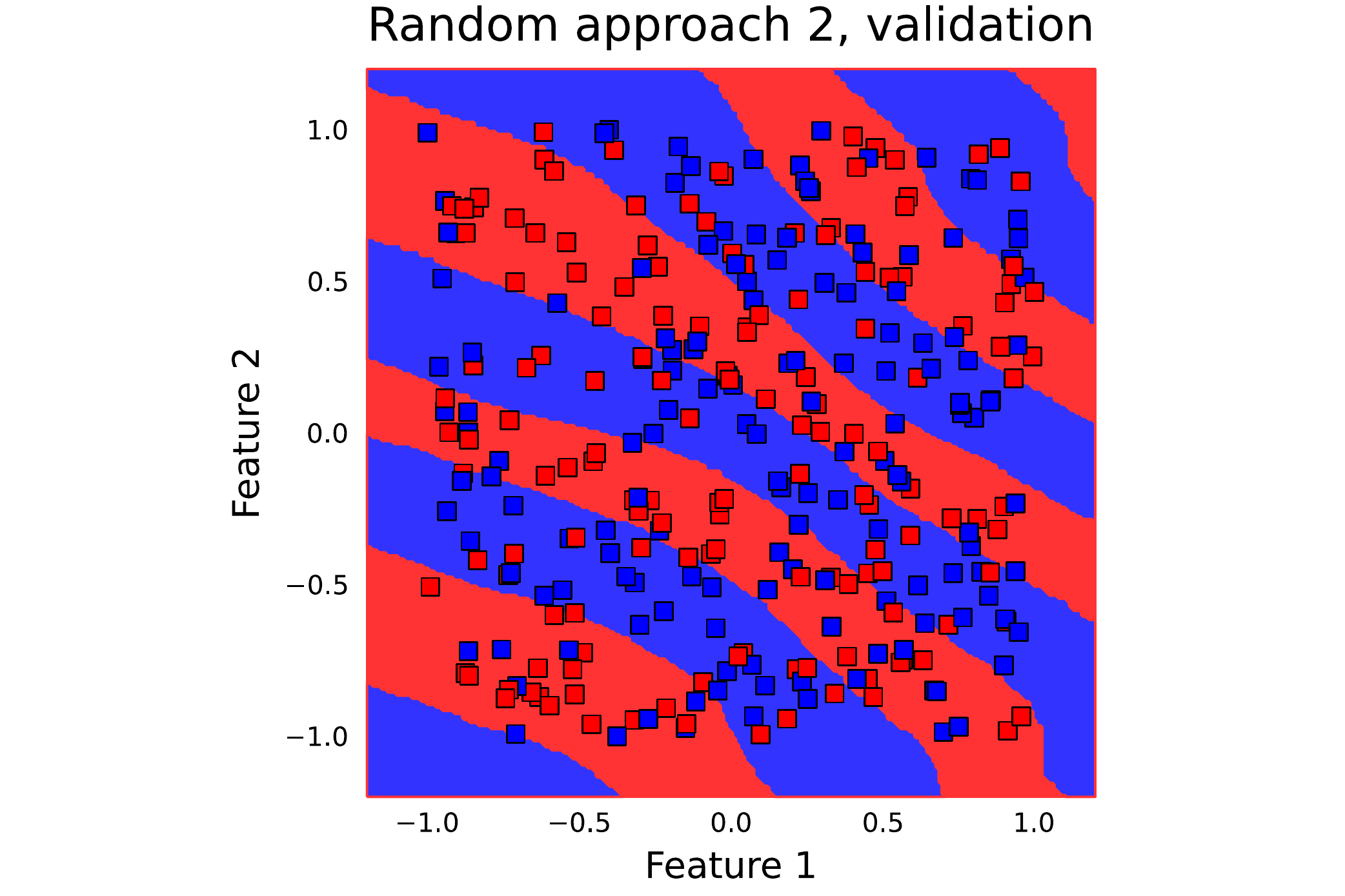}
\end{subfigure}%
\begin{subfigure}{0.33\paperwidth}
	\includegraphics[width=\columnwidth]{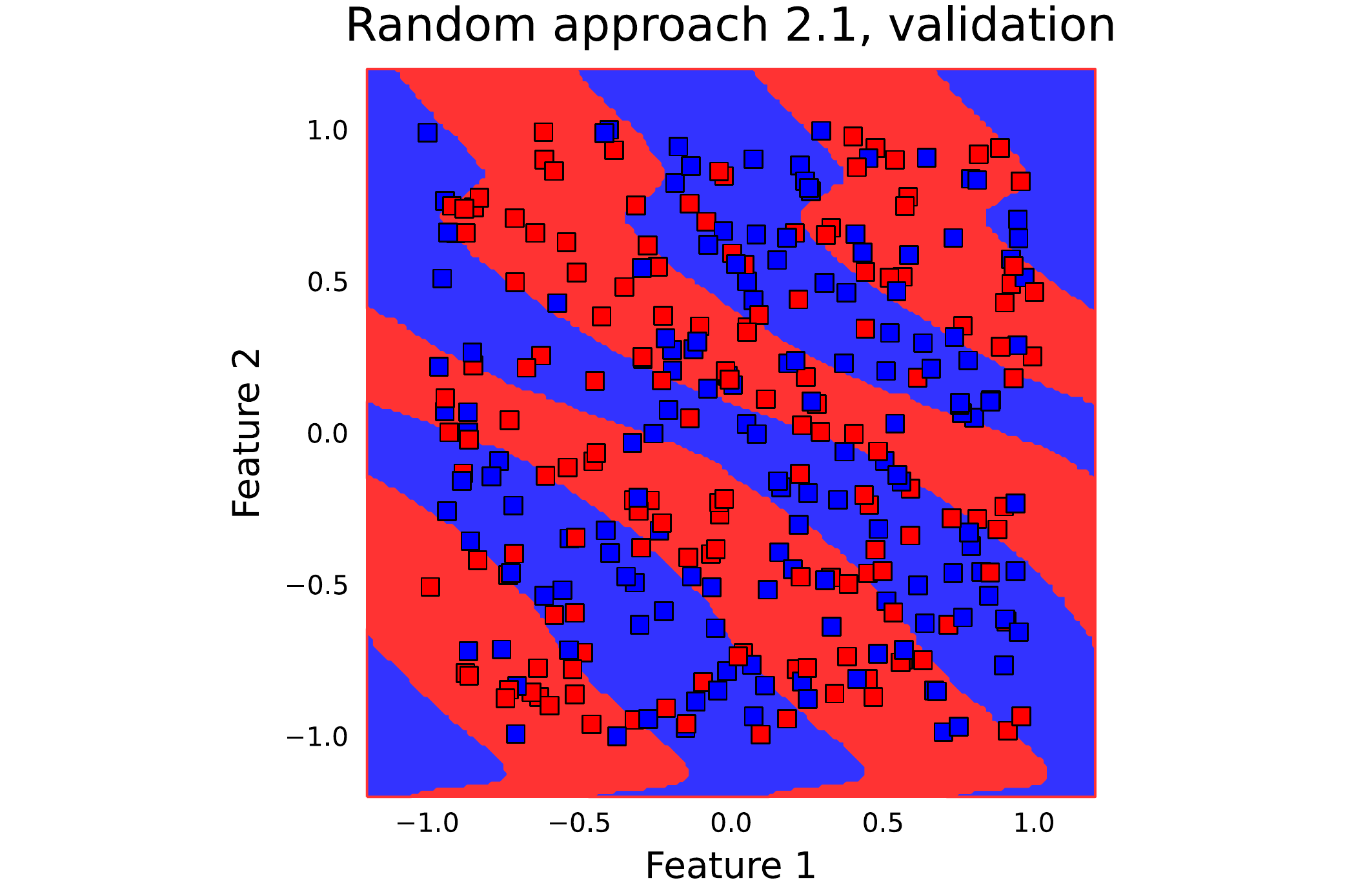}
\end{subfigure}%
\begin{subfigure}{0.33\paperwidth}
	\includegraphics[width=\columnwidth]{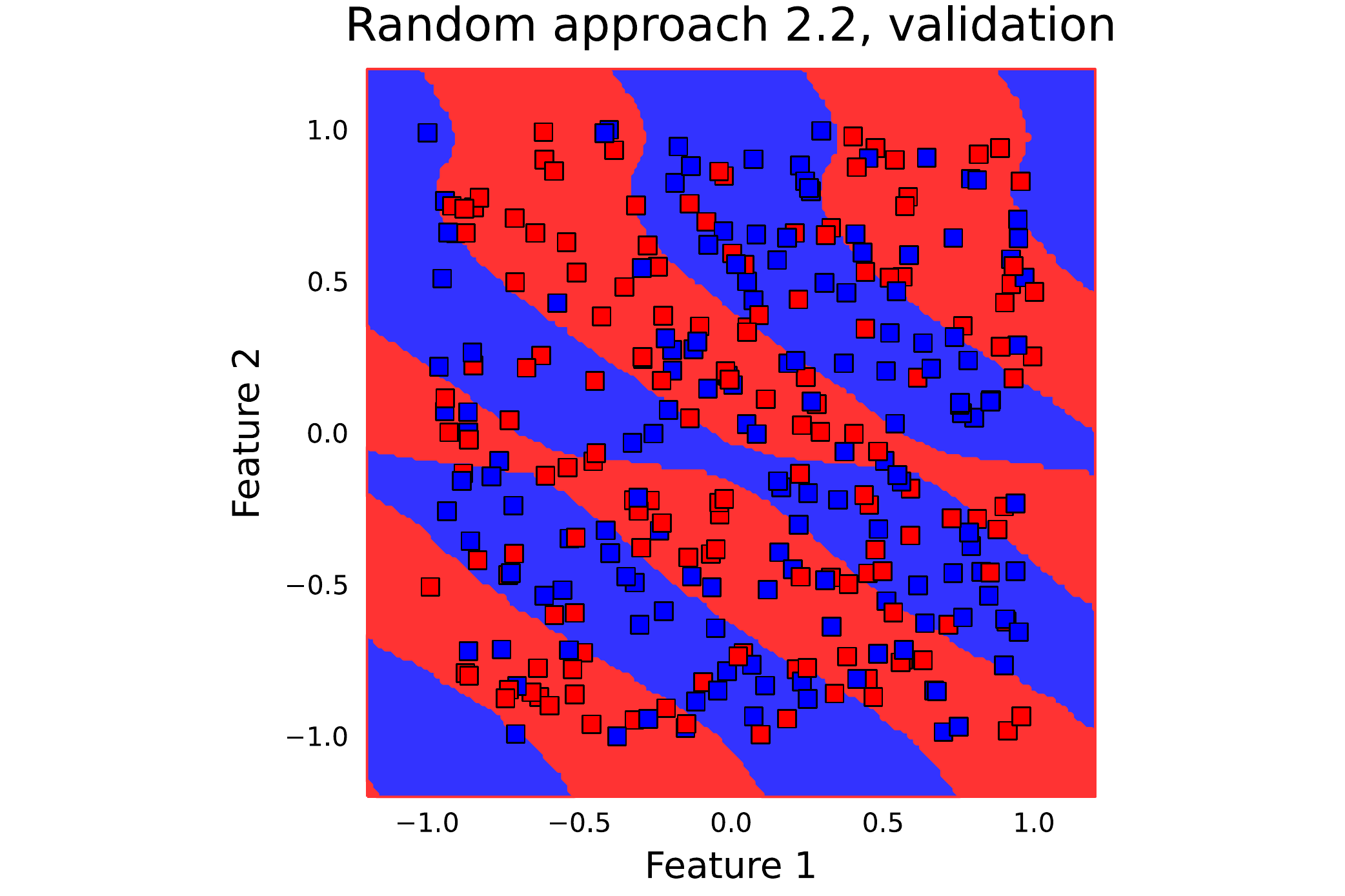}
\end{subfigure}
\end{adjustbox}

\begin{adjustbox}{center}
\begin{subfigure}{0.33\paperwidth}
	\includegraphics[width=\columnwidth]{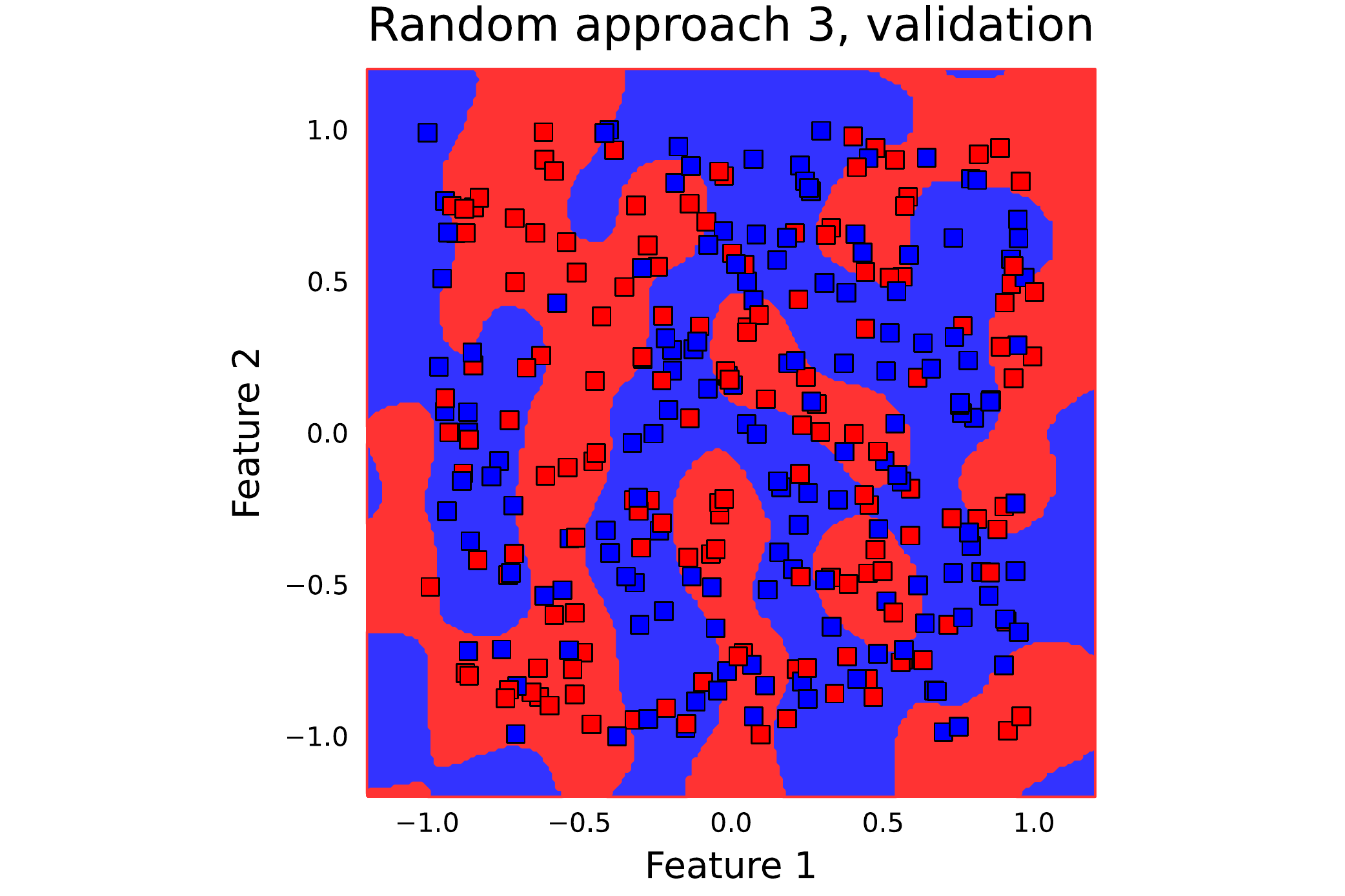}
\end{subfigure}%
\begin{subfigure}{0.33\paperwidth}
	\includegraphics[width=\columnwidth]{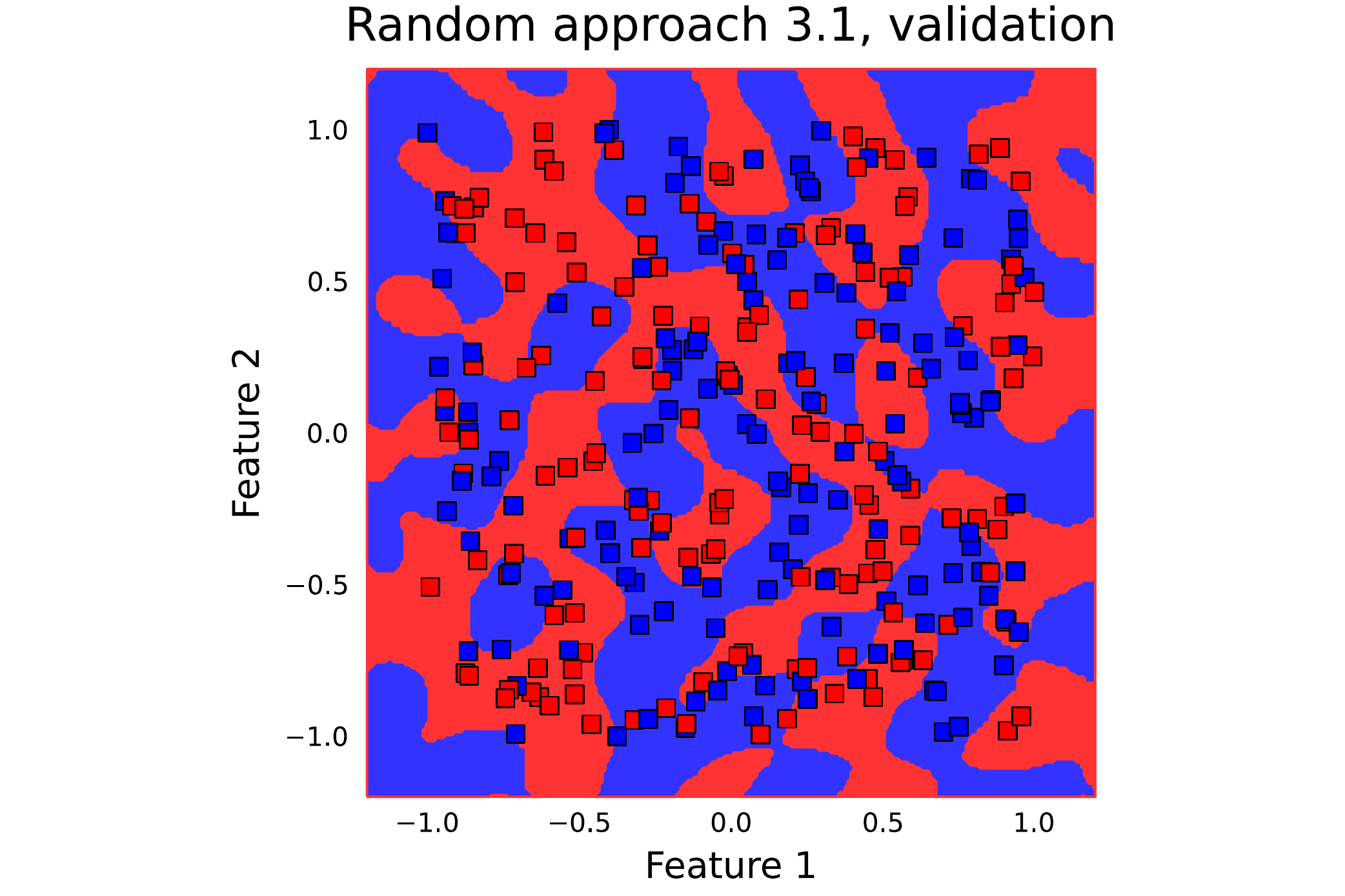}
\end{subfigure}%
\begin{subfigure}{0.33\paperwidth}
	\includegraphics[width=\columnwidth]{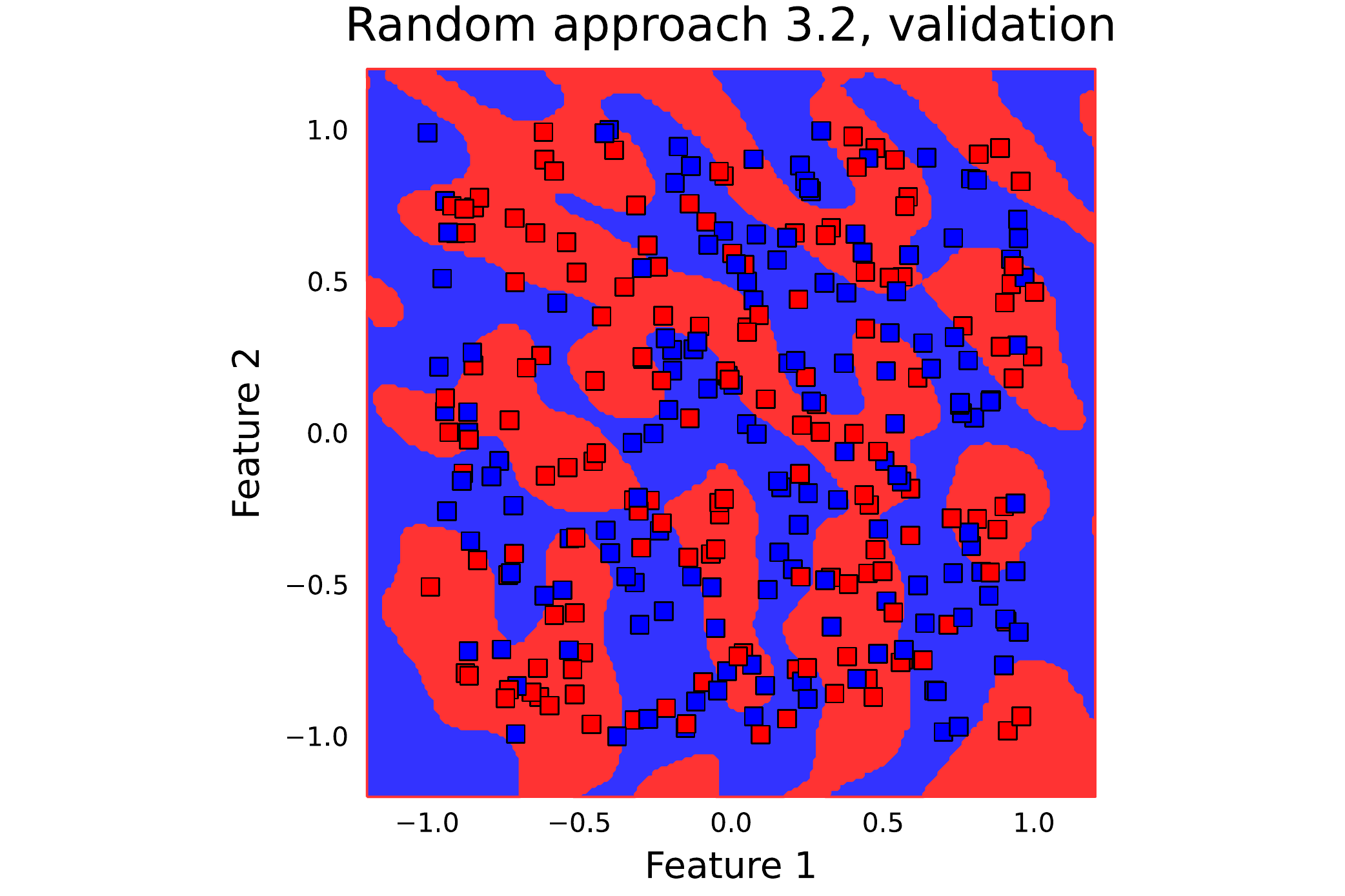}
\end{subfigure}
\end{adjustbox}
\caption{Decision boundaries for each approach on the Random validation data. In the case of the Random dataset, the validation data does not consist of unseen randomly generated points as would be the case in other datasets. Under the assumption that the models cannot generalise to new pseudorandom data, the validation dataset instead shows the concatenation of the training and testing points to give an idea of the different approaches' capacity for memorizing a random assignment of labels to random points.}
\label{figure:decision_boundaries_random_appendix}
\end{figure}

Figures \ref{figure:decision_boundaries_moons_appendix} - \ref{figure:decision_boundaries_random_appendix} show the decision boundaries resulting from the best feature map produced by each approach, for the datasets with two dimensional feature vectors. These can be inspected to see the improvements made by parameter training as well as the complexity of the decision rules produced by each approach.


\begin{figure}[h]
\includegraphics[width=0.9\columnwidth]{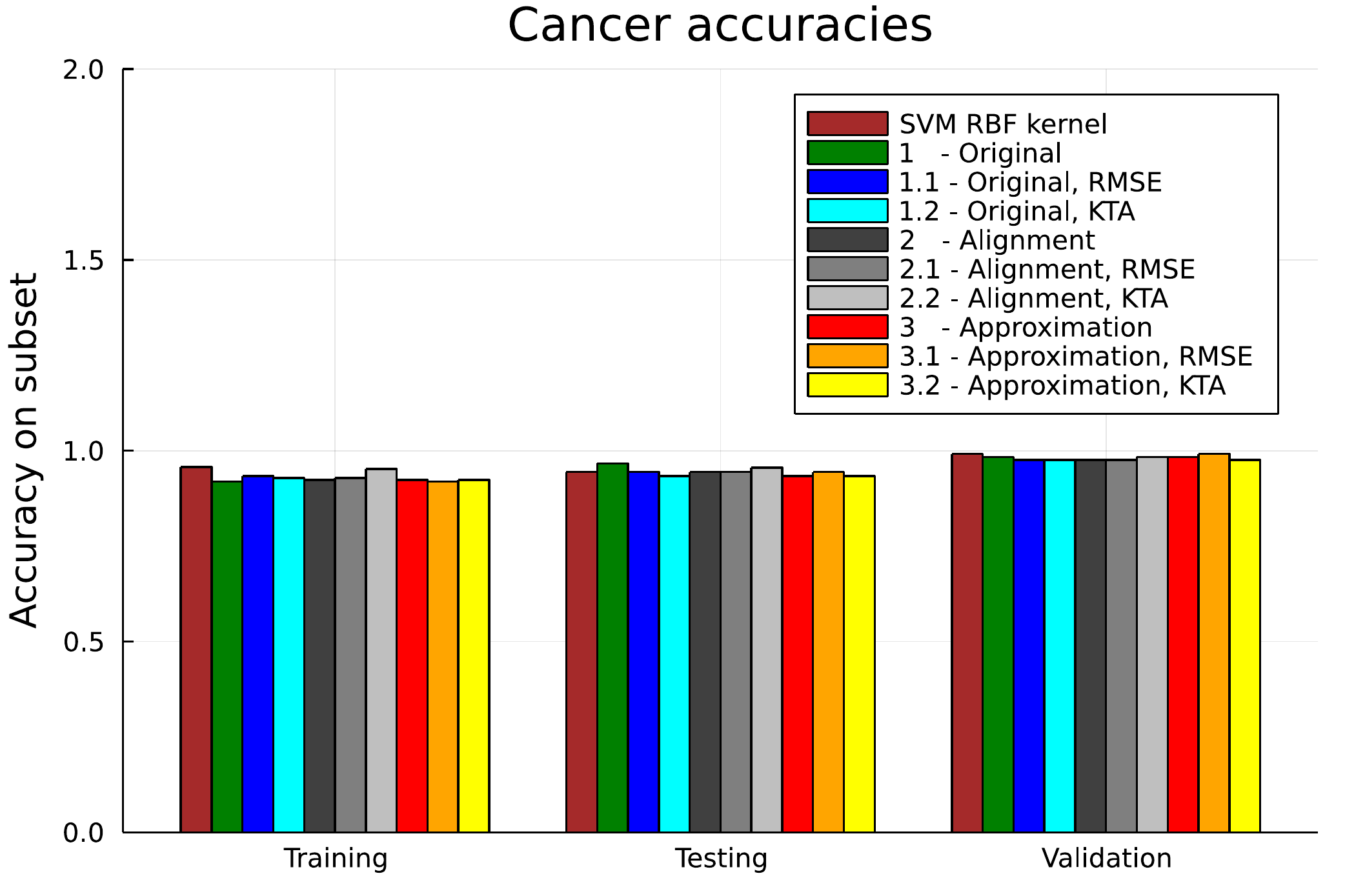}
\caption{A graph showing the classification accuracies of the best models produced by various approaches of quantum feature map design on the Cancer dataset, compared with a classical RBF kernel for reference.}
\end{figure}
\begin{figure}[h]
\includegraphics[width=0.9\columnwidth]{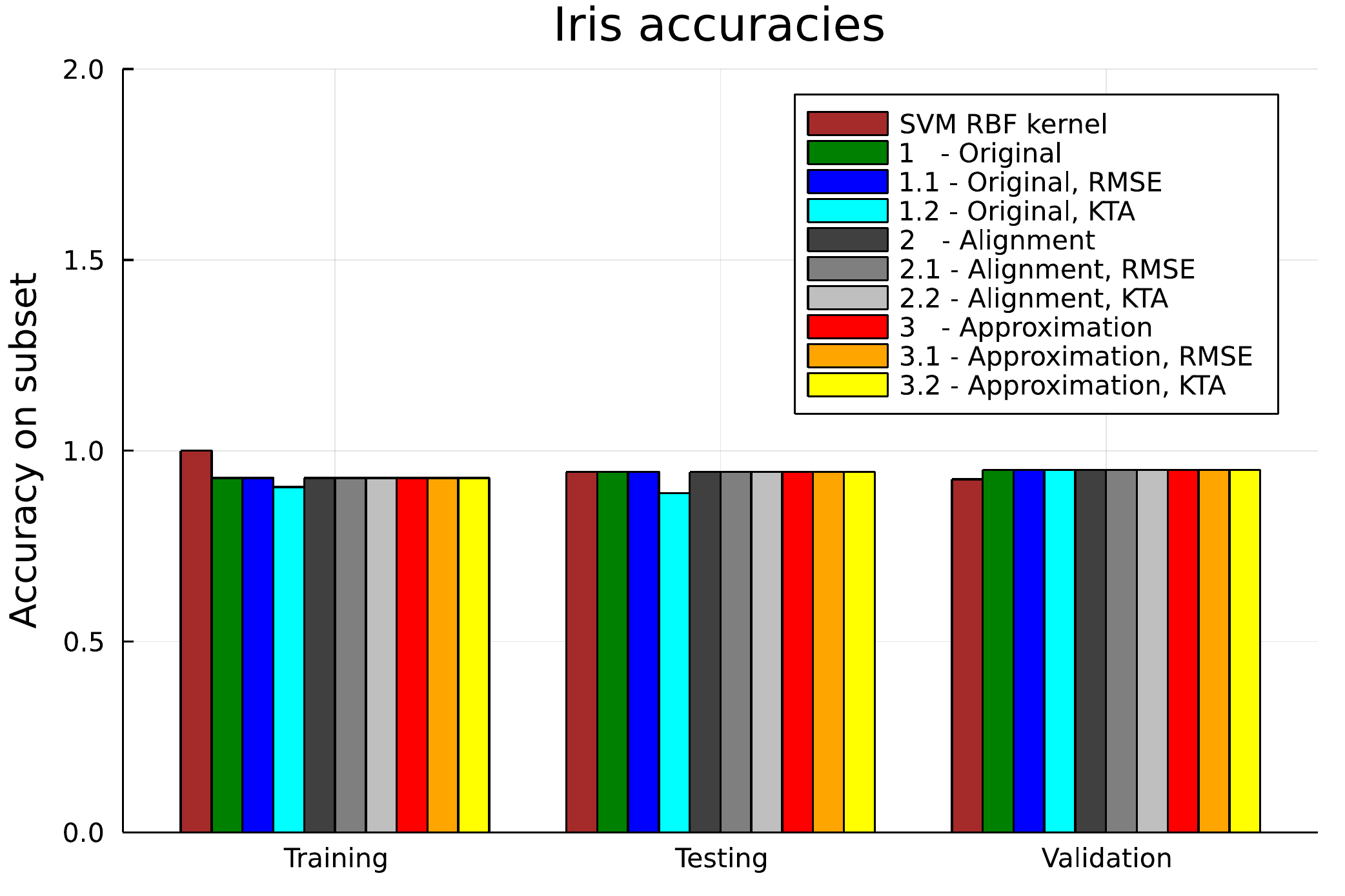}
\caption{A graph showing the classification accuracies of the best models produced by various approaches of quantum feature map design on the Iris dataset, compared with a classical RBF kernel for reference.}
\end{figure}
\begin{figure}[h]
\includegraphics[width=0.9\columnwidth]{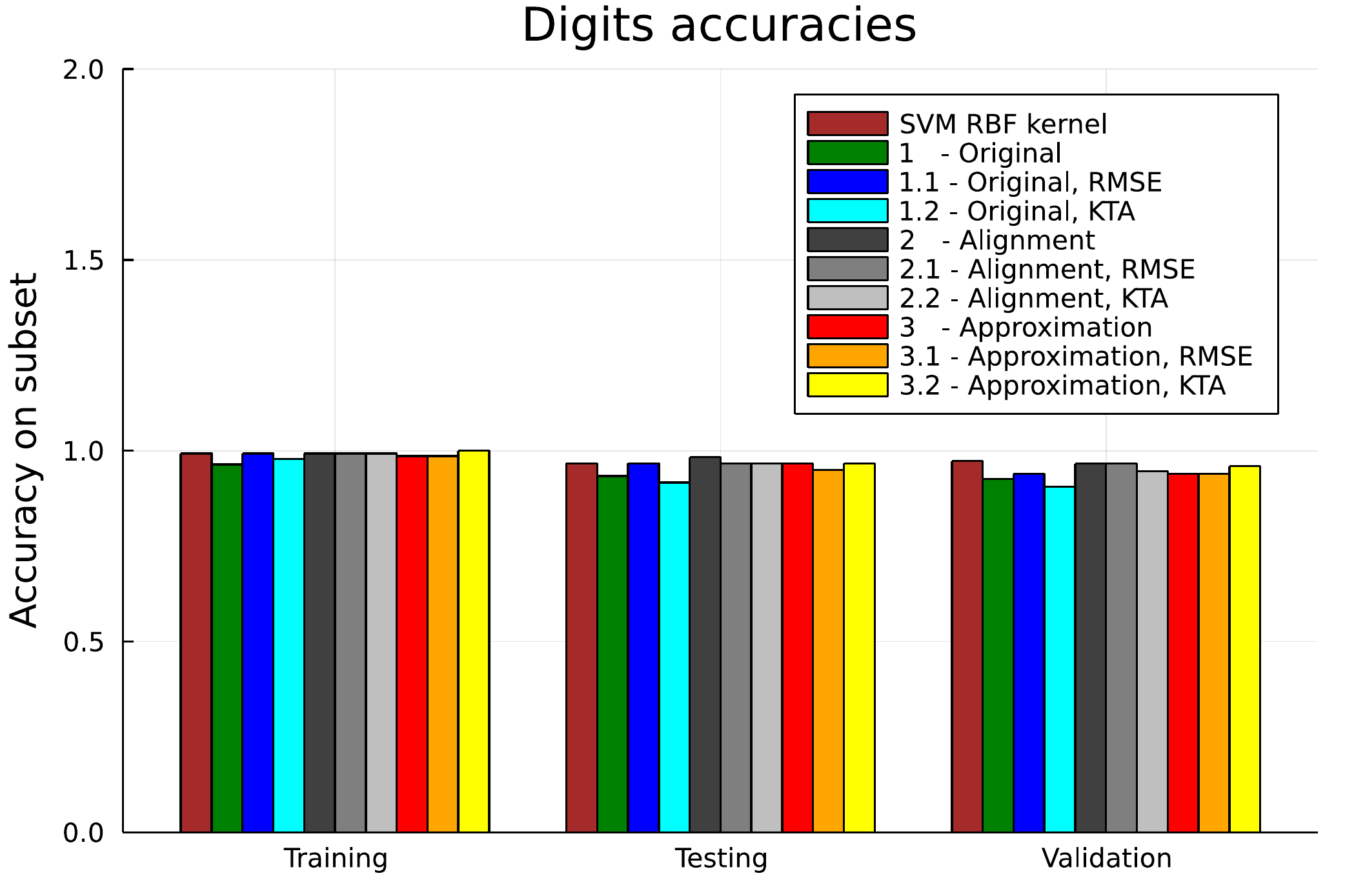}
\caption{A graph showing the classification accuracies of the best models produced by various approaches of quantum feature map design on the Digits dataset, compared with a classical RBF kernel for reference.}
\end{figure}
\begin{figure}[h]
\includegraphics[width=0.9\columnwidth]{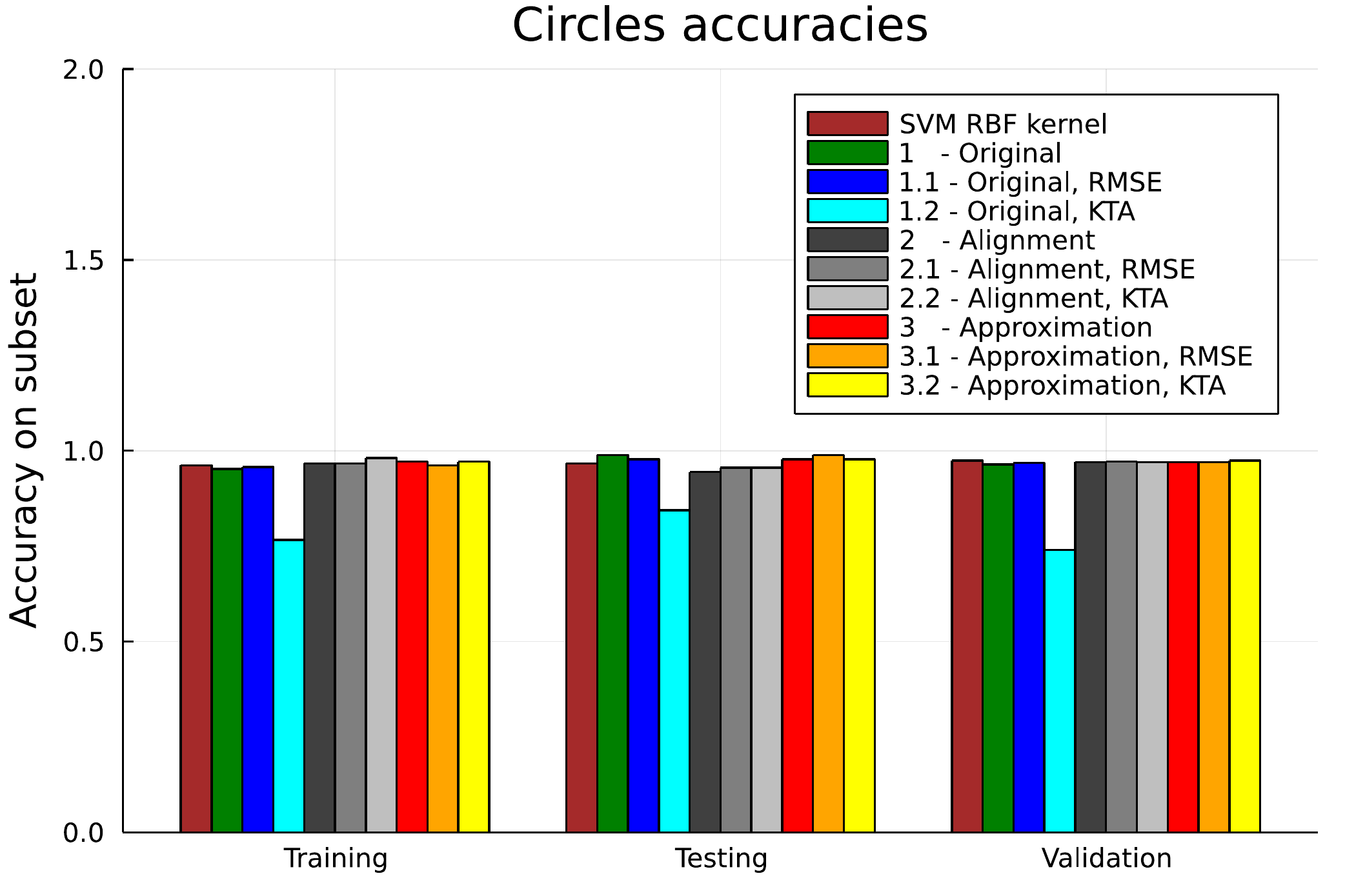}
\caption{A graph showing the classification accuracies of the best models produced by various approaches of quantum feature map design on the Circles dataset, compared with a classical RBF kernel for reference.}
\end{figure}
\begin{figure}[h]
\includegraphics[width=0.85\columnwidth]{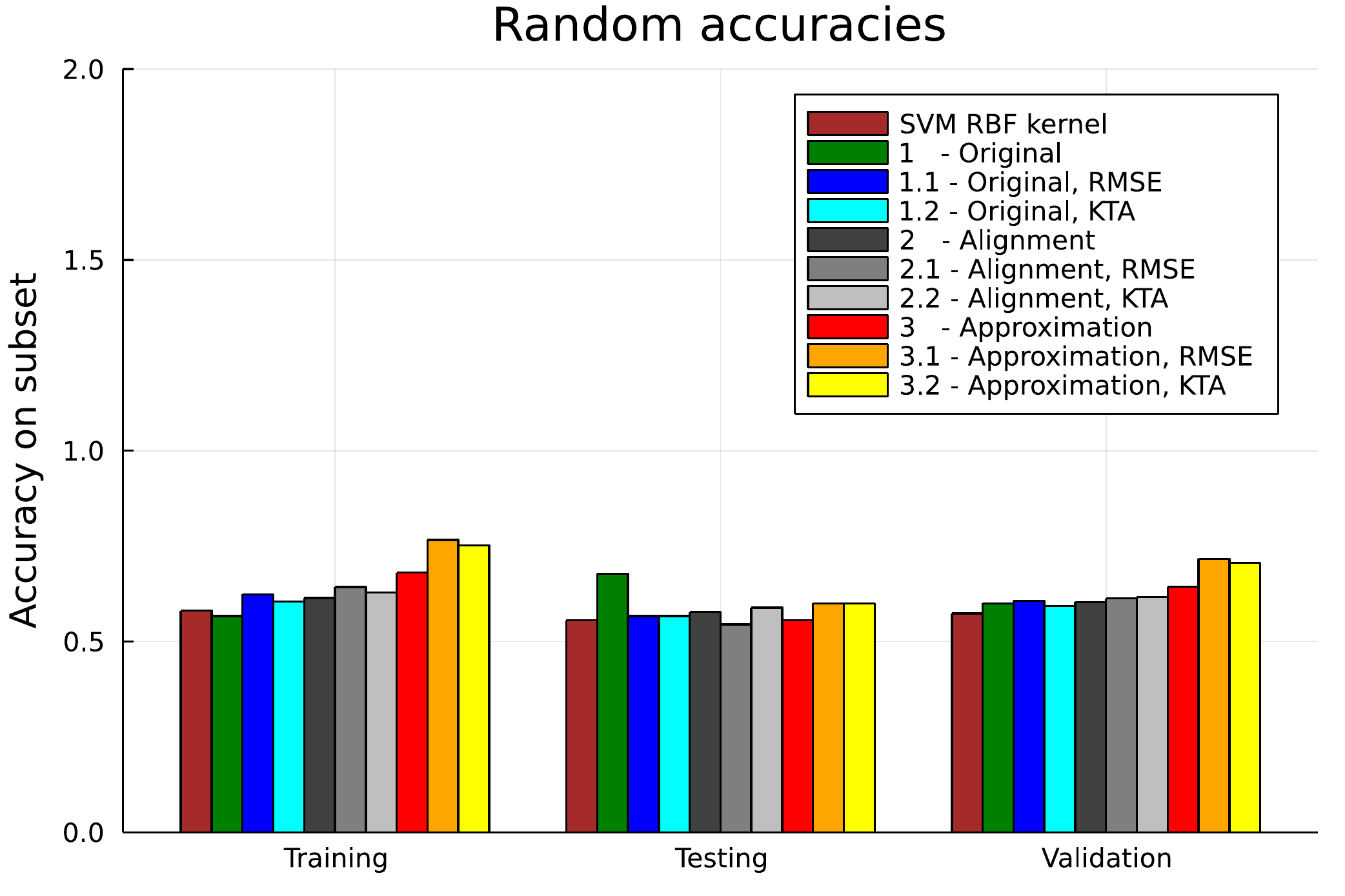}
\caption{A graph showing the classification accuracies of the best models produced by various approaches of quantum feature map design on the Random dataset, compared with a classical RBF kernel for reference. In the case of this dataset, the validation set is the union of the training and testing points. This gives an idea of the extent to which the approach was able to memorize all the points.}
\end{figure}
\begin{figure}[h]
\includegraphics[width=0.9\columnwidth]{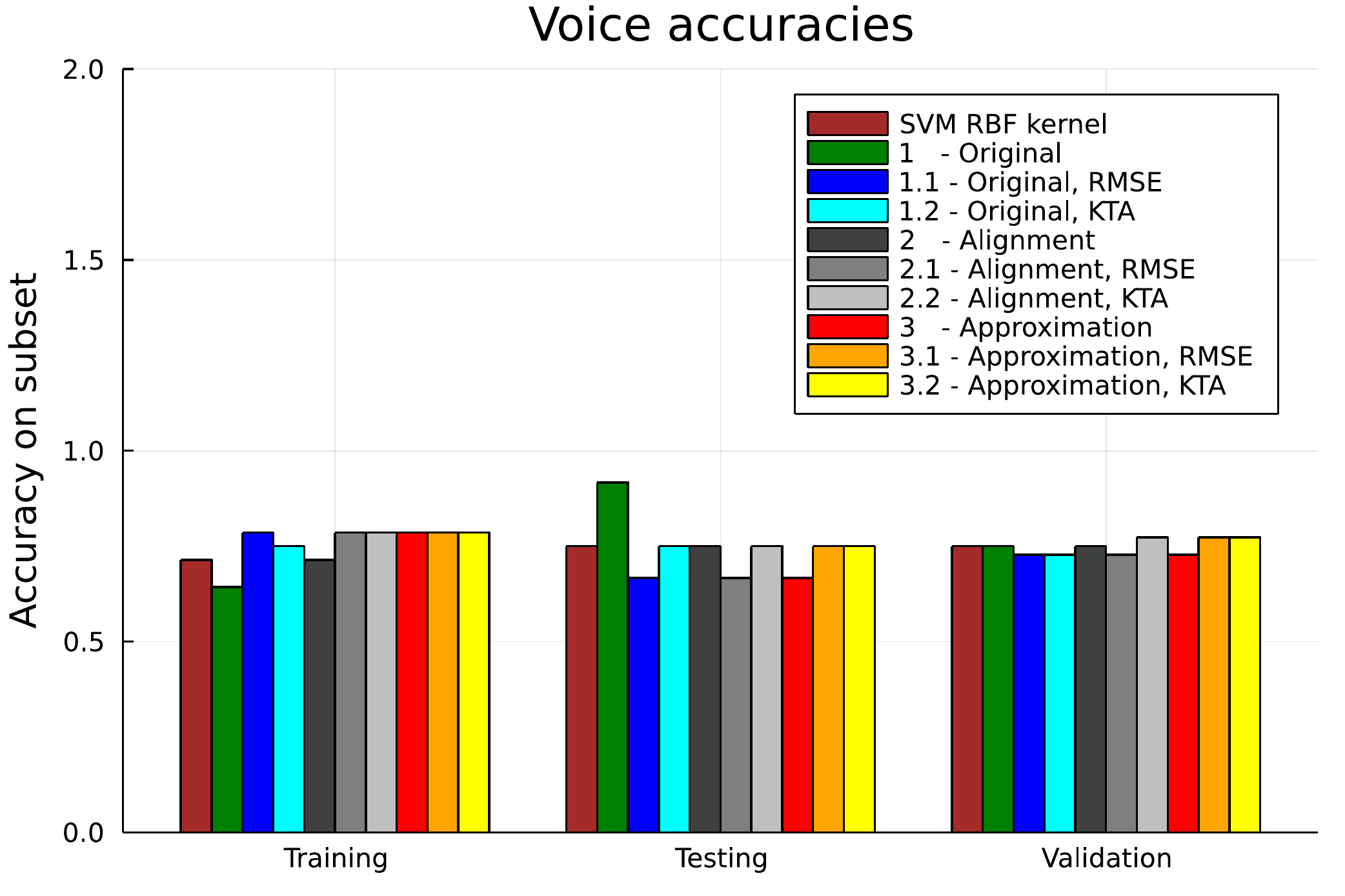}
\caption{A graph showing the classification accuracies of the best models produced by various approaches of quantum feature map design on the Voice dataset, compared with a classical RBF kernel for reference.}
\end{figure}
\begin{figure}[h]
\includegraphics[width=0.9\columnwidth]{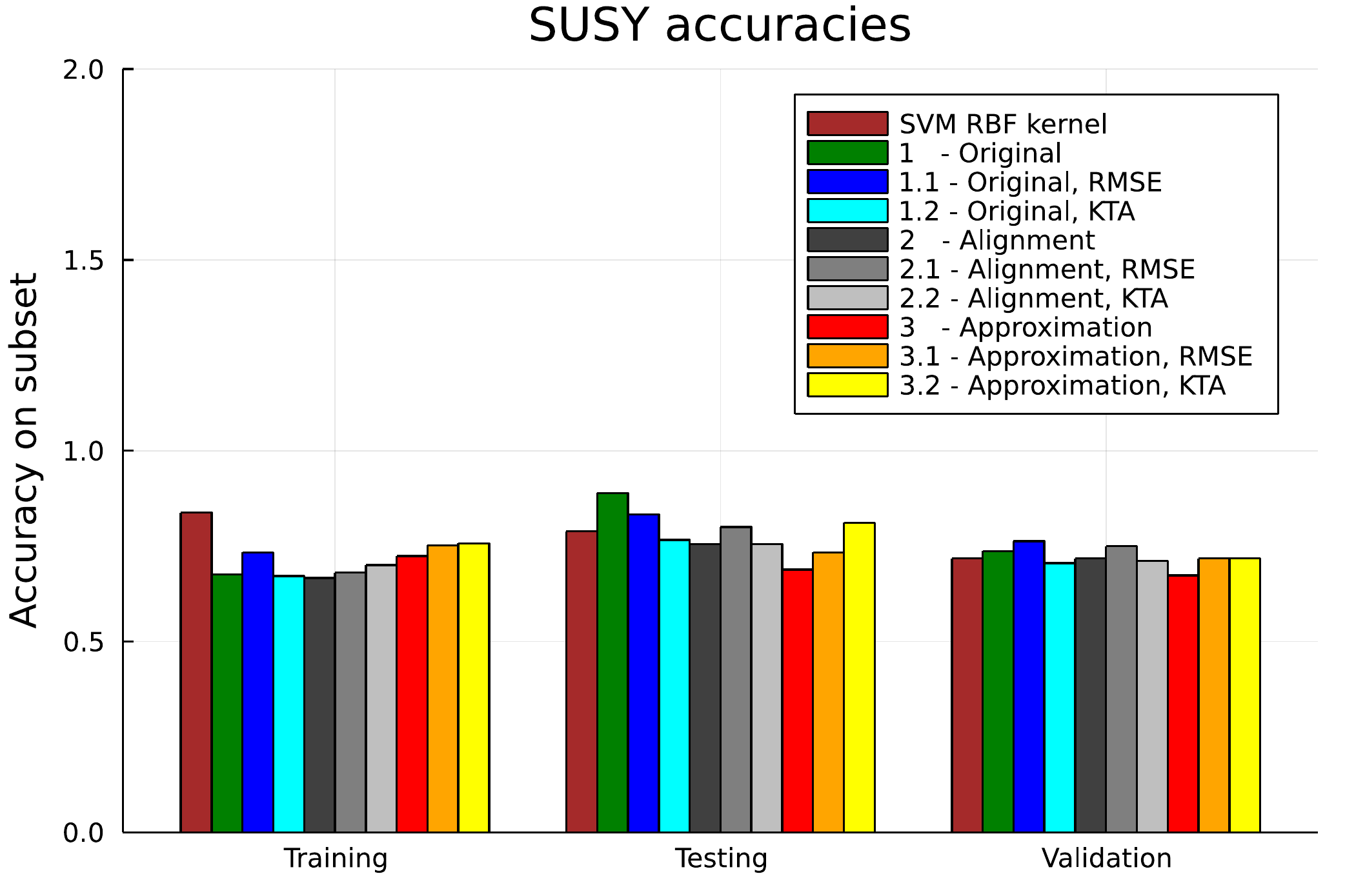}
\caption{A graph showing the classification accuracies of the best models produced by various approaches of quantum feature map design on the SUSY dataset, compared with a classical RBF kernel for reference.}
\end{figure}

\begin{figure}[h]
\includegraphics[width=0.9\columnwidth]{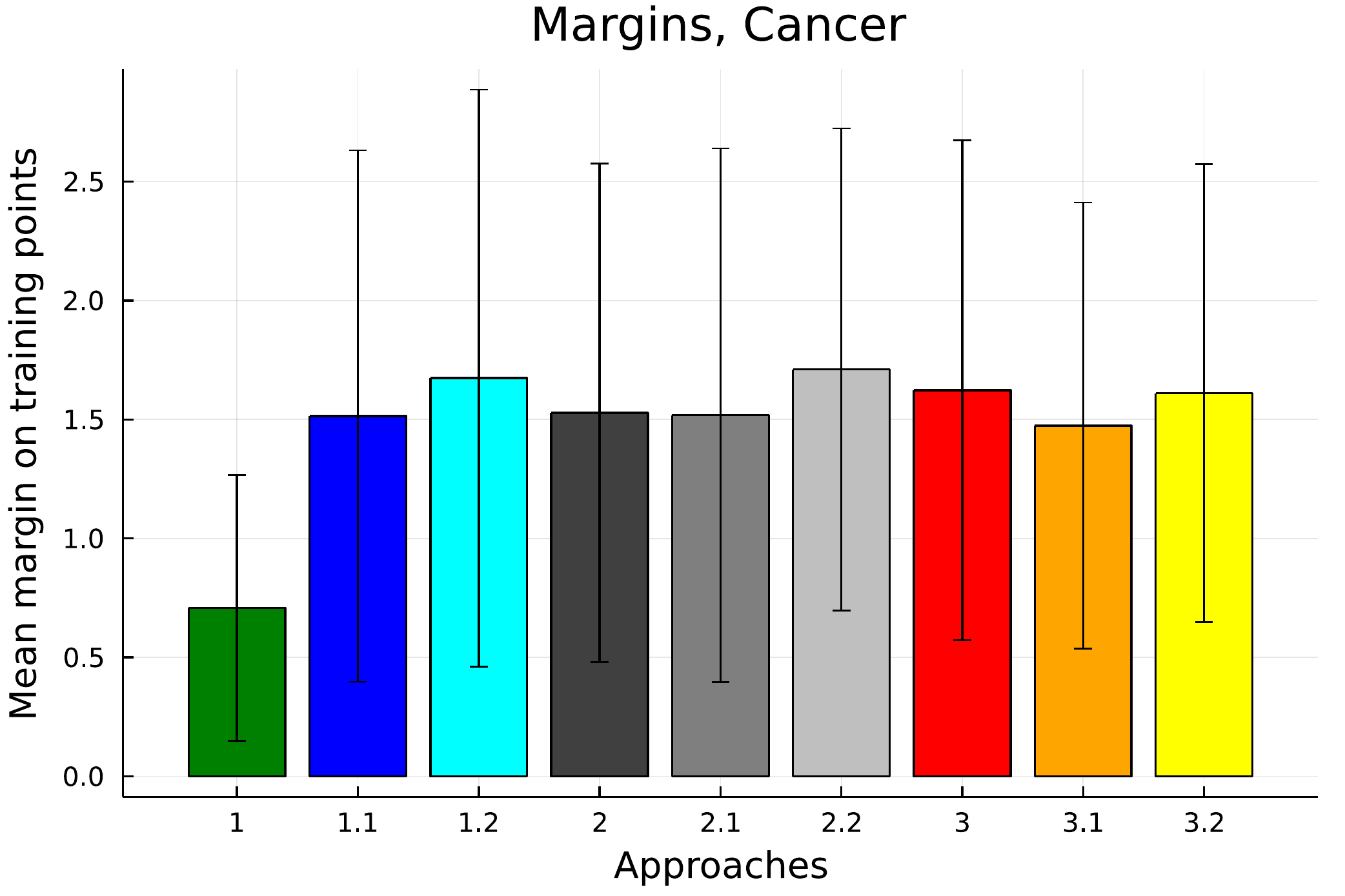}
\caption{A graph showing the mean margin of the Cancer training set points for the best classifiers produced by each approach, with errors bars showing standard deviation.}
\end{figure}
\begin{figure}[h]
\includegraphics[width=0.9\columnwidth]{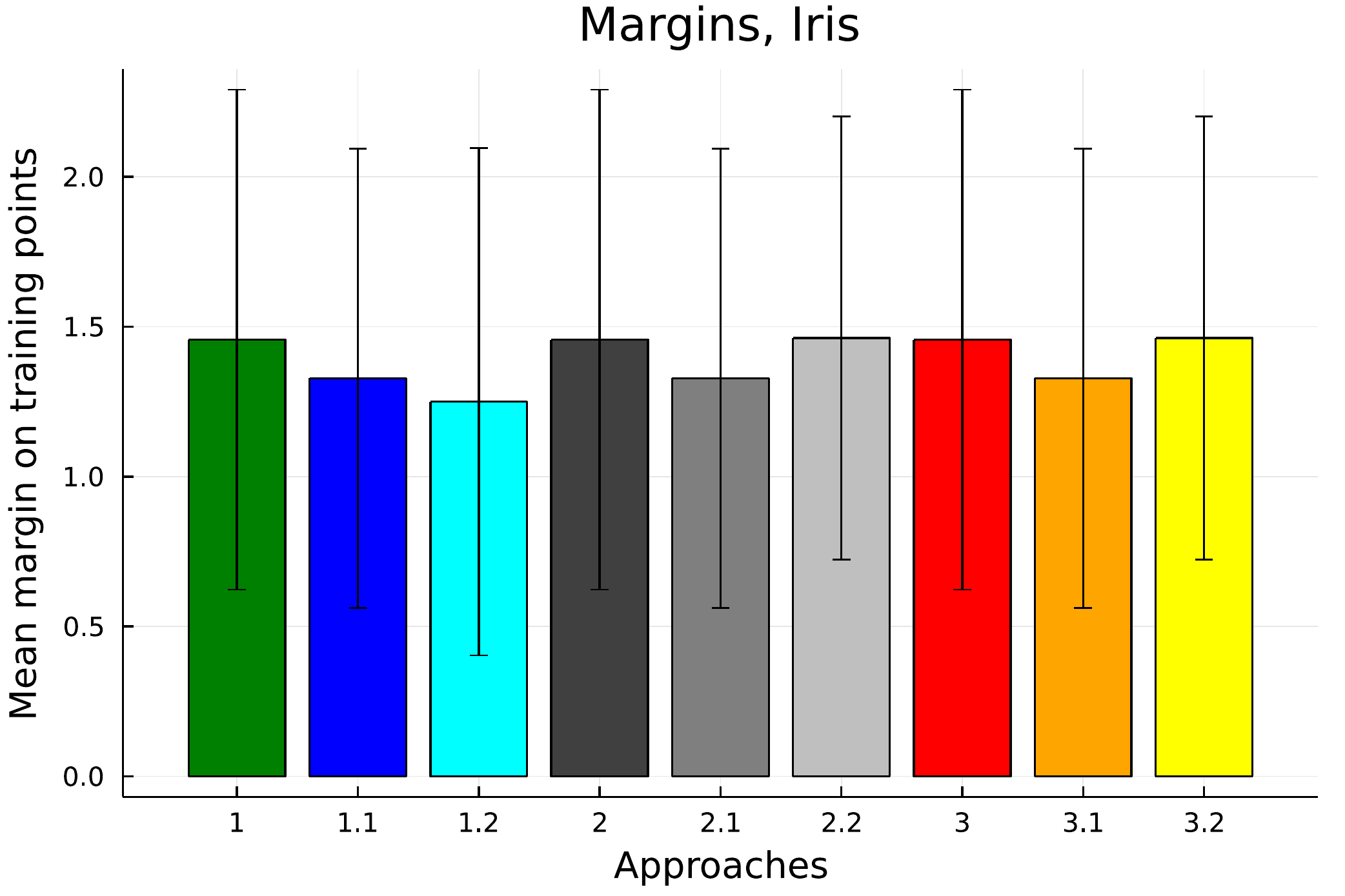}
\caption{A graph showing the mean margin of the Iris training set points for the best classifiers produced by each approach, with errors bars showing standard deviation.}
\end{figure}
\begin{figure}[h]
\includegraphics[width=0.9\columnwidth]{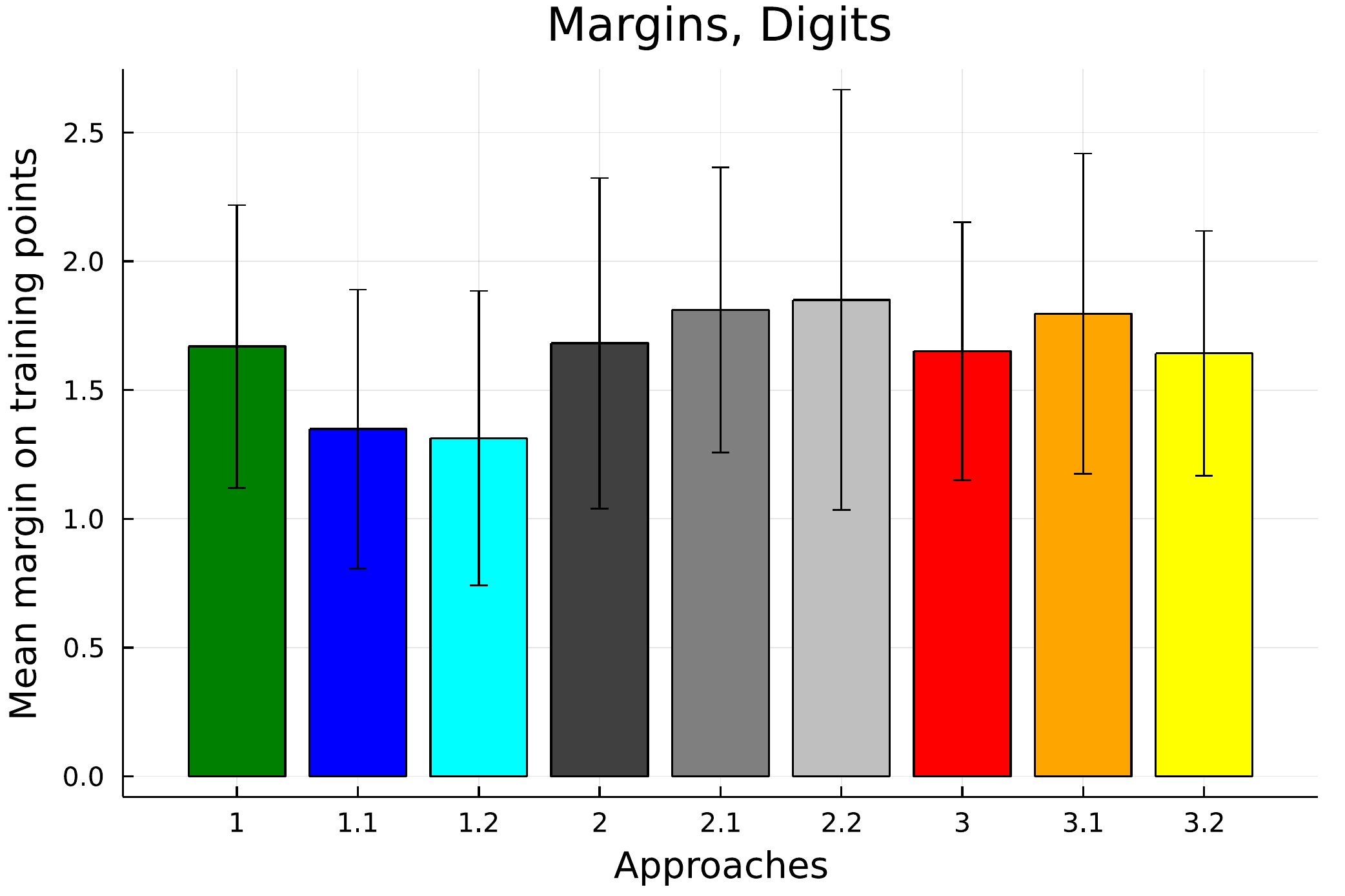}
\caption{A graph showing the mean margin of the Digits training set points for the best classifiers produced by each approach, with errors bars showing standard deviation.}
\end{figure}
\begin{figure}[h]
\includegraphics[width=0.9\columnwidth]{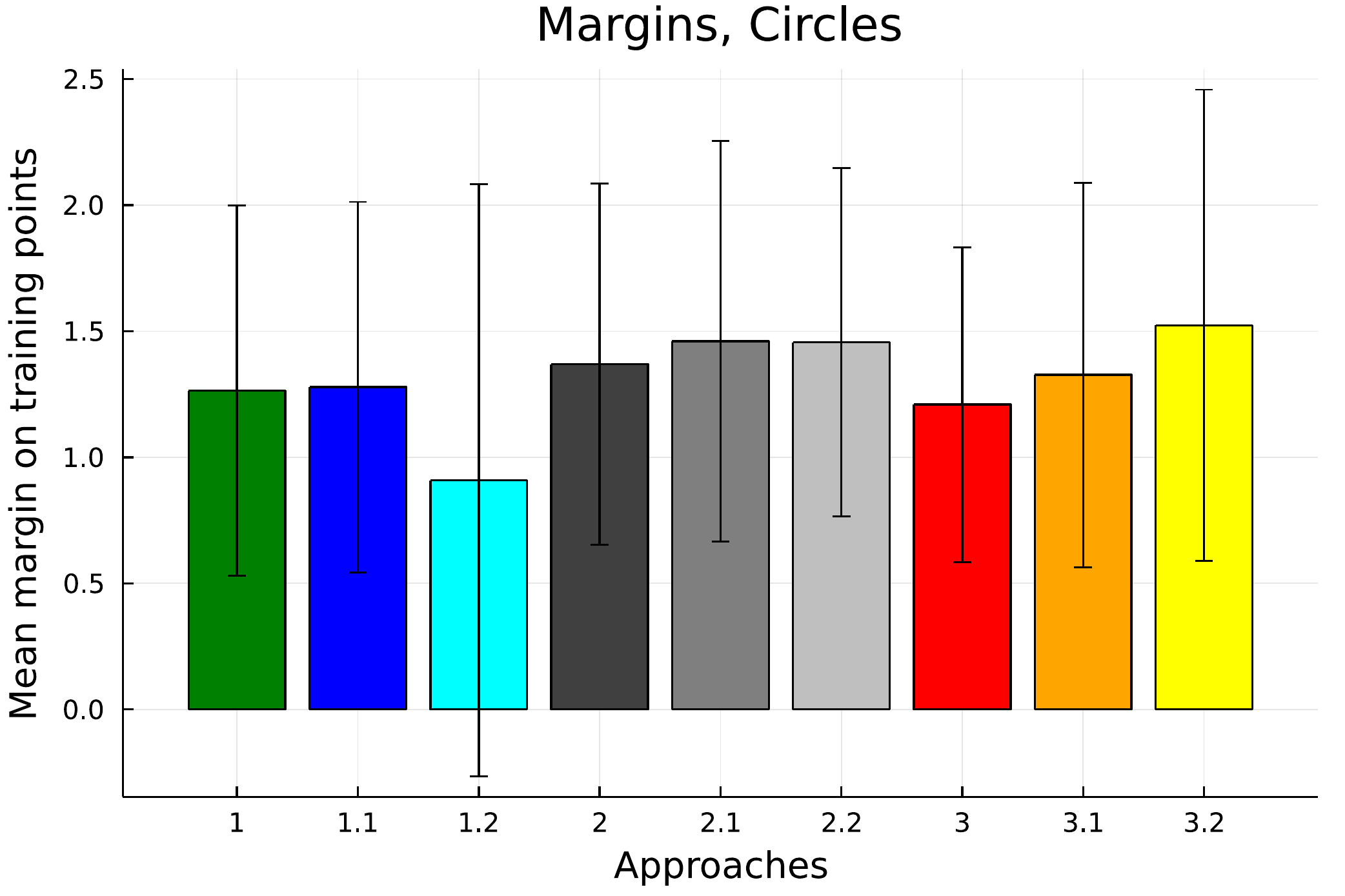}
\caption{A graph showing the mean margin of the Circles training set points for the best classifiers produced by each approach, with errors bars showing standard deviation.}
\end{figure}
\begin{figure}[h]
\includegraphics[width=0.9\columnwidth]{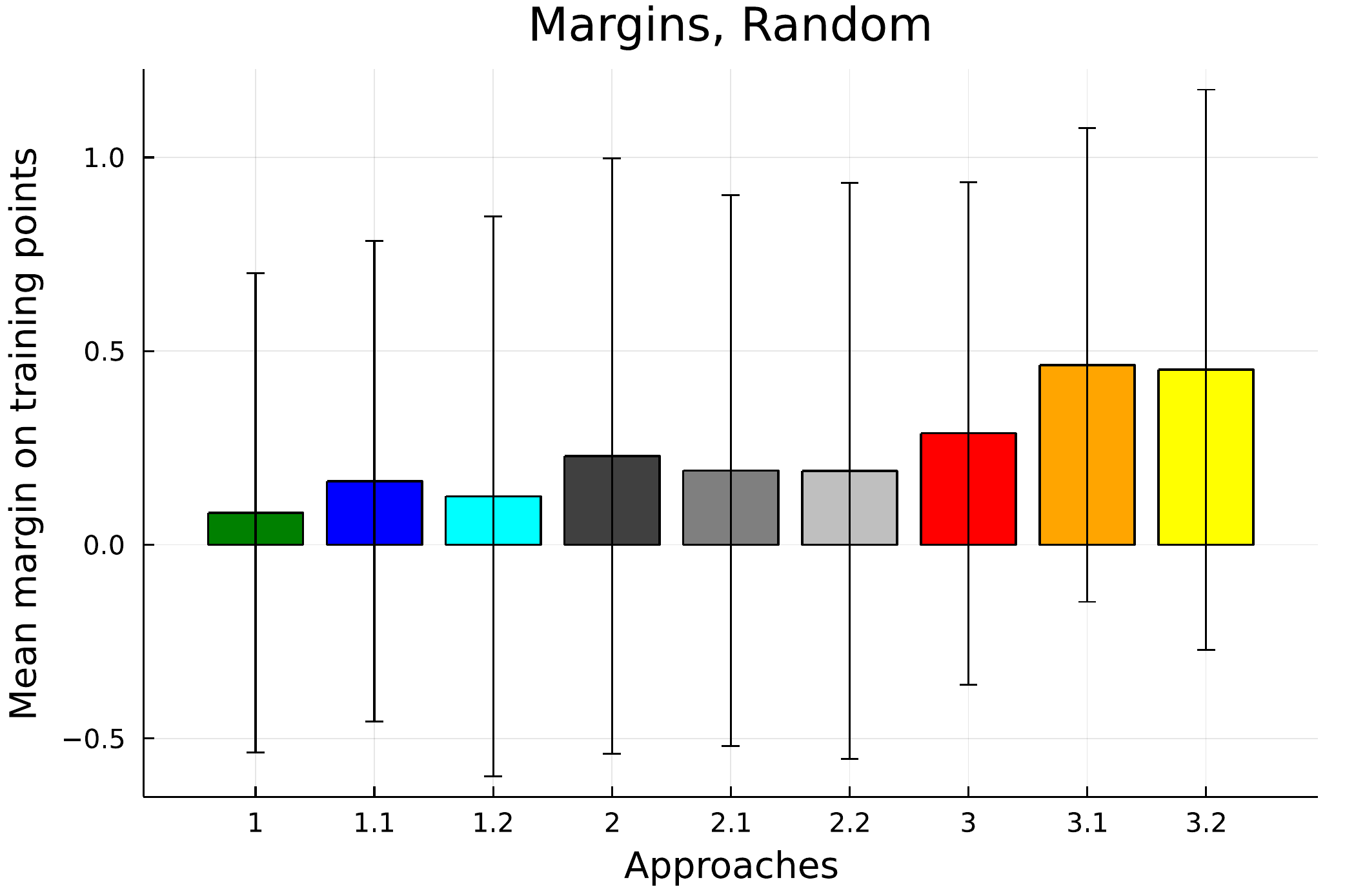}
\caption{A graph showing the mean margin of the Random training set points for the best classifiers produced by each approach, with errors bars showing standard deviation.}
\end{figure}
\begin{figure}[h]
\includegraphics[width=0.9\columnwidth]{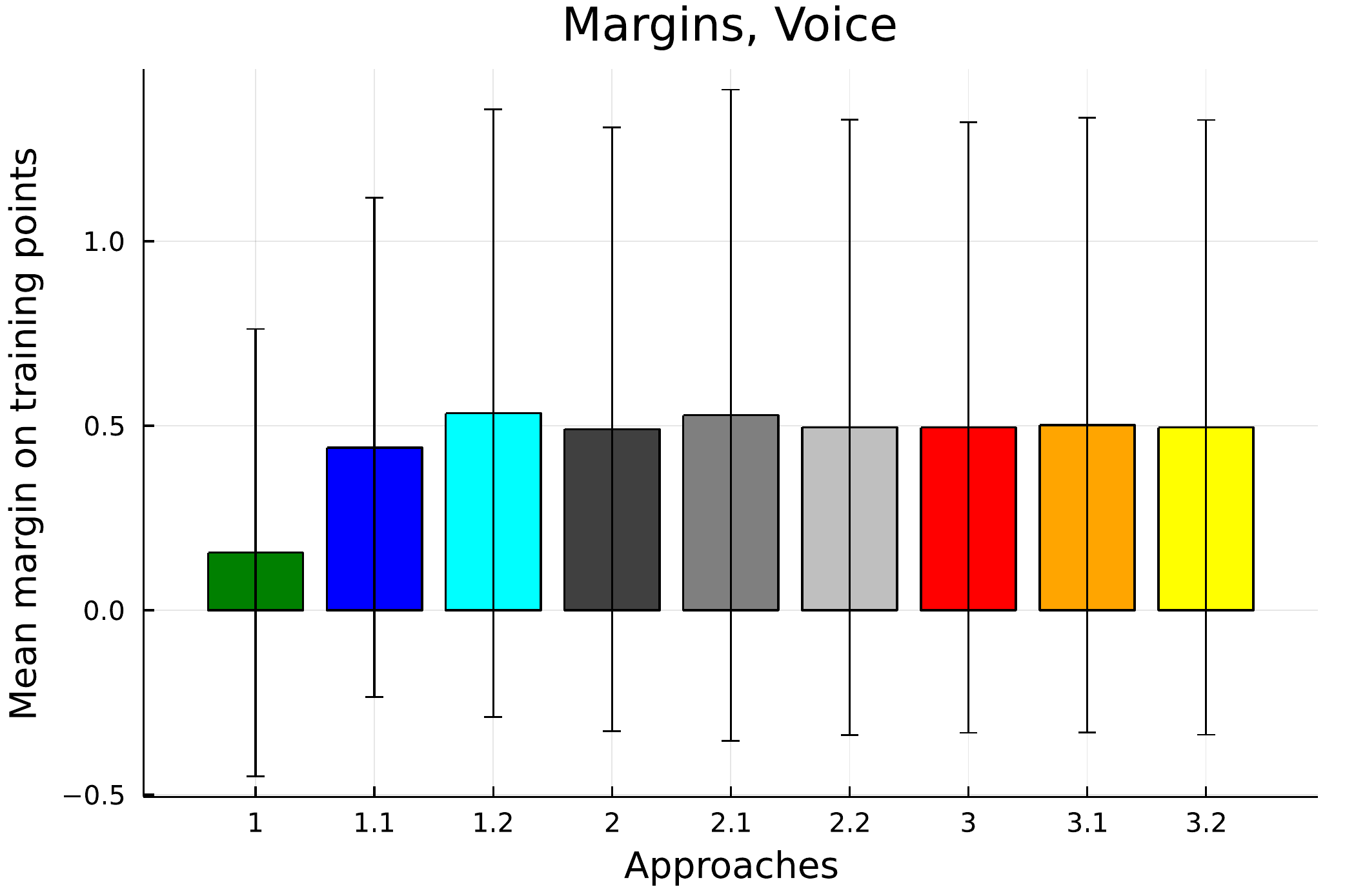}
\caption{A graph showing the mean margin of the Voice training set points for the best classifiers produced by each approach, with errors bars showing standard deviation.}
\end{figure}
\begin{figure}[h]
\includegraphics[width=0.9\columnwidth]{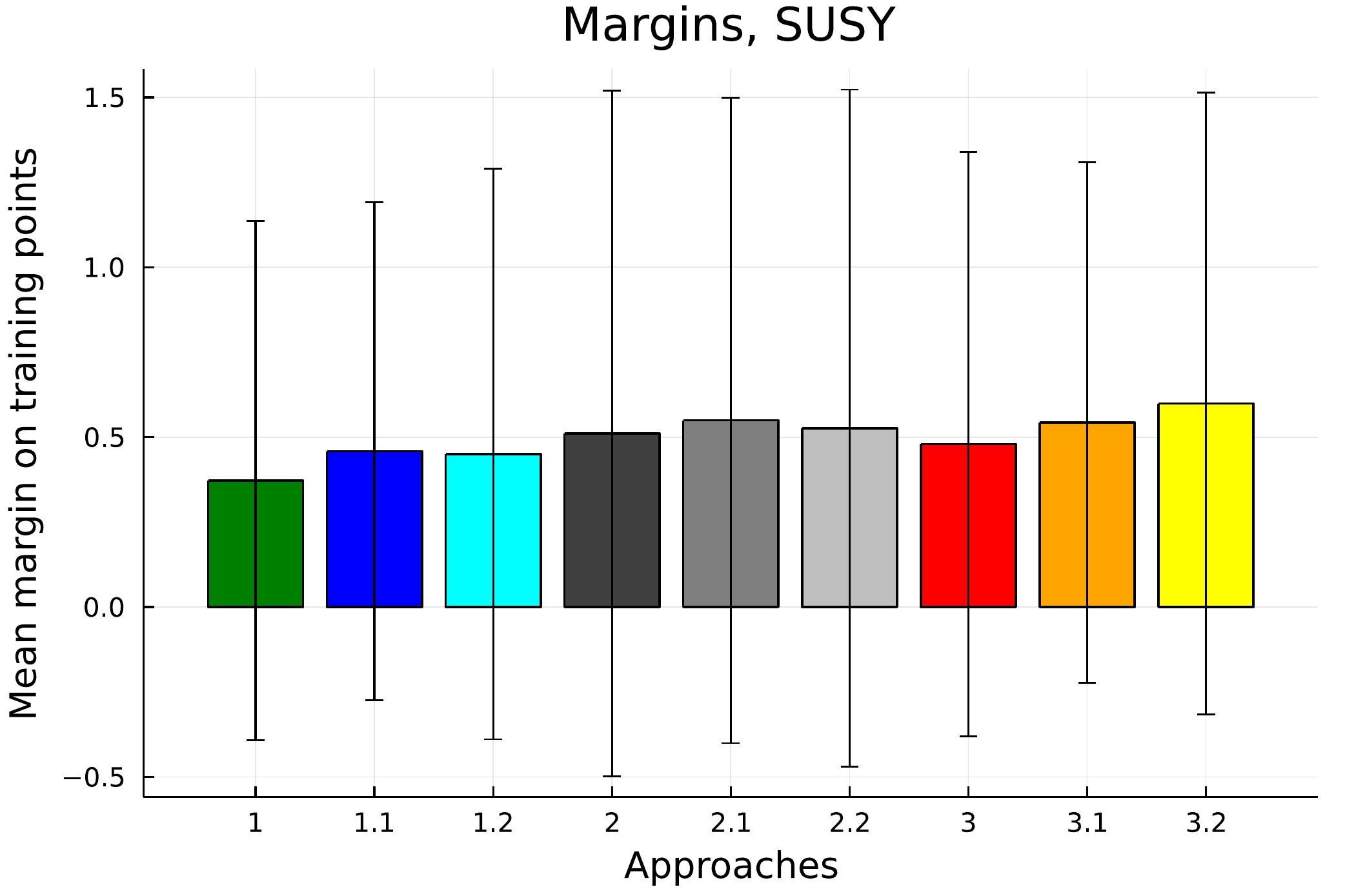}
\caption{A graph showing the mean margin of the SUSY training set points for the best classifiers produced by each approach, with errors bars showing standard deviation.}
\end{figure}

\begin{figure}
\includegraphics[width=0.9\columnwidth]{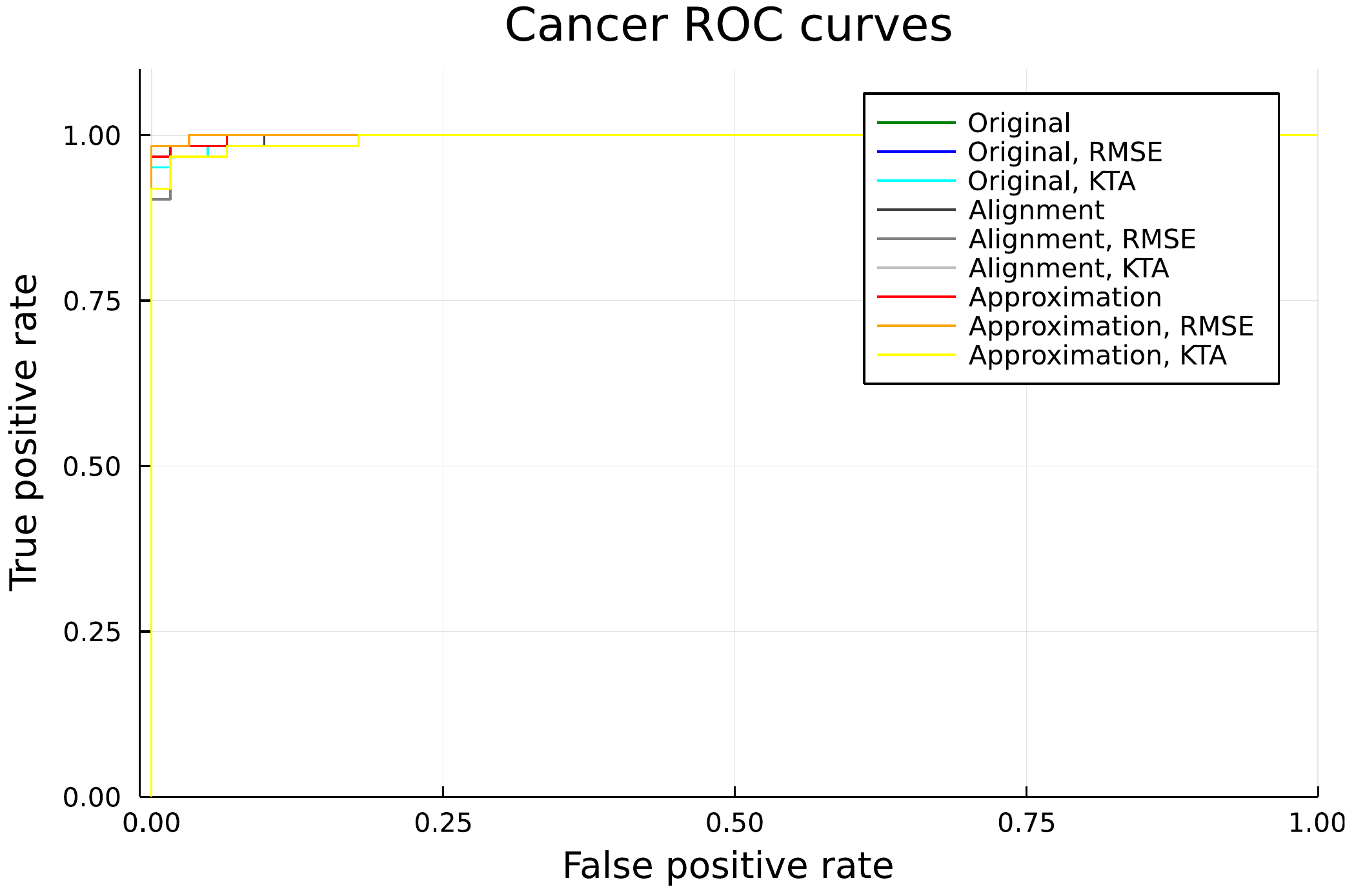}
\caption{A graph showing the ROC curves of the best models produced by various approaches of quantum feature map design on the Cancer dataset.}
\end{figure}
\begin{figure}
\includegraphics[width=0.9\columnwidth]{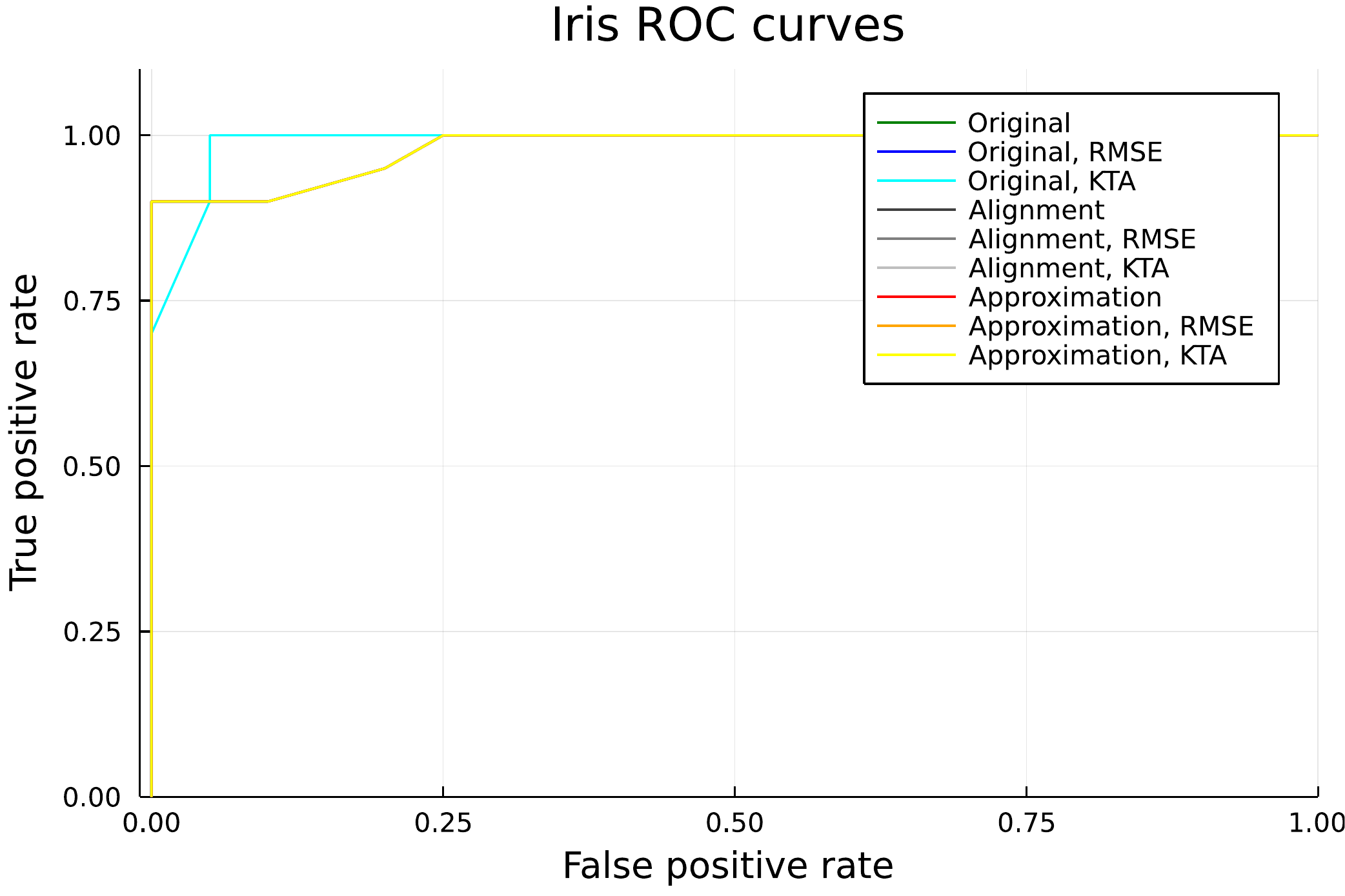}
\caption{A graph showing the ROC curves of the best models produced by various approaches of quantum feature map design on the Iris dataset.}
\end{figure}
\begin{figure}
\includegraphics[width=0.9\columnwidth]{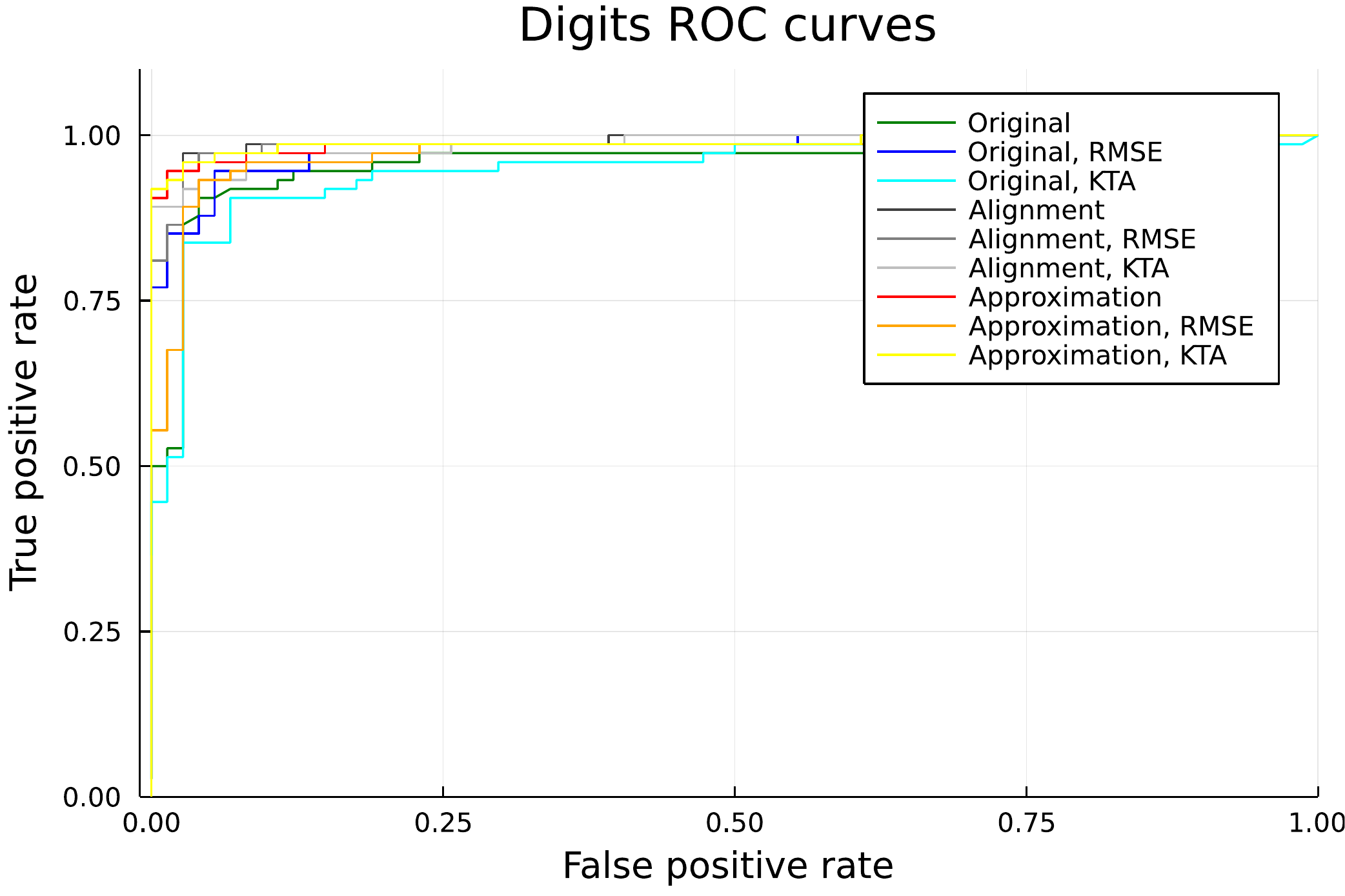}
\caption{A graph showing the ROC curves of the best models produced by various approaches of quantum feature map design on the Digits dataset.}
\end{figure}
\begin{figure}
\includegraphics[width=0.9\columnwidth]{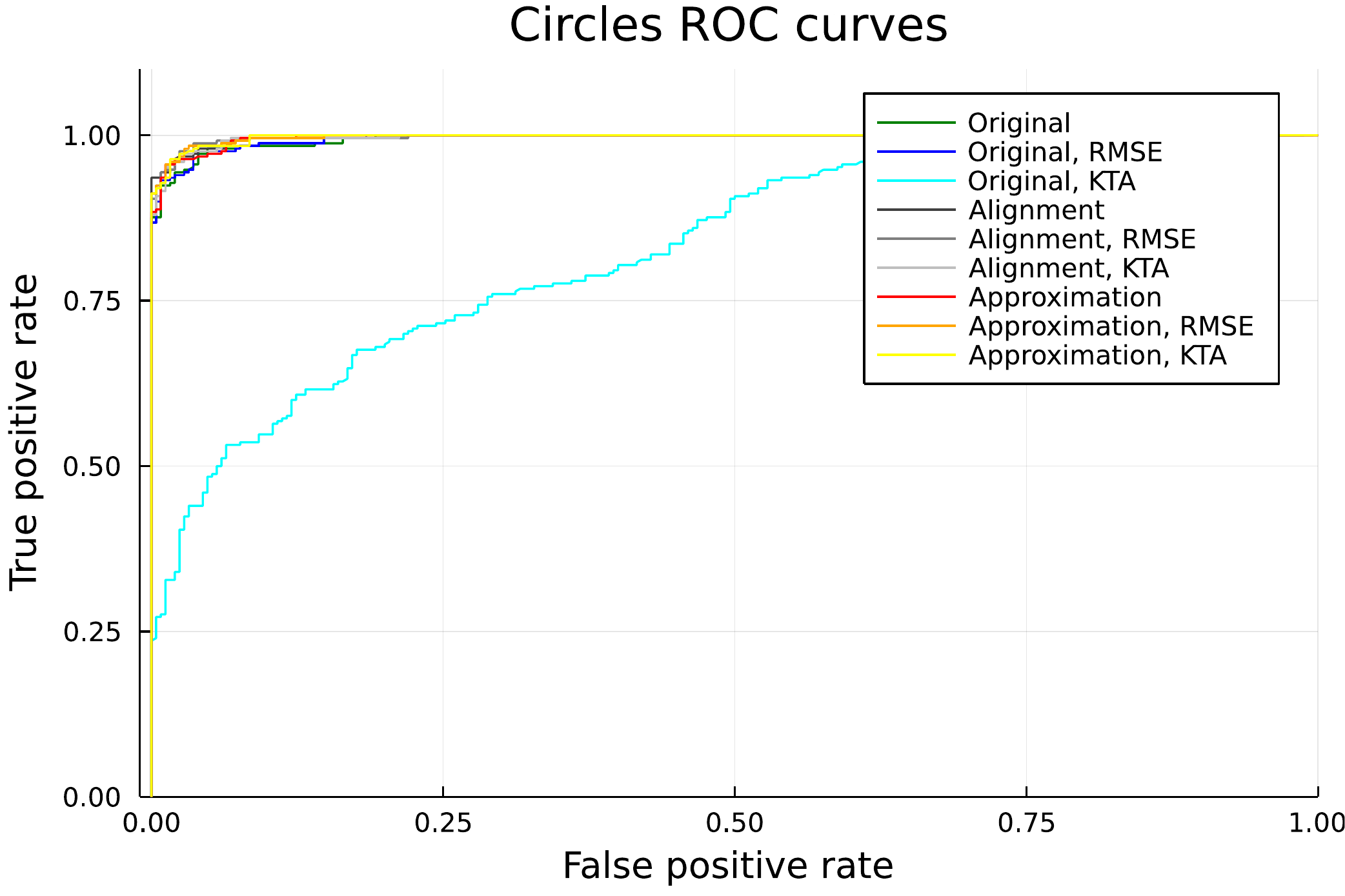}
\caption{A graph showing the ROC curves of the best models produced by various approaches of quantum feature map design on the Circles dataset.}
\end{figure}
\begin{figure}
\includegraphics[width=0.9\columnwidth]{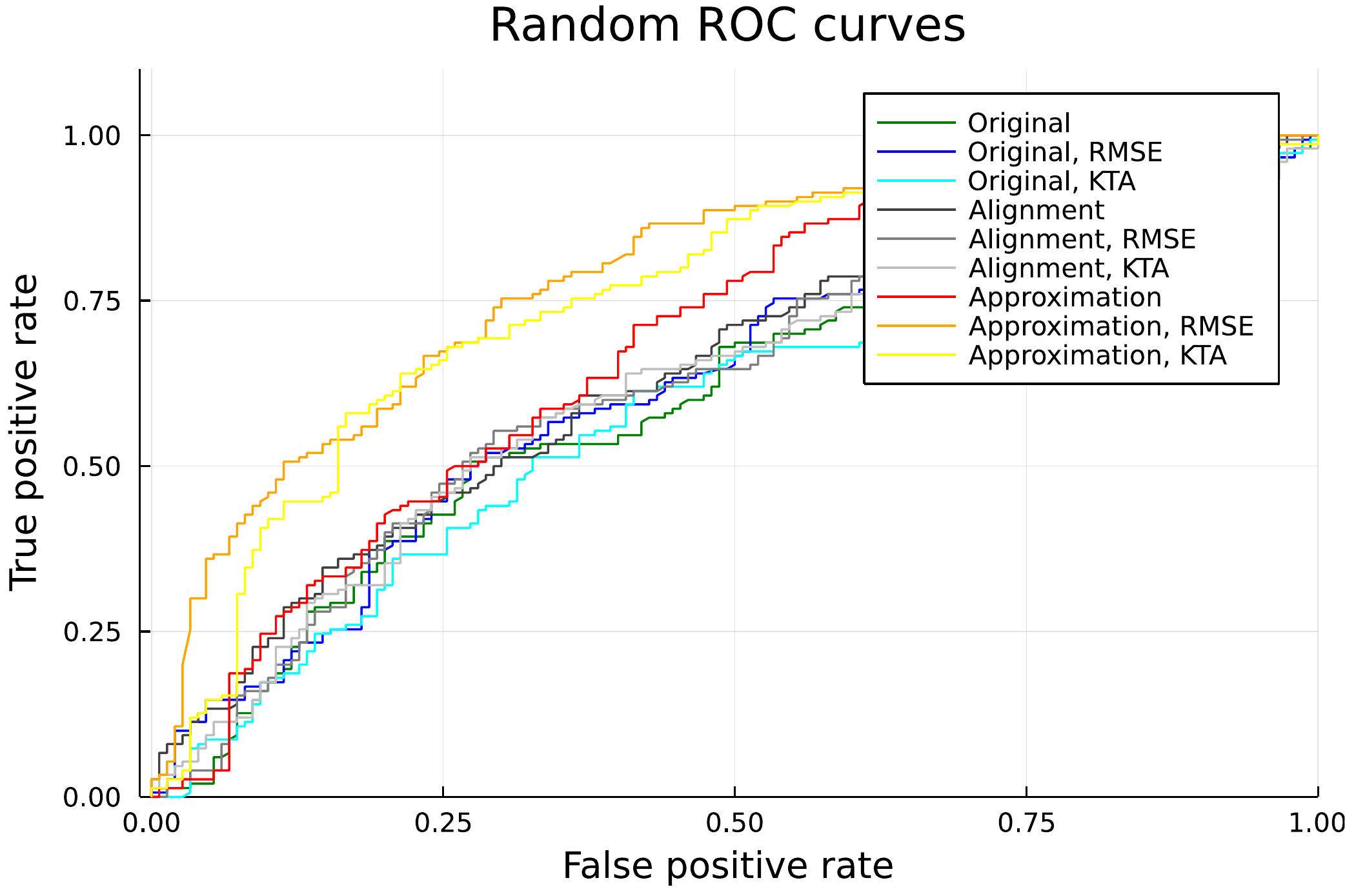}
\caption{A graph showing the ROC curves of the best models produced by various approaches of quantum feature map design on the Random dataset.}
\end{figure}
\begin{figure}
\includegraphics[width=0.9\columnwidth]{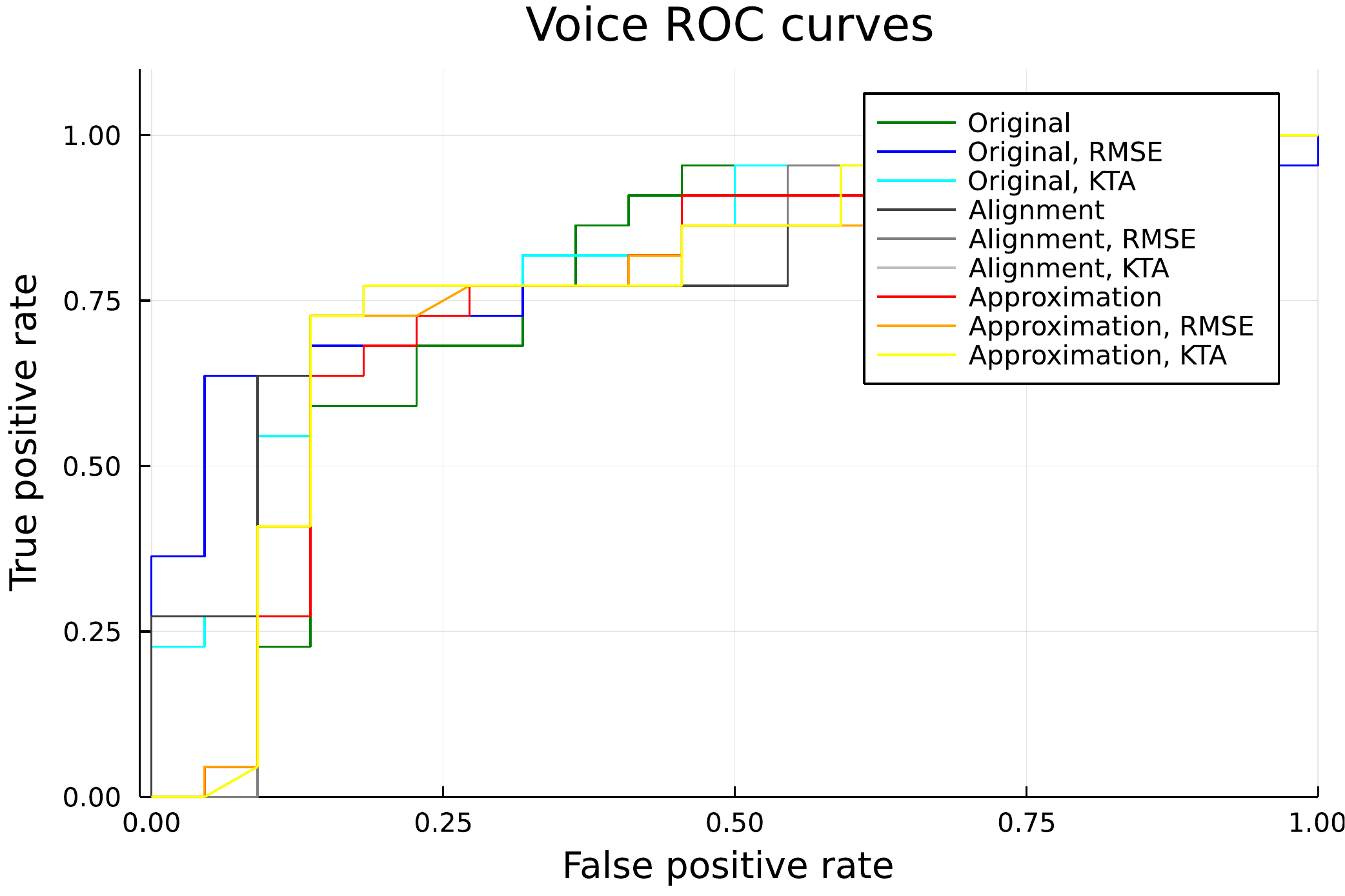}
\caption{A graph showing the ROC curves of the best models produced by various approaches of quantum feature map design on the Voice dataset.}
\end{figure}
\begin{figure}
\includegraphics[width=0.9\columnwidth]{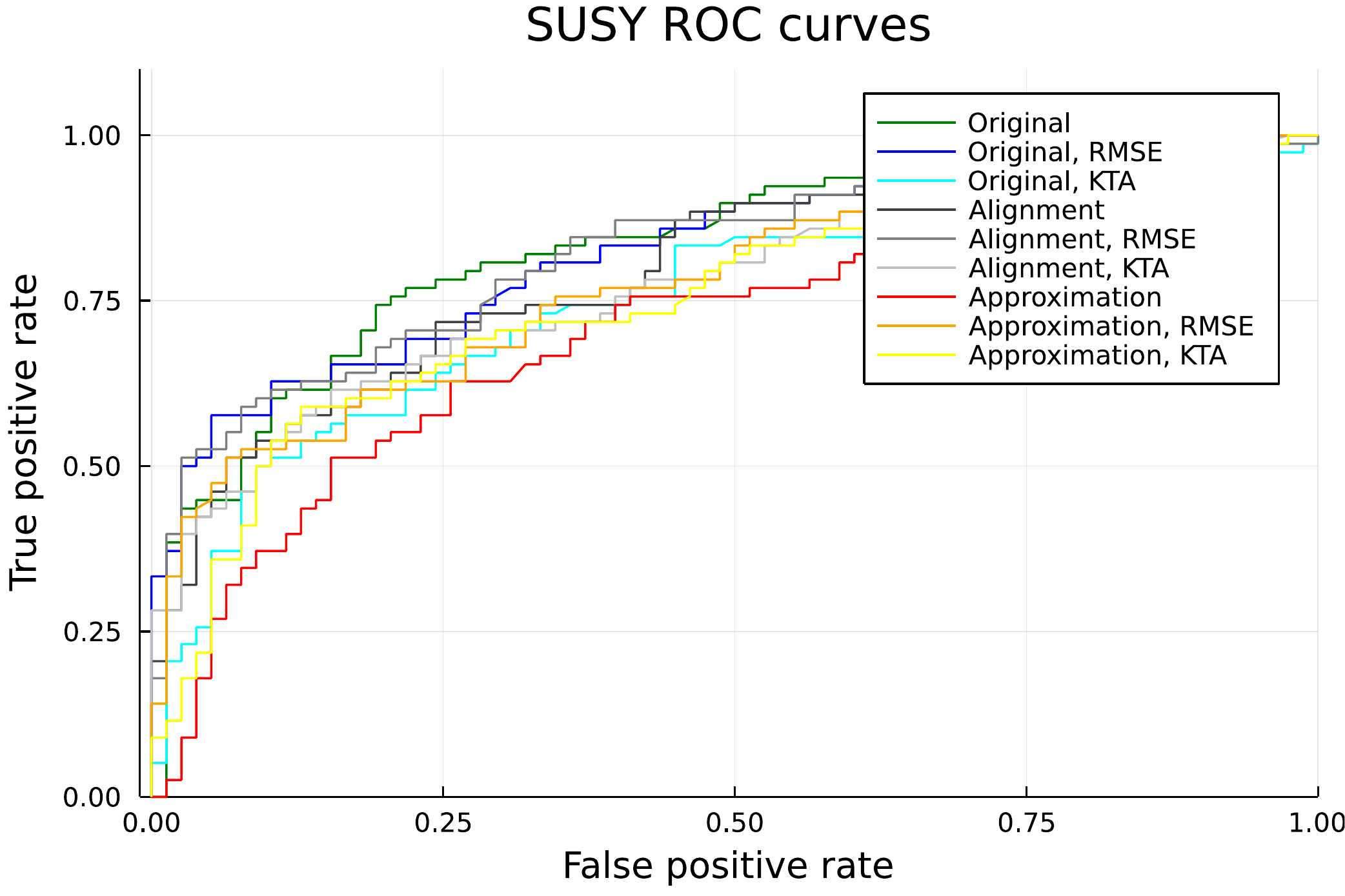}
\caption{A graph showing the ROC curves of the best models produced by various approaches of quantum feature map design on the SUSY dataset.}
\end{figure}


\begin{figure}
\begin{adjustbox}{center}
\begin{subfigure}{0.33\paperwidth}
	\includegraphics[width=\columnwidth]{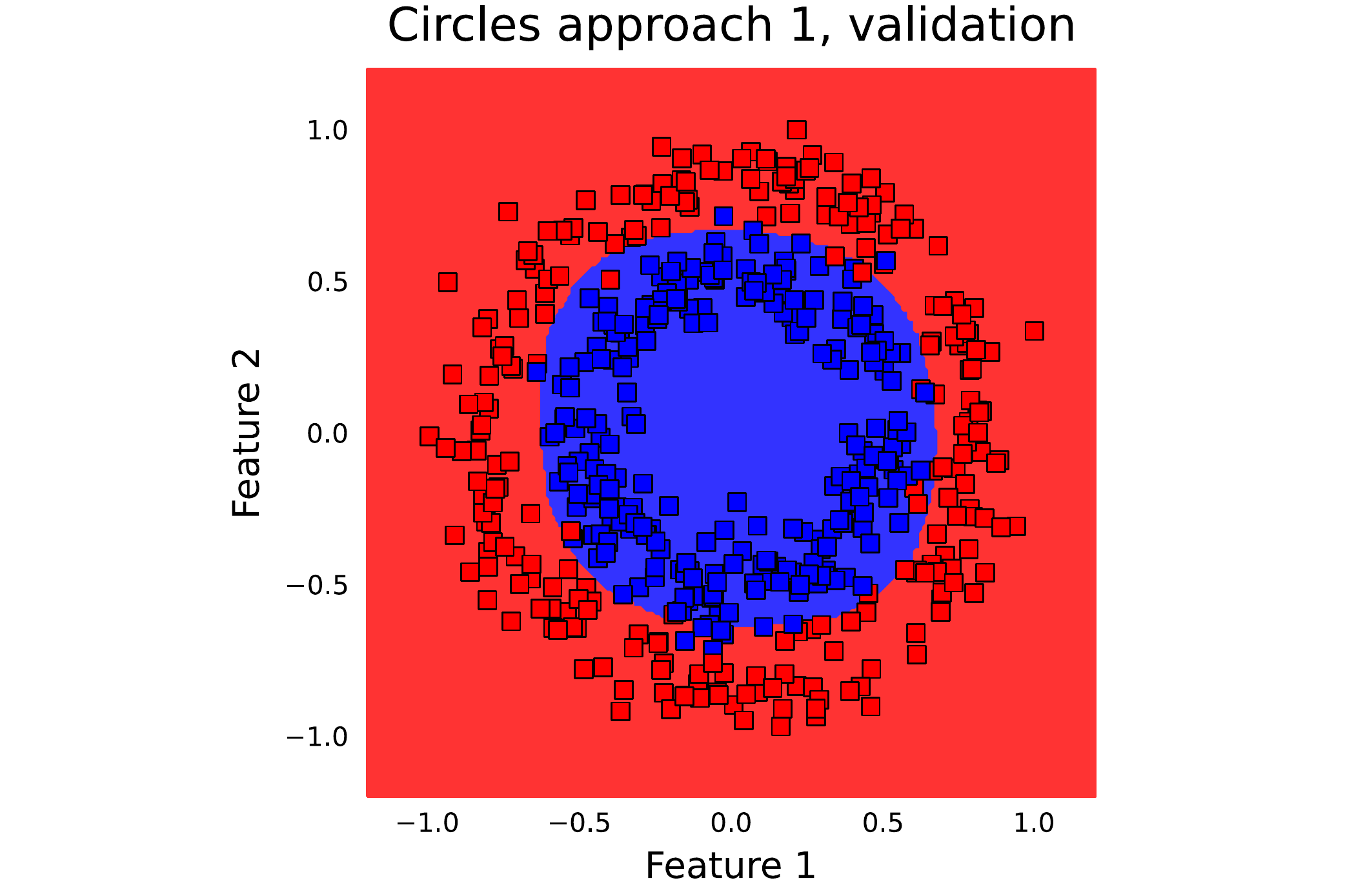}
\end{subfigure}%
\begin{subfigure}{0.33\paperwidth}
	\includegraphics[width=\columnwidth]{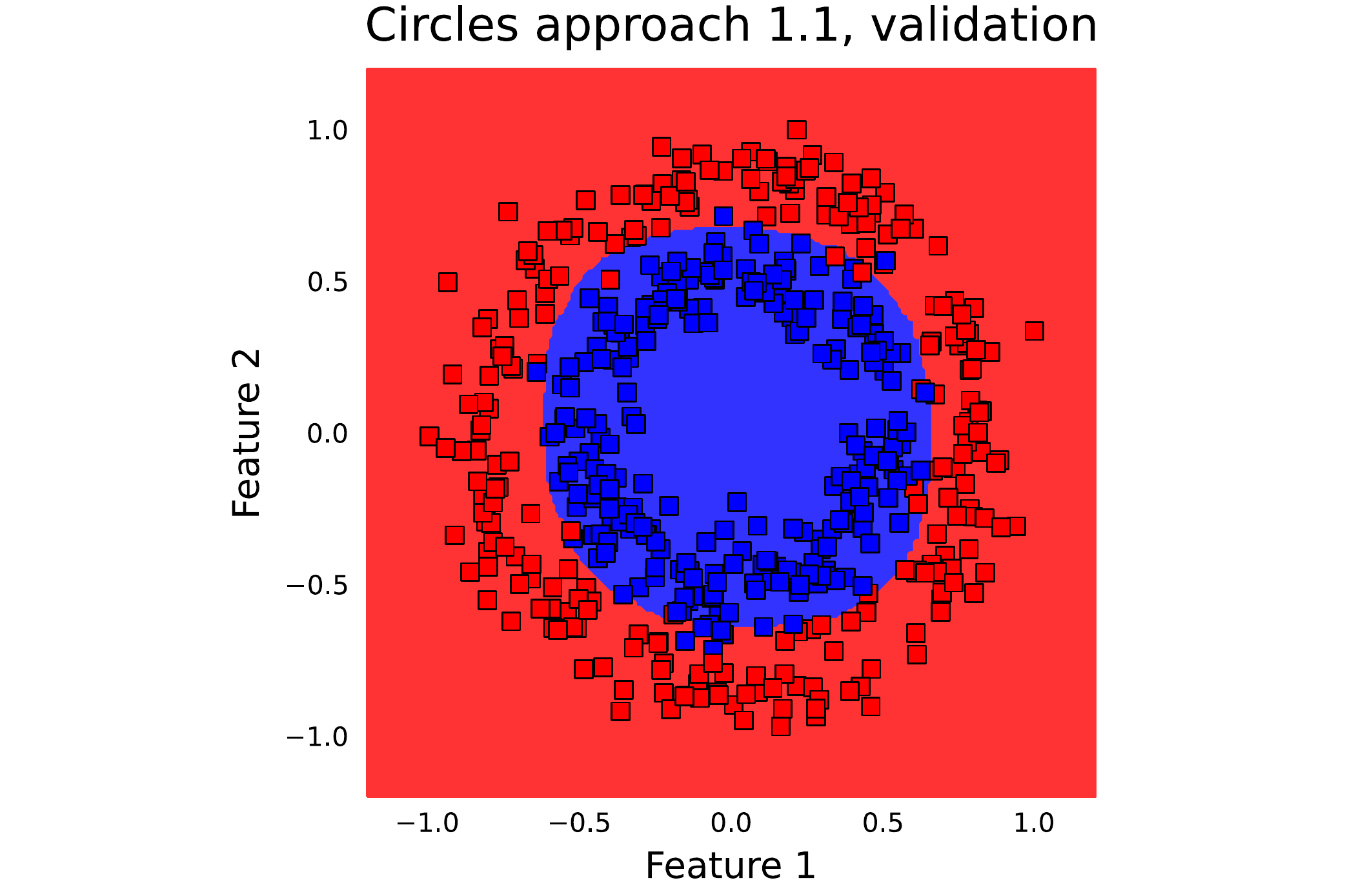}
\end{subfigure}%
\begin{subfigure}{0.33\paperwidth}
	\includegraphics[width=\columnwidth]{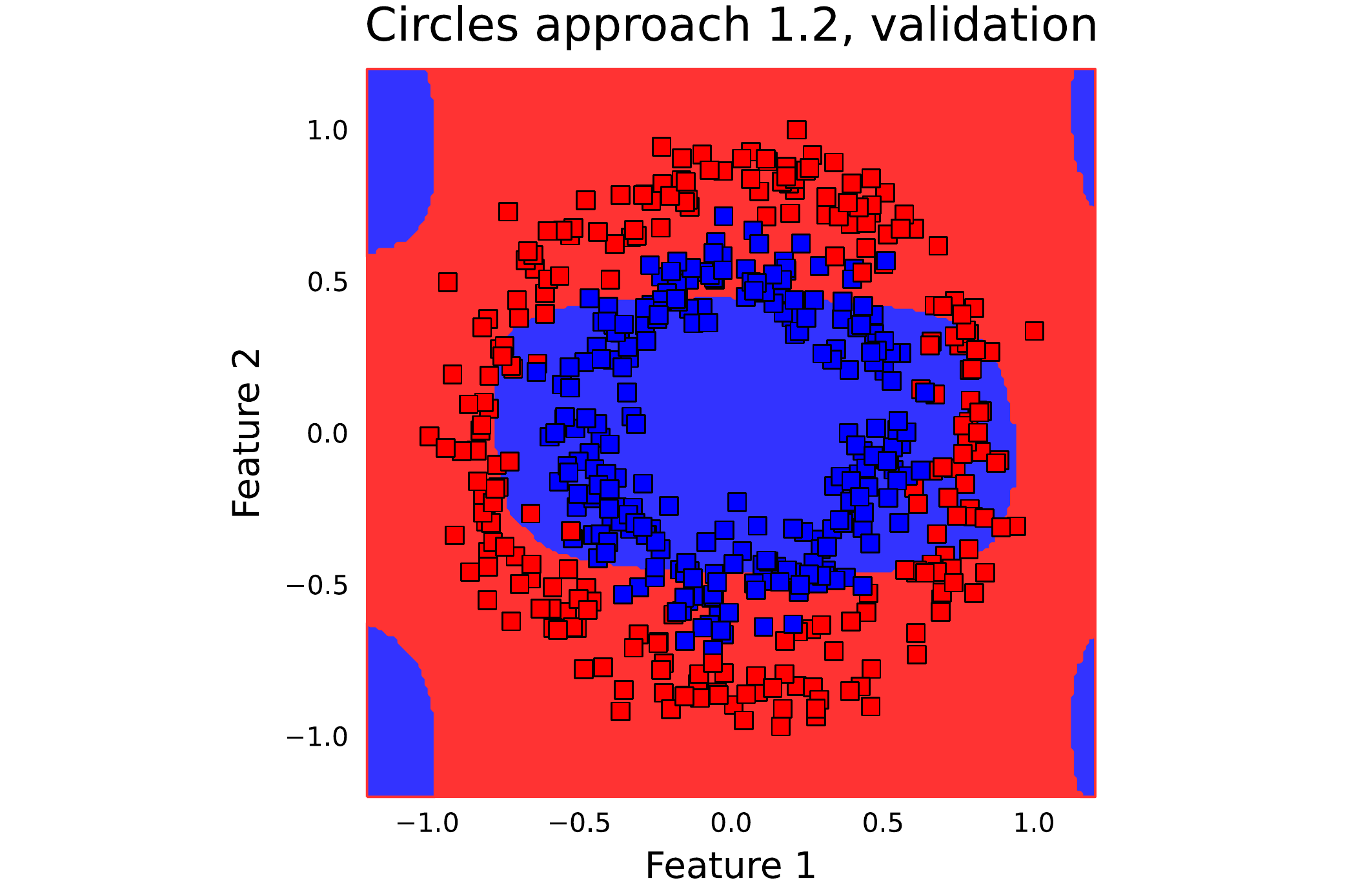}
\end{subfigure}
\end{adjustbox}

\begin{adjustbox}{center}
\begin{subfigure}{0.33\paperwidth}
	\includegraphics[width=\columnwidth]{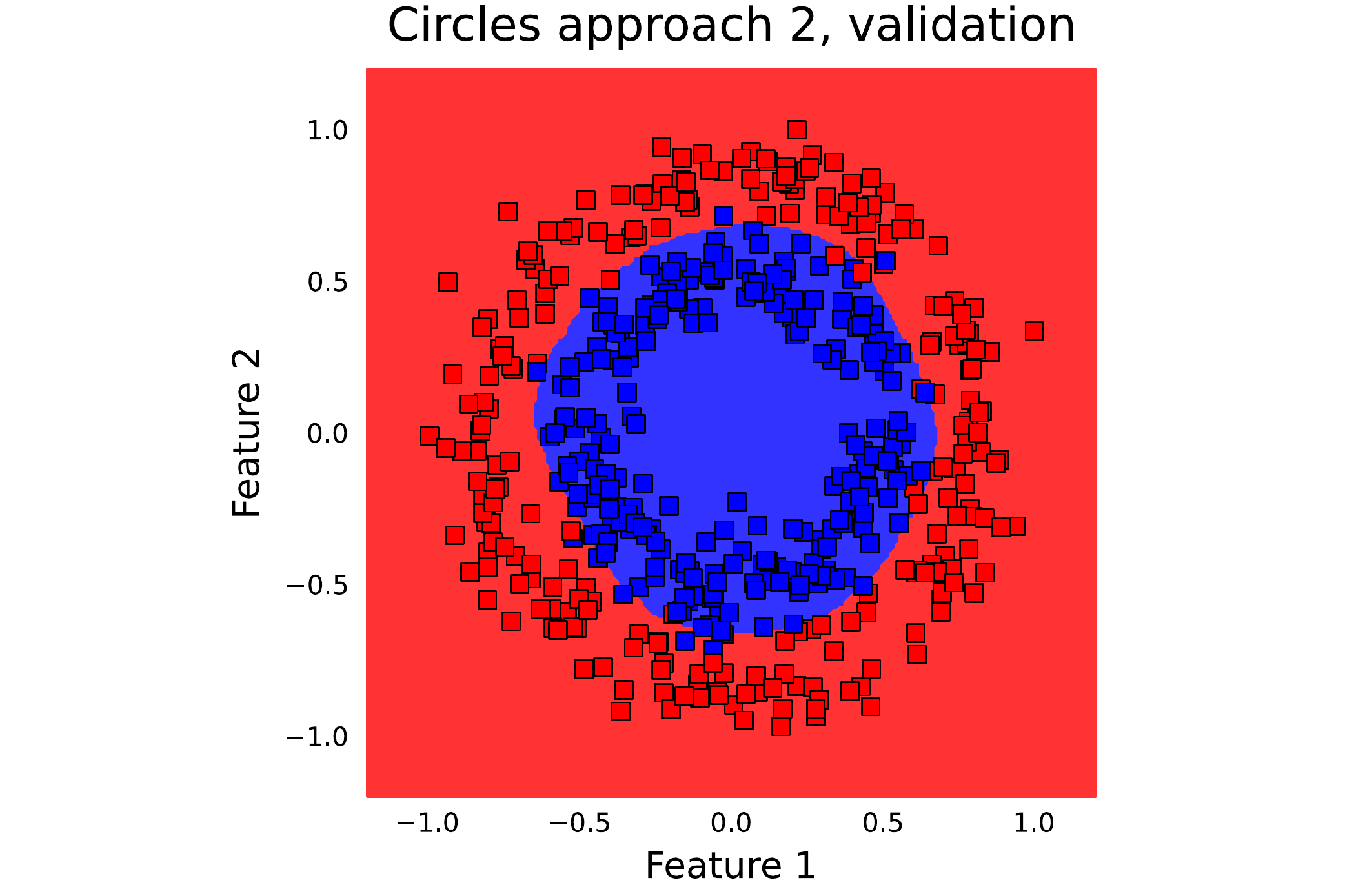}
\end{subfigure}%
\begin{subfigure}{0.33\paperwidth}
	\includegraphics[width=\columnwidth]{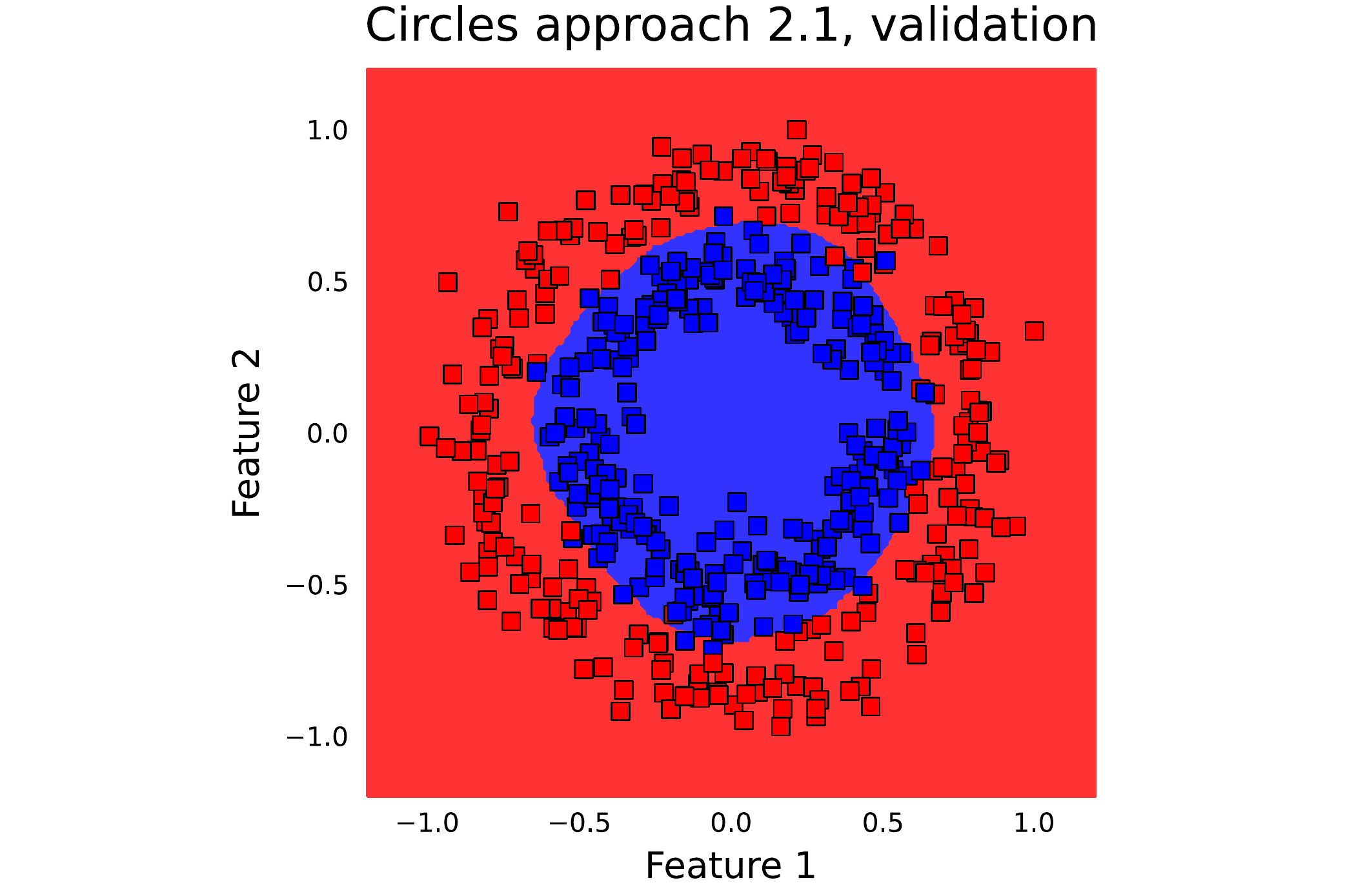}
\end{subfigure}%
\begin{subfigure}{0.33\paperwidth}
	\includegraphics[width=\columnwidth]{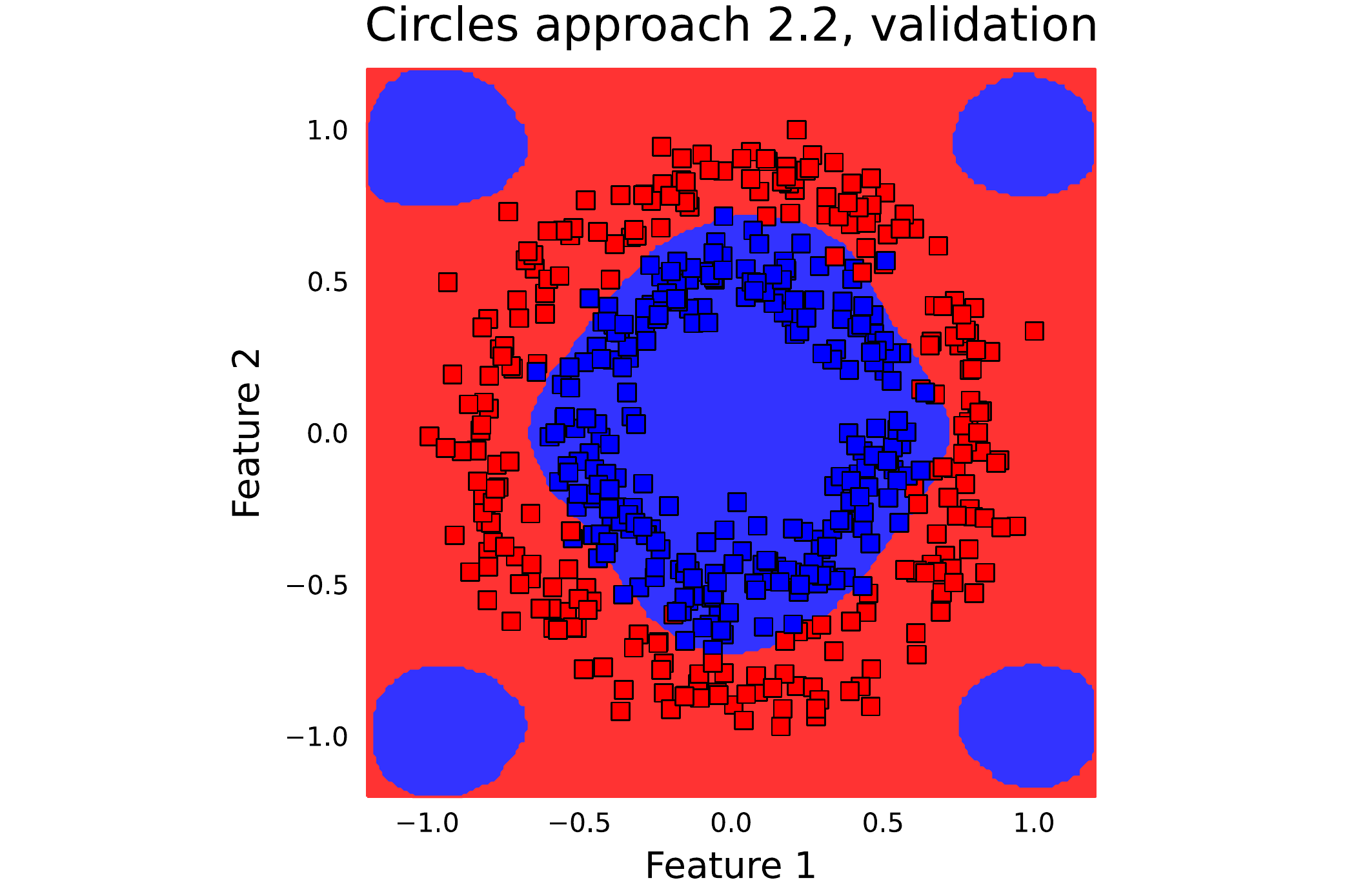}
\end{subfigure}
\end{adjustbox}

\begin{adjustbox}{center}
\begin{subfigure}{0.33\paperwidth}
	\includegraphics[width=\columnwidth]{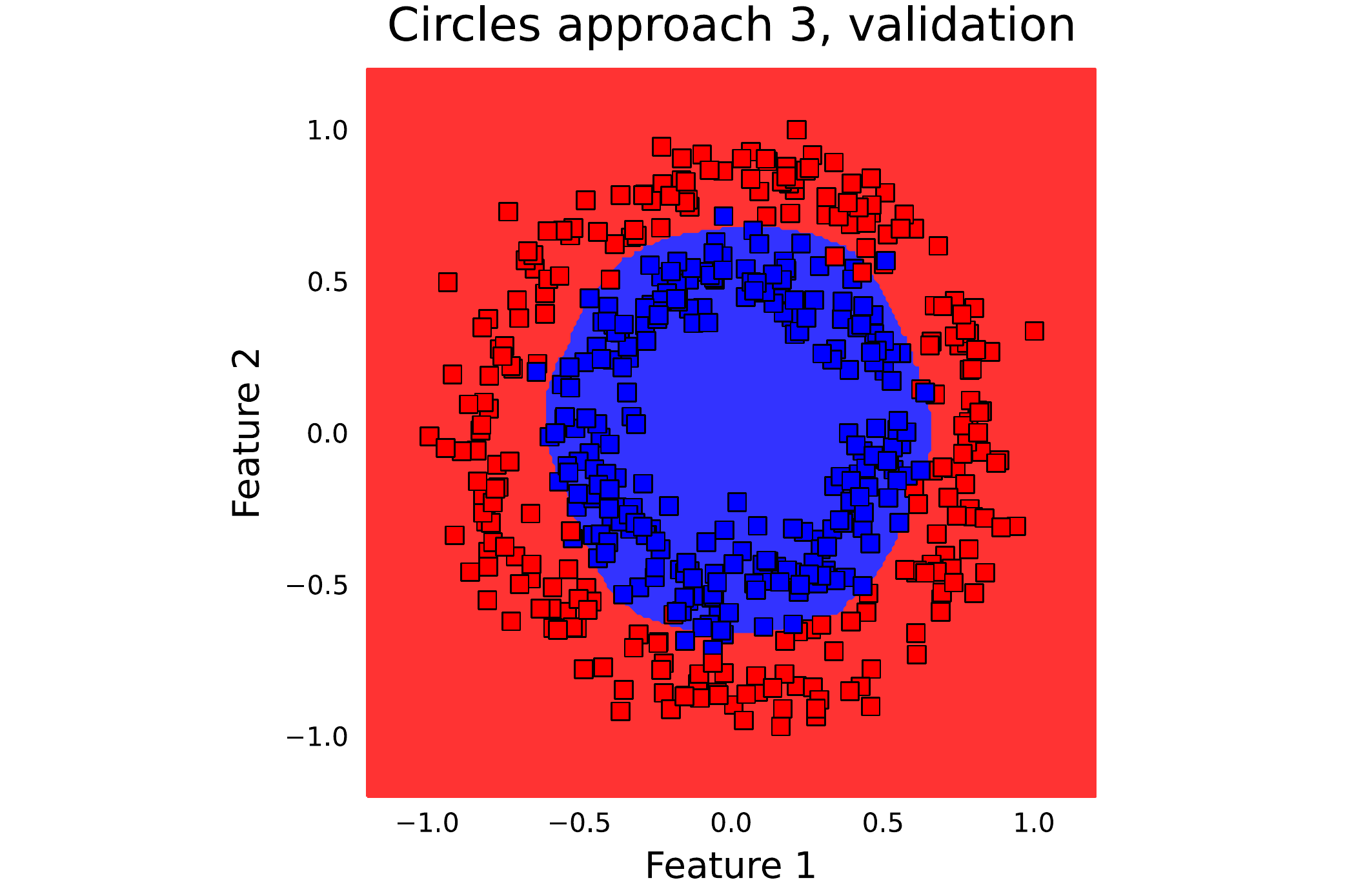}
\end{subfigure}%
\begin{subfigure}{0.33\paperwidth}
	\includegraphics[width=\columnwidth]{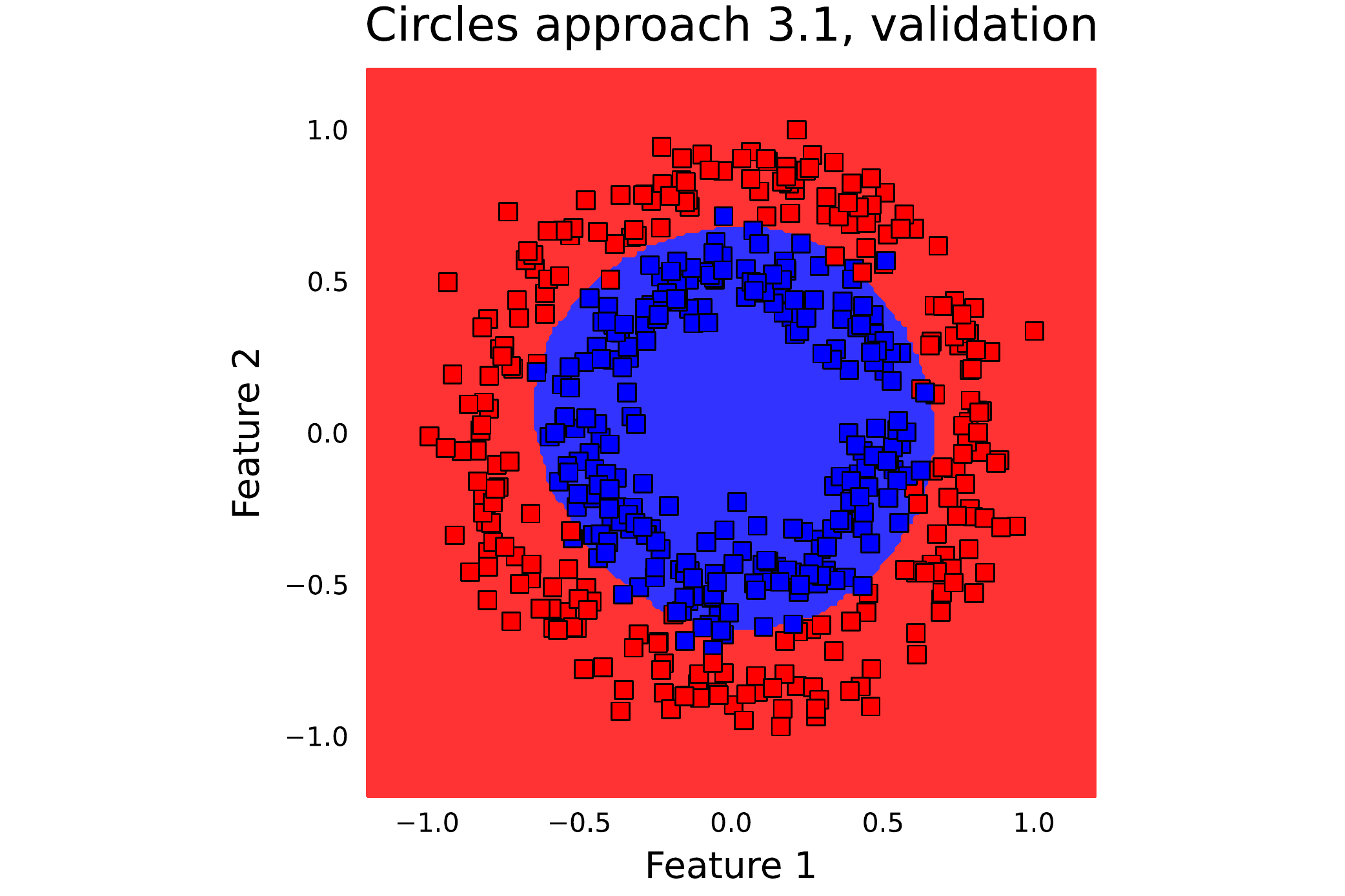}
\end{subfigure}%
\begin{subfigure}{0.33\paperwidth}
	\includegraphics[width=\columnwidth]{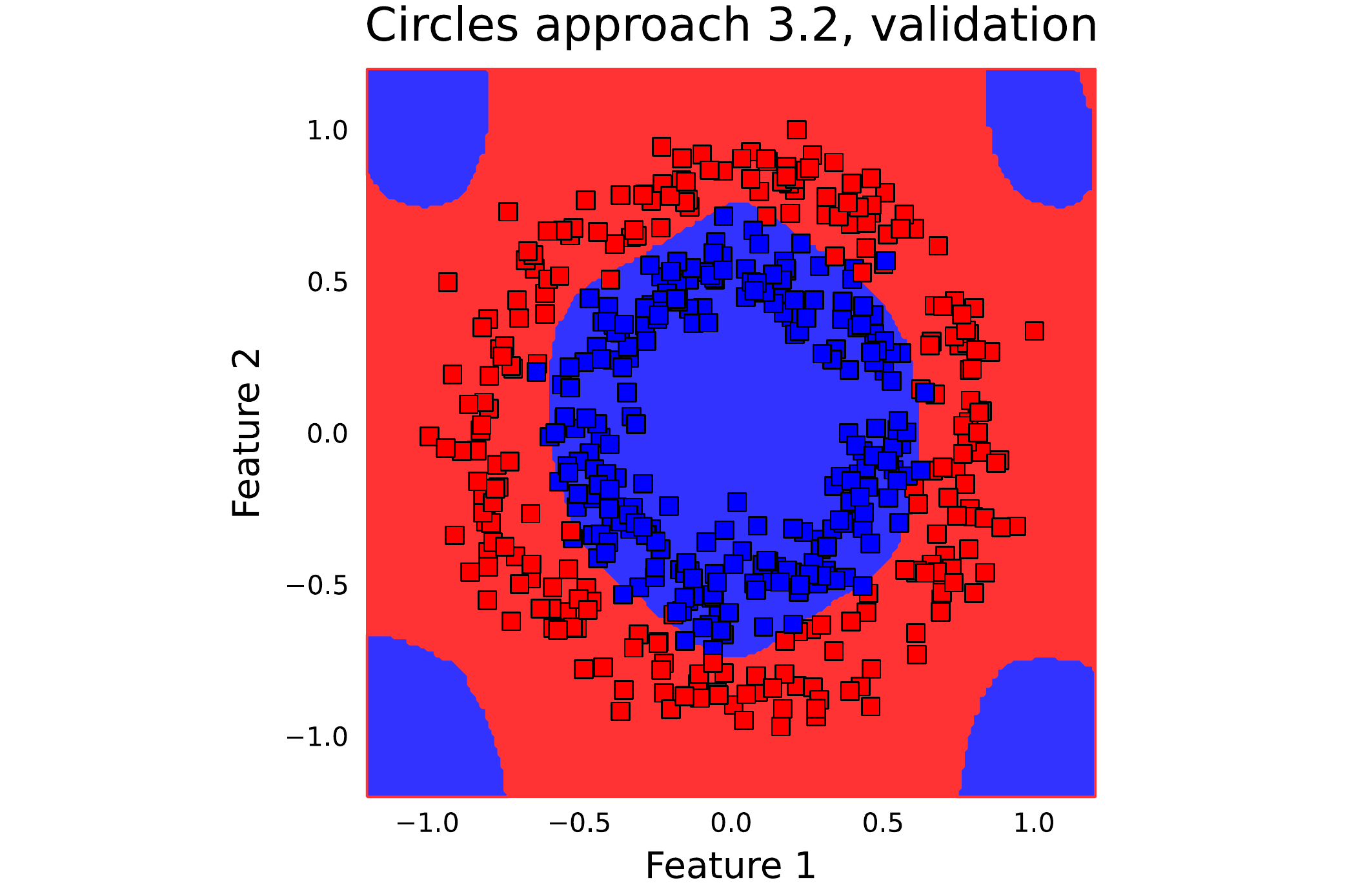}
\end{subfigure}
\end{adjustbox}
\caption{Decision boundaries for each approach on the Circles validation data.}
\end{figure}

\end{appendices}


\end{document}